\documentclass[prd,aps,nofootinbib,eqsecnum, superscriptaddress]{revtex4}

\usepackage[latin9]{inputenc}
\usepackage{amsmath,amssymb,bm}
\usepackage[colorlinks]{hyperref}
\usepackage[all]{hypcap}
\usepackage{graphicx,psfrag,float,epsfig}
\usepackage{mathrsfs,wasysym}
\usepackage{epstopdf}
\usepackage{comment}
\usepackage{pbox}
\usepackage{array}
\usepackage{xspace}
\usepackage[usenames,dvipsnames]{color}

\def\addUPitt{Pittsburgh Particle Physics Astrophysics and Cosmology Center, Department of Physics and Astronomy, University of Pittsburgh, Pittsburgh, PA 15260, USA}

\def\addCMU{Department of Physics, Carnegie Mellon University, Pittsburgh, PA 15213, USA}

\makeatother

\begin{document}

\title{Second post-Newtonian order radiative dynamics of inspiralling compact binaries in the Effective Field Theory approach}

\author{Adam K.~Leibovich}

\affiliation{\addUPitt}

\author{Nat\'alia T.~Maia}

\affiliation{\addUPitt}

\author{Ira Z.~Rothstein}

\affiliation{\addCMU}

\author{Zixin~Yang}

\affiliation{\addUPitt}
\begin{abstract}
We use the Effective Field Theory (EFT) framework to compute the mass quadrupole
moment, the equation of motion, and the power loss of inspiralling
compact binaries at the second order in the Post-Newtonian (PN) approximation.
We present expressions for the stress-energy pseudo-tensor components of the binary system in higher PN orders.
The 2PN correction to the mass quadrupole moment as well as to the acceleration computed in the linearized harmonic gauge presented here
are the  ingredients needed 
for the calculation of  the next-to-next-to leading order radiation reaction force, which will be presented elsewhere.  While this paper reproduces known results,  it supplies the building blocks necessary for future higher order calculations in the EFT methodology.
\end{abstract}
\maketitle


\section{Introduction}

The successful detections of gravitational waves by LIGO 
and Virgo \cite{Abbott:2016blz, 2016htt, 2016pea, PhysRevLett.118.221101, PhysRevLett.119.141101, PhysRevLett.119.161101, Abbott_2017, PhysRevX.9.031040} and the consequent advent of Multimessenger Astronomy \cite{Abbott_2017_mult, Abbott_2017_multi, Abbott_2019} have expedited the need for  precise  theoretical descriptions of the dynamics of binary inspirals. While numerical
techniques are required for the late stages of inspirals, 
the early stage admits a
perturbative treatment via the PN approximation, which is an expansion
in $v^{2}/c^{2}$, and   can be matched onto numerical results 
for later stages of the inspiral.
Generating higher order PN corrections will allow for more accurate parameter estimations. 

In this paper, we will utilize
the EFT approach called Non-Relativistic General Relativity (NRGR), proposed in \cite{nrgr}  (for reviews see~\cite{nrgrLH,Rothstein:2014sra,Foffa:2013qca,Porto:2016pyg,Levi:2018nxp}), as our calculation framework. 
To date, most of the results in the non-spinning sector of the EFT formalism have been geared towards the potential sector culminating in the present state of the art 4PN results \cite{Foffa:2019rdf,Foffa:2019yfl}, which 
agree with results previously derived using other methods \cite{Bini:2013zaa,Damour:2014jta,Bernard:2015njp,Bernard:2016wrg}. In the radiation sector, the EFT results have only\footnote{For spinning constituents the relevant multipole moments at 3PN for the flux \cite{Porto:2010zg} and 2.5 for the amplitude \cite{Porto:2012as}.} been calculated to 1PN \cite{Goldberger:2009qd} as compared to the 3PN results
calculated using more traditional GR methods \cite{Blanchet:2001ax}. Therefore, this paper is the next step in the calculation of higher order radiative effects in NRGR. In particular, in a separate paper we will use the results herein to calculate 
the next-to-next-to leading order radiation reaction force via the generation of an effective action.

The radiation sector of NRGR, the topic of this paper, was first
studied in \cite{nrgr}.  The effective action that
describes radiative effects is determined by the underlying symmetries
- reparameterization and diffeomorphism invariances - and is applicable
to arbitrary gravitational wave sources in the long wavelength approximation.
The Wilson coefficients of the action, the multipole moments, cannot
be determined by the symmetries and need to be fixed by a matching
procedure. The expression for the effective action to all orders in
the multipole expansion and the exact expressions of the multipole moments
in terms of the components of the stress-energy tensor were presented
in \cite{andirad2}. The NRGR framework provides a systematic
way to compute the multipole moments of a binary system by integrating
out the modes of the gravitational field that live in the near zone. 
The stress energy tensor, whose moments are our targeted goal,  is determined  by calculating the radiation graviton one point function
in the presence of the background potentials using Feynman diagrams.
The number of Feynman diagrams grows rapidly with PN order.

The  goal of this  paper is to determine the 2PN  correction
to the mass quadrupole moment, which comes from various moments of the  stress-energy pseudo-tensor. Each such contribution starts at different order in the PN expansion and
only a few of these contributions can be derived from known
quantities. We also derive the equation of motion of the binary system at 2PN order in the appendix
\ref{sec:AppB}.  Note that this acceleration was calculated previously in the EFT approach in \cite{nrgr2pn}, 
 where the authors worked with Kaluza-Klein variables \cite{KS} in conjunction 
with harmonic coordinates. The 2PN acceleration derived here, on the other hand, is written in the linearized (background) harmonic gauge,  which leaves a gauge invariant effective action  for the radiation field after the potential modes are integrated out, and can be used in combination with previous results obtained in the EFT approach where the linearized harmonic gauge was used.
Our results constitute the final missing part necessary for the computation of the next-to-next-to-leading order  radiation reaction force as well as for the construction of spinning templates at 2.5 PN for the phase and 3PN for the amplitude. These computations  are ongoing
and will be reported in a subsequent publication. 

This paper is organized as follows. In section \ref{sec:EFT-setup},
we provide a summary of NRGR for binary systems of compact bodies
with emphasis in the radiation sector, where we explicitly show how
the mass quadrupole moment depends on the components of the pseudotensor
in different PN orders. The contributions to the quadrupole that
come from higher PN order components of the pseudotensor are computed
in section \ref{sec:Higher-order}, while the contributions coming
from the lower PN order components are obtained in section \ref{sec:Lower-order}.
We use the results obtained in these sections to write down, in section
\ref{sec:Consistency-tests}, the components of the pseudotensor that
can be used to compute the multipole moments, which are  shown to agree
with the literature. The assembly of all contributions constitutes
the 2PN correction to the mass quadrupole moment, presented
in section \ref{sec:Mass-quadrupole}, in terms of the worldlines
of the compact bodies and also in the center of mass frame. In section
\ref{sec:Final-remarks} we present our final remarks on the results
presented in this paper. Appendix \ref{sec:AppA} is intended for readers interested in computing radiation effects in NRGR to higher orders. The necessary ingredients for the computation of the
higher PN order components of the pseudotensor are presented therein. In the appendix
\ref{sec:AppB} we show the result for the acceleration at 2PN order computed in
the linearized harmonic gauge, which is necessary  to
compare the quadrupole moment obtained
in this paper with there result in \cite{Blanchet:2001ax},  as well as to compute
the power loss at the second PN order. 

We use the following definitions throughout this paper: $m=m_{1}+m_{2}$,
$\nu\equiv m_{1}m_{2}/m^{2}$, and $\mu=m\nu$. The relative position
is defined as $\mathbf{r}\equiv\mathbf{x}_{1}-\mathbf{x}_{2}$, while $\mathbf{v}\equiv\mathbf{v}_{1}-\mathbf{v}_{2}$ and $\mathbf{a}\equiv\mathbf{a}_{1}-\mathbf{a}_{2}$ are
the relative velocity and acceleration, respectively. We adopt the
mostly minus signature convention for $\eta^{\alpha\beta}$ and Latin indices
are contracted with the Euclidean metric. We use $c=1$ units and
the Planck mass is defined as $m_{Pl}\equiv1/\sqrt{32\pi G}$.


\section{EFT setup\label{sec:EFT-setup}}

During the inspiral stage, the physics of a binary system of compact
bodies is naturally separated into three length scales: the typical
size of the bodies  of order of the Schwarzchild radius $r_{s}$,
the orbital distance between the two bodies given by $r$, and the
wavelength $\lambda_{GW}$ of the gravitational radiation. As the
relative velocity $v$ of the bodies is small, those three length
scales together constitute a hierarchical structure
\begin{equation}
r_{s}\ll r\ll\lambda_{GW}.
\end{equation}
 The first step is to ``integrate out" the scale associated with
the bodies size.\footnote{Finite size effects are accounted for by inserting higher-dimensional
operators in the effective action, respecting the symmetries of the
system.} Hence, the binary system can be initially described by the action
\begin{equation}
S=S_{EH}+S_{GF}+S_{pp},\label{eq:Sinitial}
\end{equation}
such that gravity is described by the Einstein-Hilbert (EH) action
$S_{EH}=-2m_{Pl}^{2}\int d^{4}x\sqrt{-g}g_{\mu\nu}R^{\mu\nu}$ with
a gauge fixing term $S_{GF}$, while the massive bodies are described
by the point particle action $S_{pp}=-\sum_{a}m_{a}\int d\tau_{a}.$
The index $a=1,2$ distinguishes the two bodies.

Next, the two different modes of
the gravitational field are separated in a diffeomorphism invariant way\footnote{Double counting subtleties arise at 4PN but can
be systematically disentangled \cite{Porto:2017dgs}.} via
\begin{equation}
g_{\mu \nu}=\eta_{\mu \nu}+h_{\mu\nu}\left(x\right)=\eta_{\mu \nu}+\bar{h}_{\mu\nu}\left(x\right)+H_{\mu\nu}\left(x\right).
\end{equation}

The off-shell potential 
mode  $H$  obeys $\partial_{0}H_{\mu\nu}\sim\left(\frac{v}{r}\right)H_{\mu\nu}$
and $\partial_{i}H_{\mu\nu}\sim\left(\frac{1}{r}\right)H_{\mu\nu}$
whereas the on-shell radiation mode obeys $\partial_{\alpha}\bar{h}_{\mu\nu}\sim\left(\frac{v}{r}\right)\bar{h}_{\mu\nu}.$
Moreover, the radiation field $\bar{h}_{\mu\nu}\left(x\right)$ has
to be Taylor expanded around a point inside the source (for instance
center of mass of the binary system) at the level of the action in
order to achieve a uniform power counting in the parameter $v^{2}\sim\frac{r_{s}}{r}$ \cite{benira}.
With these considerations, the action in \eqref{eq:Sinitial} is then
given as an expansion in the fields $\bar{h}_{\mu\nu}\left(x\right)$
and $H_{\mu\nu}\left(x\right)$, each of which scale homogeneously in  $v^{2}$.

To describe the dynamics associated to gravitational waves, the potential
mode of the gravitational field is integrated  leaving an effective action that will depend
only on the radiation field and the worldlines. This action will be diffeomorphism invariant  if one chooses  the linearized harmonic gauge when integrating out the potential field, via the gauge fixing action
\begin{equation}
S_{GF}=\int d^{4}x\sqrt{-\bar{g}}\bar{\Gamma}_{\mu}\bar{\Gamma}^{\mu},\label{eq:GF}
\end{equation}
where $\bar{\Gamma}_{\mu}=\bar{\nabla}_{\alpha}H_{\mu}^{\alpha}-\frac{1}{2}\bar{\nabla}_{\mu}H_{\alpha}^{\alpha},$
with $\bar{\nabla}_{\mu}$ representing the covariant derivative associated
to the background metric $\bar{g}_{\mu\nu}\left(x\right)=\eta_{\mu\nu}+\frac{\bar{h}_{\mu\nu}\left(x\right)}{m_{Pl}}$.

Moreover, as a result of the ``elimination'' of the degrees of freedom
that live in the orbital scale, the binary system is then regarded
as a single point particle coupled to its gravitational field and
whose internal dynamics is described by a set of multipole moments.
We present a brief review of the EFT radiation sector in the next
section.


\subsection{Radiation sector}

 The radiation action, which describes arbitrary gravitational
wave sources in the long wavelength approximation, can be written in a diffeomorpshim invariant way  in terms of  multipoles.
 Specifically, it is a derivative expansion where higher
order terms are suppressed by powers of the ratio between the size
of the binary system over the wavelength of the radiation emitted.
In the center-of-mass frame, the action of the radiation sector is
\cite{Goldberger:2009qd}
\begin{equation}
S_{rad}\left[\bar{h},x_{a}\right]=-\int dt\sqrt{\bar{g}_{00}}\left[m+\frac{1}{2}L_{ij}\omega_{0}^{ij}+\sum_{l=2}^{\infty}\left(\frac{1}{l!}I^{L}\nabla_{L-2}E_{i_{l-1}i_{l}}-\frac{2l}{\left(2l+1\right)!}J^{L}\nabla_{L-2}B_{i_{l-1}i_{l}}\right)\right],\label{eq:Srad}
\end{equation}
where a multi-index representation $L=i_{1}...i_{l}$ is used. The
first two terms generate the Kerr background in which the gravitational
waves propagate. The multipole moments, which constitute the source
of radiation, are coupled to the electric and the magnetic components
of the Weyl tensor.

To determine the moments, one performs a matching
between the effective action \eqref{eq:Srad} in the long wavelength
limit and the action valid below the orbital scale \eqref{eq:Sinitial},
which depends on both radiation and potential modes of the gravitational
field. The latter action is used in order to compute the one-graviton
emission amplitude.
As a result, by definition the resulting action takes the form
\begin{equation}
\Gamma\left[\bar{h}\right]=-\frac{1}{2m_{Pl}}\int d^{4}xT^{\mu\nu}\bar{h}_{\mu\nu},\label{eq:Sonegrav}
\end{equation}
where $T^{\mu\nu}$ is the stress-energy pseudotensor of the system.
Relations from the Ward
identity $\partial_{\mu}T^{\mu\nu}=0$ as well as the on-shell equations
of motion can be used to bring both actions \eqref{eq:Srad} and
\eqref{eq:Sonegrav} in a comparable form. After that, a general form
for the mass quadrupole moment is obtained in terms of the components
of the stress-energy pseudotensor and its derivatives,
\begin{align}
I^{ij} & =\sum_{p=0}^{\infty}\frac{5!!}{(2p)!!(5+2p)!!}\left\{ \left(1+\frac{2p(3+p)}{3}\right)\left[\int d^{3}\mathbf{x}\partial_{0}^{2p}T^{00}\mathbf{x}^{2p}\mathbf{x}^{i}\mathbf{x}^{j}\right]_{TF}\right.\nonumber \\
 & +\left(1+\frac{p}{3}\right)\left[\int d^{3}\mathbf{x}\partial_{0}^{2p}T^{ll}\mathbf{x}^{2p}\mathbf{x}^{i}\mathbf{x}^{j}\right]_{TF}-\frac{4}{3}\left(1+\frac{p}{2}\right)\left[\int d^{3}\mathbf{x}\partial_{0}^{2p+1}T^{0l}\mathbf{x}^{2p}\mathbf{x}^{l}\mathbf{x}^{i}\mathbf{x}^{j}\right]_{TF}\nonumber \\
 & \left.+\frac{1}{6}\left[\int d^{3}\mathbf{x}\partial_{0}^{2p+2}T^{kl}\mathbf{x}^{2p}\mathbf{x}^{k}\mathbf{x}^{l}\mathbf{x}^{i}\mathbf{x}^{j}\right]_{TF}\right\} ,
\end{align}
where TF stands for trace-free\footnote{More precisely, the multipole moments are symmetric trace-free (STF)
quantities, but we are supressing the ``S'' in the label to avoid
redundancy since the general expression for the quadrupole moment
is explicitly written as a symmetric tensor already.}. For the exact expressions for the multipole moments in all orders in the PN expansion, see \cite{andirad2}. The leading-order contribution
to the mass quadrupole moment comes from just one term
\begin{align}
I_{0PN}^{ij} & =\left[\int d^{3}\mathbf{x}T_{0PN}^{00}\mathbf{x}^{i}\mathbf{x}^{j}\right]_{TF}=\sum_{a} m_{a}\left[\mathbf{x}_{a}^{i}\mathbf{x}_{a}^{j}\right]_{TF},\label{eq:mq0PN}
\end{align}
while its 1PN correction \cite{Goldberger:2009qd} is given by four different contributions
of the components of the stress-energy pseudotensor:
\begin{align}
I_{1PN}^{ij} & =\left[\int d^{3}\mathbf{x}T_{1PN}^{00}\mathbf{x}^{i}\mathbf{x}^{j}\right]_{TF}+\left[\int d^{3}\mathbf{x}T_{0PN}^{ll}\mathbf{x}^{i}\mathbf{x}^{j}\right]_{TF}\nonumber \\
 & -\frac{4}{3}\left[\int d^{3}\mathbf{x}\partial_{0}T_{0PN}^{0l}\mathbf{x}^{l}\mathbf{x}^{i}\mathbf{x}^{j}\right]_{TF}+\frac{11}{42}\left[\int d^{3}\mathbf{x}\partial_{0}^{2}T_{0PN}^{00}\mathbf{x}^{2}\mathbf{x}^{i}\mathbf{x}^{j}\right]_{TF}\nonumber \\
 & =\sum_{a}m_{a}\left[\left(\frac{3}{2}\mathbf{v}_{a}^{2}-\sum_{b\neq a}\frac{Gm_{b}}{r}\right)\mathbf{x}_{a}^{i}\mathbf{x}_{a}^{j}+\frac{11}{42}\frac{d^{2}}{dt^{2}}\left(\mathbf{x}_{a}^{2}\mathbf{x}_{a}^{i}\mathbf{x}_{a}^{j}\right)-\frac{4}{3}\frac{d}{dt}\left(\mathbf{x}_{a}\cdot\mathbf{v}_{a}\mathbf{x}_{a}^{i}\mathbf{x}_{a}^{j}\right)\right]_{TF}.\label{eq:mq1PN}
\end{align}

The 2PN correction to the leading order mass quadrupole moment is given by 
\begin{align}
I_{2PN}^{ij} & =\left[\int d^{3}\mathbf{x}T_{2PN}^{00}\mathbf{x}^{i}\mathbf{x}^{j}\right]_{TF}+\left[\int d^{3}\mathbf{x}T_{1PN}^{ll}\mathbf{x}^{i}\mathbf{x}^{j}\right]_{TF}-\frac{4}{3}\left[\int d^{3}\mathbf{x}\partial_{0}T_{1PN}^{0l}\mathbf{x}^{l}\mathbf{x}^{i}\mathbf{x}^{j}\right]_{TF}\nonumber \\
 & +\frac{1}{6}\left[\int d^{3}\mathbf{x}\partial_{0}^{2}T_{0PN}^{kl}\mathbf{x}^{k}\mathbf{x}^{l}\mathbf{x}^{i}\mathbf{x}^{j}\right]_{TF}+\frac{11}{42}\left[\int d^{3}\mathbf{x}\partial_{0}^{2}T_{1PN}^{00}\mathbf{x}^{2}\mathbf{x}^{i}\mathbf{x}^{j}\right]_{TF}\nonumber \\
 & +\frac{2}{21}\left[\int d^{3}\mathbf{x}\partial_{0}^{2}T_{0PN}^{ll}\mathbf{x}^{2}\mathbf{x}^{i}\mathbf{x}^{j}\right]_{TF}-\frac{1}{7}\left[\int d^{3}\mathbf{x}\partial_{0}^{3}T_{0PN}^{0l}\mathbf{x}^{2}\mathbf{x}^{l}\mathbf{x}^{i}\mathbf{x}^{j}\right]_{TF}\nonumber \\
 & +\frac{23}{1512}\left[\int d^{3}\mathbf{x}\partial_{0}^{4}T_{0PN}^{00}\mathbf{x}^{4}\mathbf{x}^{i}\mathbf{x}^{j}\right]_{TF}+I_{1PN}^{ij}\left(\mathbf{a}_{1PN}\right).\label{eq:I2}
\end{align}
Notice that the last term in the expression above arises from
two terms in the second line of \eqref{eq:mq1PN} after using  the equations of motion. While $T_{0PN}^{00}$ and $T_{0PN}^{0l}$
are trivial, the higher PN order components $T_{2PN}^{00}$, $T_{1PN}^{0i}$,
$T_{1PN}^{ll}$ have yet to be obtained in the EFT formalism.


\section{Higher order stress-energy tensors\label{sec:Higher-order}}

Introducing the partial Fourier transform of the stress-energy pseudotensor
$T^{\mu\nu}\left(t,\mathbf{k}\right)=\int d^{3}xT^{\mu\nu}\left(t,\mathbf{x}\right)e^{-i\mathbf{k}\cdot\mathbf{x}}$,
we consider the long wavelength limit $\mathbf{k}\to0$ to write
\begin{equation}
T^{\mu\nu}\left(t,\mathbf{k}\right)=\sum_{n=0}^{\infty}\frac{\left(-i\right)^{n}}{n!}\left(\int d^{3}\mathbf{x}T^{\mu\nu}\left(t,\mathbf{x}\right)\mathbf{x}^{i_{1}}...\mathbf{x}^{i_{n}}\right)\mathbf{k}_{i_{1}}...\mathbf{k}_{i_{n}},\label{eq:Texp}
\end{equation}
where each term in this expansion corresponds to a sum of Feynman
diagrams that scale as a definite power of the parameter $v$. This
partial Fourier transform is convenient since Feynman graphs are more
easily handled in momentum space and, with the pseudotensor written
in this way, we can read off the contributions to the mass quadrupole
moment \eqref{eq:I2}, the ultimate goal of this paper.

\subsection{2PN correction to $T^{00}$\label{subsec:2PN-T00}}

The leading order and the next-to-leading order temporal components
of the pseudotensor, obtained in \cite{Goldberger:2009qd} using the EFT techniques
summarized in the previous section, are given by
\begin{align}
T_{0PN}^{00}\left(t,\mathbf{k}\right) & =\sum_{a}m_{a}e^{-i\mathbf{k}\cdot\mathbf{x}_{a}},\label{eq:T00LO}\\
T_{1PN}^{00}\left(t,\mathbf{k}\right) & =\left[\sum_{a}\frac{1}{2}m_{a}\mathbf{v}_{a}^{2}-\sum_{a\neq b}\frac{Gm_{a}m_{b}}{2r}+O\left(\mathbf{k}\right)+...\right]e^{-i\mathbf{k}\cdot\mathbf{x}_{a}}.\label{eq:T00NLO}
\end{align}
If we take into account the zeroth order term of the exponential expanded
in the radiation momentum $\mathbf{k}$, we see that the leading order
pseudotensor provides the total mass whereas the next-to-leading order
represents the Newtonian energy of a dynamical two-body system. These
quantities scale as $mv^{0}$ and $mv^{2}$, respectively. Hence,
to obtain the 2PN correction to the leading order $T^{00}$, we have
to calculate all Feynman diagrams that contribute to the one-graviton
$\bar{h}_{00}$ emission and enter at  order $v^{4}$. 
\begin{figure}[H]
\label{T00_fig1}
\begin{centering}
\includegraphics[scale=0.35]{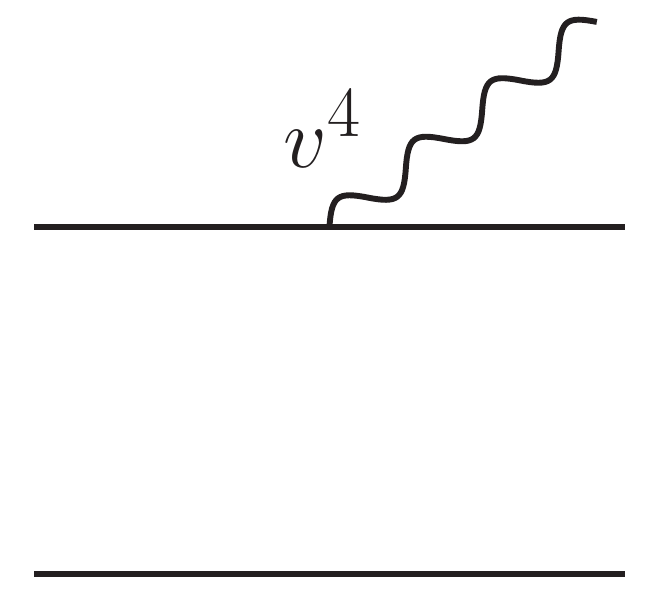}
\par\end{centering}
\caption{No graviton exchange between the two particles, one external $\bar{h}^{00}$
momentum.}
\end{figure}

The simplest contribution to the second PN correction for the temporal
component of the stress-energy pseudotensor is illustrated in Fig.~\ref{T00_fig1}
and comes from the source action term \eqref{eq:pph00v4}. Comparing
this diagram against \eqref{eq:Sonegrav}, we extract the following
contribution to the pseudotensor
\begin{equation}
T_{Fig1}^{00}\left(t,\mathbf{k}\right)=\sum_{a}\frac{3}{8}m_{a}\mathbf{v}_{a}^{4}e^{-i\mathbf{k}\cdot\mathbf{x}_{a}}.
\end{equation}
By expanding the exponential up to the second order in the radiation
momentum $\mathbf{k}$, we read off the contribution for the mass
quadrupole moment:

\begin{equation}
\int d^{3}\mathbf{x}T_{Fig1}^{00}\left[\mathbf{x}^{i}\mathbf{x}^{j}\right]_{TF}=\sum_{a}\frac{3}{8}m_{a}\mathbf{v}_{a}^{4}\left[\mathbf{x}_{a}^{i}\mathbf{x}_{a}^{j}\right]_{TF}.\label{eq:1}
\end{equation}

\begin{figure}[H]
\label{T00_fig2a-2d}
\begin{centering}
\includegraphics[scale=0.35]{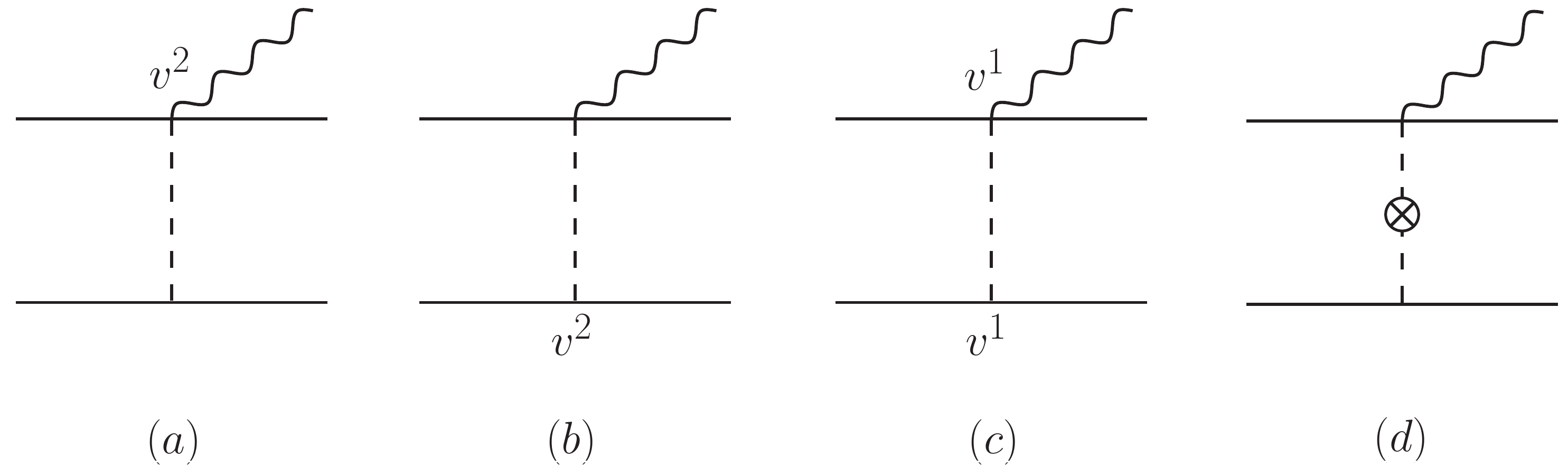}
\par\end{centering}
\caption{One-graviton exchange with external $\bar{h}^{00}$ momentum.}

\end{figure}

The diagrams that contain the exchange of one potential graviton
are shown in Fig.~\ref{T00_fig2a-2d} and are composed by the couplings between
the source action terms (\ref{eq:ppH00v0}-\ref{eq:pph00v2}) and
also the propagator \eqref{eq:prop} and its correction \eqref{eq:corrprop}. 
Notice that we need not separate out all of the various terms that arise in the Feynman rules into different orders in the PN expansion
as is done in the appendix. We also calculated covariantly vertices, as is done when calculating in the Post-Minkowskian (PM) expansion (see e.g. \cite{Cheung:2018wkq}), 
and then expand in $v$, as a calculational check. However, for
pedagogical purposes we have separated Feynman rules into give orders in the PN expansion.
The results from  Fig.~\ref{T00_fig2a-2d} is given by
\begin{align}
T_{Fig2a}^{00}\left(t,\mathbf{k}\right) & =\sum_{a\neq b}\frac{5}{2}\frac{Gm_{a}m_{b}}{r}\mathbf{v}_{a}^{2}e^{-i\mathbf{k}\cdot\mathbf{x}_{a}},\\
T_{Fig2b}^{00}\left(t,\mathbf{k}\right) & =\sum_{a\neq b}\frac{3}{2}\frac{Gm_{a}m_{b}}{r}\mathbf{v}_{b}^{2}e^{-i\mathbf{k}\cdot\mathbf{x}_{a}},\\
T_{Fig2c}^{00}\left(t,\mathbf{k}\right) & =-\sum_{a\neq b}4\frac{Gm_{a}m_{b}}{r}\mathbf{v}_{a}\cdot\mathbf{v}_{b}e^{-i\mathbf{k}\cdot\mathbf{x}_{a}},\\
T_{Fig2d}^{00}\left(t,\mathbf{k}\right) & =\sum_{a\neq b}\frac{Gm_{a}m_{b}}{2r}\left(-\mathbf{a}_{b}^{i}\mathbf{r}^{i}+\mathbf{v}_{b}^{2}-\left(\mathbf{v}_{b}\cdot\mathbf{n}\right)^{2}\right)e^{-i\mathbf{k}\cdot\mathbf{x}_{a}}.
\end{align}
Leaving
\begin{align}
\int d^{3}\mathbf{x}T_{Fig2a-2d}^{00}\left[\mathbf{x}^{i}\mathbf{x}^{j}\right]_{TF} & =\sum_{a\neq b}\frac{Gm_{a}m_{b}}{2r}\left[\left(5\mathbf{v}_{a}^{2}+4\mathbf{v}_{b}^{2}-8\mathbf{v}_{a}\cdot\mathbf{v}_{b}-\mathbf{a}_{b}\cdot\mathbf{r}-\left(\mathbf{v}_{b}\cdot\mathbf{n}\right)^{2}\right)\mathbf{x}_{a}^{i}\mathbf{x}_{a}^{j}\right]_{TF}.\label{eq:2a-2d}
\end{align}
Note the implicit dependence on the indices $a,b$ in the quantities
$\mathbf{r}=\mathbf{x}_{a}-\mathbf{x}_{b}$, $r=\left|\mathbf{r}\right|$
and $\mathbf{n}=\frac{\mathbf{r}}{r}$ inside the sum.

\begin{figure}[H]\label{T00_fig3a-3e}
\begin{centering}
\includegraphics[scale=0.35]{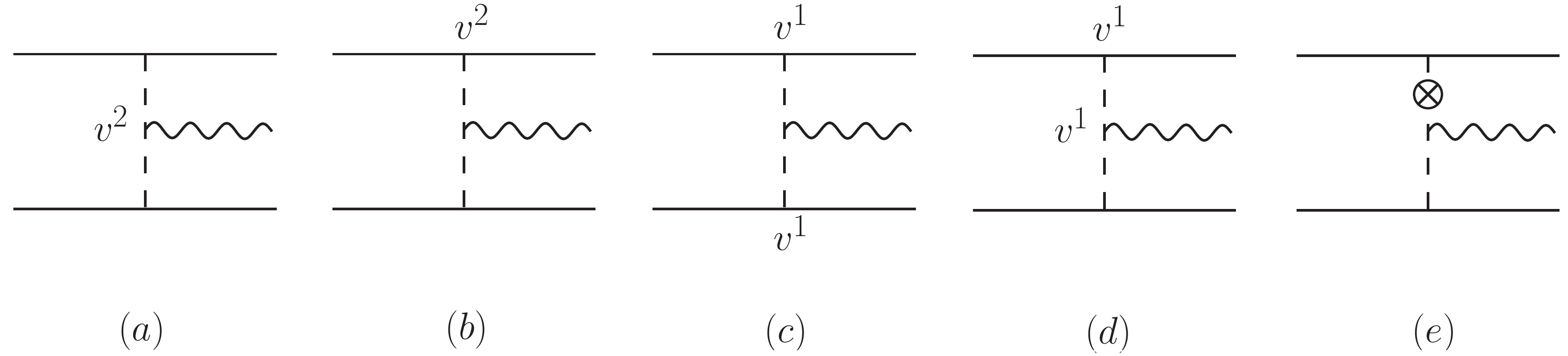}
\par\end{centering}
\caption{Diagrams with two potential gravitons coupled to $\bar{h}_{00}$. }
\end{figure}

The graphs in Fig.~\ref{T00_fig3a-3e} are composed by the source terms (\ref{eq:ppH00v0}-\ref{eq:ppH00Hijv2})
together with the vertices (\ref{eq:hHH}-\ref{eq:hHH3}) and \eqref{eq:corrprop}.
Note that we multipole expand the denominators in $\mathbf{k}/\mathbf{q}\sim v$ 
\begin{align}
\frac{1}{\mathbf{q}^{2}\left(\mathbf{q}+\mathbf{k}\right)^{2}} & =\frac{1}{\mathbf{q}^{4}}-\frac{2\left(\mathbf{q}\cdot\mathbf{k}\right)}{\mathbf{q}^{6}}+\frac{4\left(\mathbf{q}\cdot\mathbf{k}\right)^{2}}{\mathbf{q}^{8}}+...,
\end{align}
In calculating the  contributions to the mass quadrupole sourced
by the temporal components of the pseudotensor at 2PN, we are allowed to drop
terms depending on $\mathbf{k}^{2}$ in the expansion
of the denominator, since those terms contribute to the trace part
of the mass quadrupole, which is removed in the definition
of the STF moment. The results are organized
in orders of the radiation momentum, as it is shown below:
\begin{align}
T_{Fig3a}^{00}\left(t,\mathbf{k}\right) & =\sum_{a\neq b}\frac{Gm_{a}m_{b}}{4r}e^{-i\mathbf{k}\cdot\mathbf{x}_{a}}\left\{ 2\left(\mathbf{v}^{2}+\mathbf{a}\cdot\mathbf{r}-\dot{r}^{2}\right)+5\mathbf{v}_{a}\cdot\mathbf{v}_{b}-5\mathbf{v}_{a}\cdot\mathbf{n}\mathbf{v}_{b}\cdot\mathbf{n}\right.\nonumber \\
 & +i\mathbf{k}^{i}\left[\left(\mathbf{v}^{2}+\mathbf{a}\cdot\mathbf{r}-\dot{r}^{2}+\frac{5}{2}\mathbf{v}_{a}\cdot\mathbf{v}_{b}-\frac{5}{2}\mathbf{v}_{a}\cdot\mathbf{n}\mathbf{v}_{b}\cdot\mathbf{n}\right)\mathbf{r}^{i}\right.\nonumber \\
 & \left.+\left(\frac{1}{2}r\dot{r}+\frac{5}{2}\mathbf{v}_{b}\cdot\mathbf{r}\right)\mathbf{v}_{b}^{i}-\left(2r\dot{r}+\frac{5}{2}\mathbf{v}_{b}\cdot\mathbf{r}\right)\mathbf{v}_{a}^{i}-r^{2}\left(\mathbf{a}_{a}^{i}+\mathbf{a}_{b}^{i}\right)\right]\nonumber \\
 & +\frac{1}{6}\mathbf{k}^{i}\mathbf{k}^{j}\left[-\left(2\mathbf{v}^{2}+5\mathbf{v}_{a}\cdot\mathbf{v}_{b}-2\dot{r}^{2}-5\mathbf{v}_{a}\cdot\mathbf{n}\mathbf{v}_{b}\cdot\mathbf{n}+2\mathbf{a}\cdot\mathbf{r}\right)\mathbf{r}^{i}\mathbf{r}^{j}\right.\nonumber \\
 & +\left(4\mathbf{v}_{a}\cdot\mathbf{r}+\mathbf{v}_{b}\cdot\mathbf{r}\right)\mathbf{v}_{a}^{i}\mathbf{r}^{j}-\left(2\mathbf{v}_{a}\cdot\mathbf{r}+8\mathbf{v}_{b}\cdot\mathbf{r}\right)\mathbf{v}_{b}^{i}\mathbf{r}^{j}\nonumber \\
 & \left.\left.+r^{2}\left(-4\mathbf{v}^{i}\mathbf{v}^{j}-7\mathbf{v}_{a}^{i}\mathbf{v}_{b}^{j}+2\mathbf{a}_{a}^{i}\mathbf{r}^{j}+4\mathbf{a}_{b}^{i}\mathbf{r}^{j}\right)\right]\right\} +O\left(\mathbf{k}^{3}\right)+...,
\end{align}
\begin{align}
T_{Fig3b}^{00}\left(t,\mathbf{k}\right) & =-\sum_{a\neq b}\frac{Gm_{a}m_{b}}{r}e^{-i\mathbf{k}\cdot\mathbf{x}_{a}}\left[\frac{7}{4}\mathbf{v}_{b}^{2}+\frac{3}{4}\mathbf{v}_{a}^{2}-\frac{i}{2}\mathbf{k}^{i}\mathbf{v}_{b}^{2}\mathbf{r}^{i}\right.\nonumber \\
 & \left.+\frac{1}{2}\mathbf{k}^{i}\mathbf{k}^{j}\left(\frac{1}{2}\mathbf{v}_{b}^{2}\mathbf{r}^{i}\mathbf{r}^{j}+2\mathbf{r}^{2}\mathbf{v}_{b}^{i}\mathbf{v}_{b}^{j}\right)\right]+O\left(\mathbf{k}^{3}\right)+...,
\end{align}
\begin{align}
T_{Fig3c}^{00}\left(t,\mathbf{k}\right) & =\sum_{a\neq b}\frac{2Gm_{a}m_{b}}{r}e^{-i\mathbf{k}\cdot\mathbf{x}_{a}}\left(2\mathbf{v}_{a}\cdot\mathbf{v}_{b}+\mathbf{k}^{i}\mathbf{k}^{j}\mathbf{r}^{2}\mathbf{v}_{a}^{i}\mathbf{v}_{b}^{j}\right)+O\left(\mathbf{k}^{3}\right)+...,
\end{align}
\begin{align}
T_{Fig3d}^{00}\left(t,\mathbf{k}\right) & =-\sum_{a\neq b}\frac{2Gm_{a}m_{b}}{r}e^{-i\mathbf{k}\cdot\mathbf{x}_{a}}\left\{ -\frac{i}{2}\mathbf{k}^{i}\left[2\mathbf{r}^{2}\mathbf{a}_{a}^{i}+2\mathbf{v}_{a}^{i}\left(\mathbf{v}_{b}\cdot\mathbf{r}+\mathbf{v}_{a}\cdot\mathbf{r}\right)\right]\right.\nonumber \\
 & \left.-\frac{1}{2}\mathbf{k}^{i}\mathbf{k}^{j}\left[\mathbf{r}^{2}\left(\mathbf{v}_{a}^{i}\mathbf{v}_{a}^{j}-\mathbf{v}_{a}^{i}\mathbf{v}_{b}^{i}-\mathbf{r}^{j}\mathbf{a}_{a}^{i}\right)-\mathbf{v}_{a}^{i}\mathbf{r}^{j}\left(\mathbf{v}_{a}\cdot\mathbf{r}+\mathbf{v}_{b}\cdot\mathbf{r}\right)\right]\right\} +O\left(\mathbf{k}^{3}\right)+...,
\end{align}
\begin{align}
T_{Fig3e}^{00}\left(t,\mathbf{k}\right) & =-\sum_{a\neq b}\frac{Gm_{a}m_{b}}{4r}e^{-i\mathbf{k}\cdot\mathbf{x}_{a}}\left\{ 6\left(-\mathbf{a}_{b}\cdot\mathbf{r}+\mathbf{v}_{b}^{2}-\left(\mathbf{v}_{b}\cdot\mathbf{n}\right)^{2}\right)\right.\nonumber \\
 & -\frac{3i}{2}\mathbf{k}^{i}\left[\left(\mathbf{a}_{b}\cdot\mathbf{r}-\mathbf{v}_{b}^{2}+\left(\mathbf{v}_{b}\cdot\mathbf{n}\right)^{2}\right)\mathbf{r}^{i}-2\mathbf{v}_{b}\cdot\mathbf{r}\mathbf{v}_{b}^{i}+\mathbf{r}^{2}\mathbf{a}_{b}^{i}\right]\nonumber \\
 & -\frac{1}{2}\mathbf{k}^{i}\mathbf{k}^{j}\left[\left(-\mathbf{a}_{b}\cdot\mathbf{r}+\mathbf{v}_{b}^{2}-\left(\mathbf{v}_{b}\cdot\mathbf{n}\right)^{2}\right)\mathbf{r}^{i}\mathbf{r}^{j}+4\mathbf{v}_{b}\cdot\mathbf{r}\mathbf{v}_{b}^{i}\mathbf{r}^{j}\right.\nonumber \\
 & \left.\left.-2\mathbf{r}^{2}\mathbf{a}_{b}^{i}\mathbf{r}^{j}+2\mathbf{r}^{2}\mathbf{v}_{b}^{i}\mathbf{v}_{b}^{j}\right]\right\} +O\left(\mathbf{k}^{3}\right)+...\cdot
\end{align}
Together, these quantities provide us with the following contribution,
\begin{align}
\int d^{3}\mathbf{x}T_{Fig3a-3e}^{00}\left[\mathbf{x}^{i}\mathbf{x}^{j}\right]_{TF} & =\sum_{a\neq b}\frac{Gm_{a}m_{b}}{12r}\left[\left(-2\mathbf{v}_{a}^{2}-35\mathbf{v}_{b}^{2}+26\mathbf{v}_{a}\cdot\mathbf{v}_{b}-10\mathbf{v}_{a}\cdot\mathbf{n}\mathbf{v}_{b}\cdot\mathbf{n}\right.\right.\nonumber \\
 & \left.+3\left(\mathbf{v}_{a}\cdot\mathbf{n}\right)^{2}+12\left(\mathbf{v}_{b}\cdot\mathbf{n}\right)^{2}-4\dot{r}^{2}+\mathbf{a}_{a}\cdot\mathbf{r}+8\mathbf{a}_{b}\cdot\mathbf{r}\right)\mathbf{x}_{a}^{i}\mathbf{x}_{a}^{j}\nonumber \\
 & +\left(\mathbf{v}_{a}^{2}+\mathbf{v}_{a}\cdot\mathbf{v}_{b}-5\mathbf{v}_{a}\cdot\mathbf{n}\mathbf{v}_{b}\cdot\mathbf{n}+3\left(\mathbf{v}_{a}\cdot\mathbf{n}\right)^{2}-2\dot{r}^{2}+\mathbf{a}_{a}\cdot\mathbf{r}\right)\mathbf{x}_{a}^{i}\mathbf{x}_{b}^{j}\nonumber \\
 & +\left(\mathbf{v}_{a}\cdot\mathbf{r}+\mathbf{v}_{b}\cdot\mathbf{r}\right)\left(-20\mathbf{v}_{a}^{i}\mathbf{x}_{a}^{j}+26\mathbf{v}_{a}^{i}\mathbf{x}_{b}^{j}\right)\nonumber \\
 & \left.+\mathbf{r}^{2}\left(2\mathbf{v}_{a}^{i}\mathbf{v}_{a}^{j}-\mathbf{v}_{a}^{i}\mathbf{v}_{b}^{j}-22\mathbf{a}_{a}^{i}\mathbf{x}_{a}^{j}-23\mathbf{a}_{a}^{i}\mathbf{x}_{b}^{j}\right)\right]_{STF}.\label{eq:3a-3e}
\end{align}

\begin{figure}[H]
\label{T00_fig4a-4c}
\begin{centering}
\includegraphics[scale=0.35]{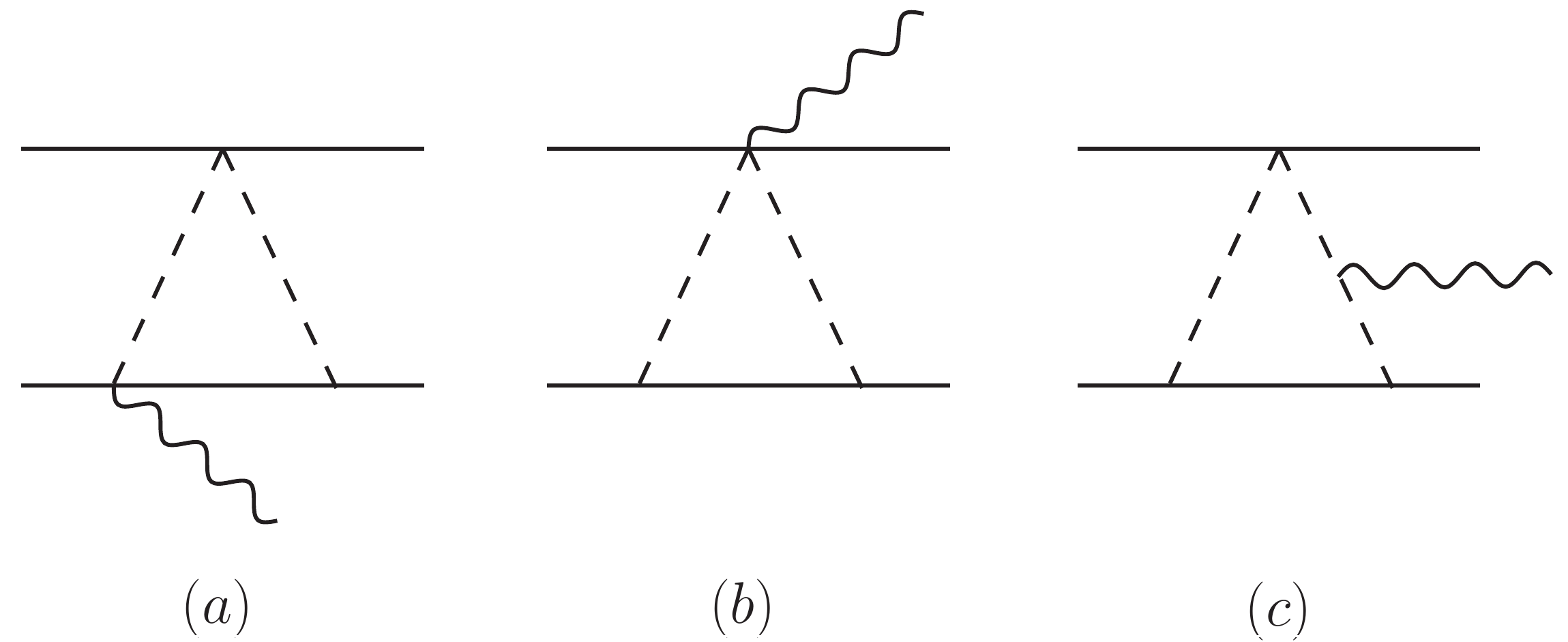}
\par\end{centering}
\caption{Two-potential-graviton exchange with external $\bar{h}^{00}$ momentum.}
\end{figure}

  Contributions from Fig.~\ref{T00_fig4a-4c}  are composed of the source terms \eqref{eq:ppH00v0},
\eqref{eq:ppH00h00}, \eqref{eq:ppHH} and \eqref{eq:ppHHh} and yield
\begin{align}
T_{Fig4a}^{00}\left(t,\mathbf{k}\right) & =\sum_{a\neq b}\frac{G^{2}m_{a}^{2}m_{b}}{r^{2}}e^{-i\mathbf{k}\cdot\mathbf{x}_{a}},\\
T_{Fig4b}^{00}\left(t,\mathbf{k}\right) & =\sum_{a\neq b}\frac{3G^{2}m_{a}m_{b}^{2}}{2r^{2}}e^{-i\mathbf{k}\cdot\mathbf{x}_{a}},\\
T_{Fig4c}^{00}\left(t,\mathbf{k}\right) & =-\sum_{a\neq b}\frac{3G^{2}m_{a}m_{b}m}{2r^{2}}e^{-i\mathbf{k}\cdot\mathbf{x}_{a}},
\end{align}
which gives us
\begin{equation}
\int d^{3}\mathbf{x}T_{Fig4a-4c}^{00}\left[\mathbf{x}^{i}\mathbf{x}^{j}\right]_{TF}=-\sum_{a\neq b}\frac{G^{2}m_{a}^{2}m_{b}}{2r^{2}}\left[\mathbf{x}_{a}^{i}\mathbf{x}_{a}^{j}\right]_{TF}.\label{eq:4a-4c}
\end{equation}

\begin{figure}[H]
\label{T00_fig5a-5e}
\begin{centering}
\includegraphics[scale=0.35]{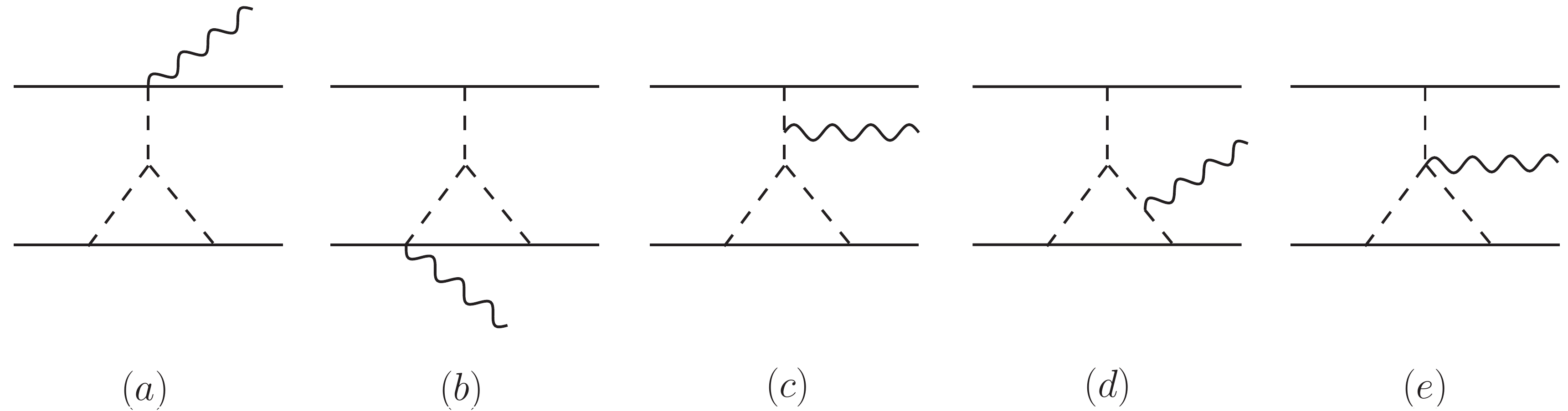}
\par\end{centering}
\caption{Three-potential-graviton exchange with external $\bar{h}^{00}$ momentum.}
\end{figure}

The diagrams illustrated in Fig.~\ref{T00_fig5a-5e}, are composed of 
the three-potential-graviton vertices (\ref{eq:3pot}-\ref{eq:comp00})
as well as the three-potential-one-radiation-graviton vertex (\ref{eq:4grav}-\ref{eq:4grav00})
in composition with \eqref{eq:ppH00v0} and \eqref{eq:ppH00h00} 
 contribute to $T_{2PN}^{00}$. These diagrams give:%
\begin{align}
T_{Fig5a}^{00}\left(t,\mathbf{k}\right) & =-\sum_{a\neq b}\frac{G^{2}m_{a}m_{b}^{2}}{r^{2}}e^{-i\mathbf{k}\cdot\mathbf{x}_{a}},\\
T_{Fig5b}^{00}\left(t,\mathbf{k}\right) & =-\sum_{a\neq b}\frac{2G^{2}m_{a}^{2}m_{b}}{r^{2}}e^{-i\mathbf{k}\cdot\mathbf{x}_{a}},\\
T_{Fig5c}^{00}\left(t,\mathbf{k}\right) & =-\sum_{a\neq b}\frac{G^{2}m_{a}^{2}m_{b}}{r^{2}}e^{-i\mathbf{k}\cdot\mathbf{x}_{a}}\left(\frac{1}{2}-\frac{7}{2}i\mathbf{k}^{i}\mathbf{r}^{i}+\frac{5}{3}\mathbf{k}^{i}\mathbf{k}^{j}\mathbf{r}^{i}\mathbf{r}^{j}\right)+O\left(\mathbf{k}^{3}\right)+...,\\
T_{Fig5d}^{00}\left(t,\mathbf{k}\right) & =\sum_{a\neq b}\frac{G^{2}m_{a}^{2}m_{b}}{r^{2}}e^{-i\mathbf{k}\cdot\mathbf{x}_{a}}\left(5-2i\mathbf{k}^{i}\mathbf{r}^{i}+\frac{2}{3}\mathbf{k}^{i}\mathbf{k}^{j}\mathbf{r}^{i}\mathbf{r}^{j}\right)+O\left(\mathbf{k}^{3}\right)+...,\\
T_{Fig5e}^{00}\left(t,\mathbf{k}\right) & =-\sum_{a\neq b}\frac{G^{2}m_{a}m_{b}^{2}}{2r^{2}}e^{-i\mathbf{k}\cdot\mathbf{x}_{a}}.
\end{align}
Keeping terms to second order in the
radiation momentum we have
\begin{equation}
\int d^{3}\mathbf{x}T_{Fig5a-5e}^{00}\left[\mathbf{x}^{i}\mathbf{x}^{j}\right]_{TF}=\sum_{a\neq b}\frac{G^{2}m_{a}m_{b}}{r^{2}}\left[\frac{3}{2}\left(m_{a}-m_{b}\right)\mathbf{x}_{a}^{i}\mathbf{x}_{a}^{j}-m_{a}\mathbf{x}_{a}^{i}\mathbf{x}_{b}^{j}+2m_{a}\mathbf{x}_{b}^{i}\mathbf{x}_{b}^{j}\right]_{TF}.\label{eq:5a-5e}
\end{equation}
Summing the contributions \eqref{eq:1}, \eqref{eq:2a-2d},
\eqref{eq:3a-3e}, \eqref{eq:4a-4c} and \eqref{eq:5a-5e},  the total contribution of $T_{2PN}^{00}$ to the mass quadrupole
is
\begin{align}
\int d^{3}\mathbf{x}T_{2PN}^{00}\left[\mathbf{x}^{i}\mathbf{x}^{j}\right]_{TF} & =\sum_{a}\frac{3}{8}m_{a}\mathbf{v}_{a}^{4}\left[\mathbf{x}_{a}^{i}\mathbf{x}_{a}^{j}\right]_{TF}+\sum_{a\neq b}\frac{Gm_{a}m_{b}}{12r}\left[\left(28\mathbf{v}_{a}^{2}-11\mathbf{v}_{b}^{2}-22\mathbf{v}_{a}\cdot\mathbf{v}_{b}-10\mathbf{v}_{a}\cdot\mathbf{n}\mathbf{v}_{b}\cdot\mathbf{n}\right.\right.\nonumber \\
 & \left.+3\left(\mathbf{v}_{a}\cdot\mathbf{n}\right)^{2}+6\left(\mathbf{v}_{b}\cdot\mathbf{n}\right)^{2}-4\dot{r}^{2}+\mathbf{a}_{a}\cdot\mathbf{r}+2\mathbf{a}_{b}\cdot\mathbf{r}+12\frac{Gm_{a}}{r}+6\frac{Gm_{b}}{r}\right)\mathbf{x}_{a}^{i}\mathbf{x}_{a}^{j}\nonumber \\
 & +\left(\mathbf{v}_{a}^{2}+\mathbf{v}_{a}\cdot\mathbf{v}_{b}-5\mathbf{v}_{a}\cdot\mathbf{n}\mathbf{v}_{b}\cdot\mathbf{n}+3\left(\mathbf{v}_{a}\cdot\mathbf{n}\right)^{2}-2\dot{r}^{2}+\mathbf{a}_{a}\cdot\mathbf{r}-12\frac{Gm_{a}}{r}\right)\mathbf{x}_{a}^{i}\mathbf{x}_{b}^{j}\nonumber \\
 & \left.+\left(\mathbf{v}_{a}\cdot\mathbf{r}+\mathbf{v}_{b}\cdot\mathbf{r}\right)\left(-20\mathbf{v}_{a}^{i}\mathbf{x}_{a}^{j}+26\mathbf{v}_{a}^{i}\mathbf{x}_{b}^{j}\right)+\mathbf{r}^{2}\left(2\mathbf{v}_{a}^{i}\mathbf{v}_{a}^{j}-\mathbf{v}_{a}^{i}\mathbf{v}_{b}^{j}-22\mathbf{a}_{a}^{i}\mathbf{x}_{a}^{j}-23\mathbf{a}_{a}^{i}\mathbf{x}_{b}^{j}\right)\right]_{STF}.\label{eq:contrT00-2PN}
\end{align}


\subsection{1PN correction to $T^{0i}$\label{subsec:1PN-T0i}}

\begin{figure}[H]
\label{T0i_fig6a-6d}
\begin{centering}
\includegraphics[scale=0.35]{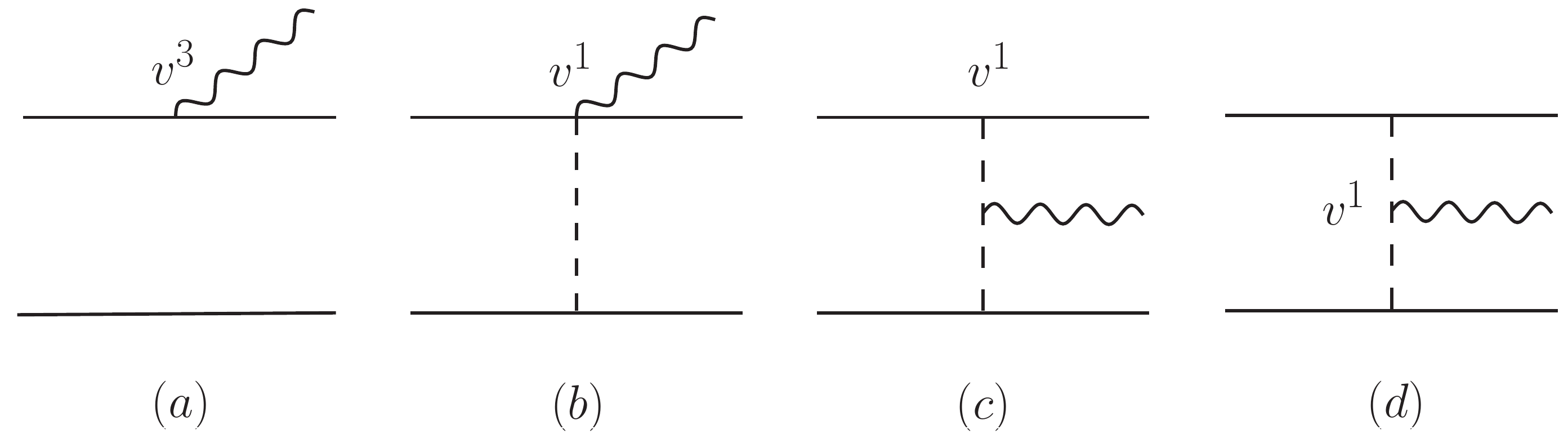}
\par\end{centering}
\caption{All diagrams that contribute to $T_{1PN}^{0i}$. }
\end{figure}

The leading order $T^{0i}$ component was obtained in \cite{Goldberger:2009qd}
is
\begin{equation}
T_{0PN}^{0i}\left(t,\mathbf{k}\right)=\sum_{a}m_{a}\mathbf{v}_{a}^{i}e^{-i\mathbf{k}\cdot\mathbf{x}_{a}}.\label{eq:T0iLO}
\end{equation}
The 1PN correction enter at $v^3$  and  are shown in Fig.~\ref{T0i_fig6a-6d}.

To  extract the $T^{0i}$ contributions to the mass quadrupole
moment, which is the third term in \eqref{eq:I2}, the expansion of
the denominator of vertices in Fig.~\ref{T0i_fig6a-6d}c-\ref{T0i_fig6a-6d}d has to be carried out  to
 third order. In addition, $\mathbf{k}^{2}$ terms can not be dropped, since they contribute terms that cannot be included in the
trace part of the quadrupole.

Comparing the diagrams illustrated in Fig.6, which  are composed of (\ref{eq:ppH00v0},\ref{eq:ppH0iv1},\ref{eq:ppHv1h0i},\ref{eq:ppv3h0i})
together with \eqref{eq:prop}, \eqref{eq:hi0HH} and \eqref{eq:hi0HHv1} 
we find
\begin{equation}
T_{Fig6a}^{0l}\left(t,\mathbf{k}\right)=\sum_{a}\frac{m_{a}}{2}\mathbf{v}_{a}^{l}\mathbf{v}_{a}^{2}e^{-i\mathbf{k}\cdot\mathbf{x}_{a}},
\end{equation}
\begin{equation}
T_{Fig6b}^{0l}\left(t,\mathbf{k}\right)=\sum_{a\neq b}\frac{Gm_{a}m_{b}}{r}\mathbf{v}_{a}^{l}e^{-i\mathbf{k}\cdot\mathbf{x}_{a}},
\end{equation}
\begin{align}
T_{Fig6c}^{0l}\left(t,\mathbf{k}\right) & =\sum_{a\neq b}\frac{Gm_{a}m_{b}}{r}e^{-i\mathbf{k}\cdot\mathbf{x}_{a}}\left[-2\mathbf{v}_{a}^{l}+2i\mathbf{k}^{i}\left(\mathbf{v}_{a}^{i}\mathbf{r}^{l}-\mathbf{r}^{i}\mathbf{v}_{a}^{l}\right)+\mathbf{k}^{i}\mathbf{k}^{j}\left(\mathbf{r}^{i}\mathbf{r}^{j}\mathbf{v}_{a}^{l}-\mathbf{v}_{a}^{i}\mathbf{r}^{j}\mathbf{r}^{l}\right)\right.\nonumber \\
 & \left.+\frac{i}{6}\mathbf{k}^{i}\mathbf{k}^{j}\mathbf{k}^{k}\left(\mathbf{r}^{2}\delta^{ij}\mathbf{v}_{a}^{k}\mathbf{r}^{l}-\mathbf{r}^{2}\delta^{il}\mathbf{v}_{a}^{j}\mathbf{r}^{k}-2\mathbf{v}_{a}^{i}\mathbf{r}^{j}\mathbf{r}^{k}\mathbf{r}^{l}+2\mathbf{r}^{i}\mathbf{r}^{j}\mathbf{r}^{k}\mathbf{v}_{a}^{l}\right)\right]+O\left(\mathbf{k}^{4}\right)+...,
\end{align}
\begin{align}
T_{Fig6d}^{0l}\left(t,\mathbf{k}\right) & =\sum_{a\neq b}\frac{Gm_{a}m_{b}}{4r}e^{-i\mathbf{k}\cdot\mathbf{x}_{a}}\left\{ \mathbf{v}_{a}^{l}+\mathbf{v}_{b}^{l}-\frac{1}{r^{2}}\left(\mathbf{v}_{a}+\mathbf{v}_{b}\right)\cdot\mathbf{r}\mathbf{r}^{l}\right.\nonumber \\
 & -\frac{i}{2}\mathbf{k}^{i}\left(3r\dot{r}\delta^{il}-\mathbf{r}^{i}\left(\mathbf{v}_{a}^{l}+\mathbf{v}_{b}^{l}\right)+\mathbf{v}^{i}\mathbf{r}^{l}+\frac{1}{r^{2}}\left(\mathbf{v}_{a}+\mathbf{v}_{b}\right)\cdot\mathbf{r}\mathbf{r}^{i}\mathbf{r}^{l}\right)\nonumber \\
 & +\frac{1}{6}\mathbf{k}^{i}\mathbf{k}^{j}\left[-5\mathbf{r}^{2}\left(\mathbf{v}_{a}^{i}+\mathbf{v}_{b}^{i}\right)\delta^{jl}+\left(4\mathbf{v}_{a}\cdot\mathbf{r}-5\mathbf{v}_{b}\cdot\mathbf{r}\right)\mathbf{r}^{i}\delta^{jl}+\left(\mathbf{v}_{a}^{i}-2\mathbf{v}_{b}^{i}\right)\mathbf{r}^{j}\mathbf{r}^{l}\right.\nonumber \\
 & \left.+\left(\mathbf{v}_{a}^{l}+\mathbf{v}_{b}^{l}\right)\left(\frac{1}{2}\delta^{ij}\mathbf{r}^{2}-\mathbf{r}^{i}\mathbf{r}^{j}\right)+\left(\mathbf{v}_{a}\cdot\mathbf{r}+\mathbf{v}_{b}\cdot\mathbf{r}\right)\left(\frac{1}{2}\delta^{ij}\mathbf{r}^{l}+\frac{1}{r^{2}}\mathbf{r}^{i}\mathbf{r}^{j}\mathbf{r}^{l}\right)\right]\nonumber \\
 & -\frac{i}{24}\mathbf{k}^{i}\mathbf{k}^{j}\mathbf{k}^{k}\left[\delta^{kl}\left(6\mathbf{r}^{2}\mathbf{v}_{a}^{i}\mathbf{r}^{j}+14\mathbf{r}^{2}\mathbf{v}_{b}^{i}\mathbf{r}^{j}-5\mathbf{v}_{a}\cdot\mathbf{r}\mathbf{r}^{i}\mathbf{r}^{j}+7\mathbf{v}_{b}\cdot\mathbf{r}\mathbf{r}^{i}\mathbf{r}^{j}\right)\right.\nonumber \\
 & +\delta^{ij}\mathbf{r}^{2}\left(3\delta^{kl}\mathbf{v}_{a}\cdot\mathbf{r}-3\delta^{kl}\mathbf{v}_{b}\cdot\mathbf{r}-\mathbf{r}^{k}\mathbf{v}_{b}^{l}-\mathbf{r}^{k}\mathbf{v}_{a}^{l}+\mathbf{v}_{a}^{k}\mathbf{r}^{l}-\mathbf{v}_{b}^{k}\mathbf{r}^{l}\right)-\delta^{ij}\left(\mathbf{v}_{a}\cdot\mathbf{r}+\mathbf{v}_{b}\cdot\mathbf{r}\right)\mathbf{r}^{k}\mathbf{r}^{l}\nonumber \\
 & \left.\left.+\mathbf{r}^{i}\mathbf{r}^{j}\mathbf{r}^{k}\left(\mathbf{v}_{a}^{l}+\mathbf{v}_{b}^{l}\right)+\left(3\mathbf{v}_{b}^{i}-\mathbf{v}_{a}^{i}\right)\mathbf{r}^{j}\mathbf{r}^{k}\mathbf{r}^{l}-\frac{1}{r^{2}}\left(\mathbf{v}_{a}+\mathbf{v}_{b}\right)\cdot\mathbf{r}\mathbf{r}^{i}\mathbf{r}^{j}\mathbf{r}^{k}\mathbf{r}^{l}\right]\right\} +O\left(\mathbf{k}^{4}\right)+...\cdot
\end{align}
Expanding the exponentials up to the third order in the radiation
momentum, we get
\begin{align}
\int d^{3}\mathbf{x}\partial_{0}T_{1PN}^{0l}\mathbf{x}^{l}\left[\mathbf{x}^{i}\mathbf{x}^{j}\right]_{TF} & =\sum_{a}\frac{d}{dt}\left[\frac{1}{2}m_{a}\mathbf{v}_{a}^{2}\mathbf{v}_{a}\cdot\mathbf{x}_{a}\mathbf{x}_{a}^{i}\mathbf{x}_{a}^{j}\right]_{TF}\nonumber \\
 & +\sum_{a\neq b}\frac{d}{dt}\left\{ \frac{Gm_{a}m_{b}}{12r}\left[\left(8\mathbf{r}^{2}-20\mathbf{r}\cdot\mathbf{x}_{b}\right)\mathbf{v}_{a}^{i}\mathbf{x}_{a}^{j}+\left(20\mathbf{r}^{2}-22\mathbf{r}\cdot\mathbf{x}_{b}\right)\mathbf{v}_{a}^{i}\mathbf{x}_{b}^{j}\right.\right.\nonumber \\
 & +\left(22\mathbf{v}_{a}\cdot\mathbf{x}_{a}-30\mathbf{v}_{b}\cdot\mathbf{x}_{a}-8\mathbf{v}_{a}\cdot\mathbf{x}_{b}+8\mathbf{v}_{b}\cdot\mathbf{x}_{b}-\frac{2}{r^{2}}\left(\mathbf{v}_{a}+\mathbf{v}_{b}\right)\cdot\mathbf{r}\mathbf{r}\cdot\mathbf{x}_{b}\right)\mathbf{x}_{a}^{i}\mathbf{x}_{a}^{j}\nonumber \\
 & \left.\left.+\left(9\mathbf{v}_{a}\cdot\mathbf{x}_{a}-7\mathbf{v}_{a}\cdot\mathbf{x}_{b}-\frac{1}{r^{2}}\left(\mathbf{v}_{a}+\mathbf{v}_{b}\right)\cdot\mathbf{r}\mathbf{r}\cdot\mathbf{x}_{b}\right)\mathbf{x}_{a}^{i}\mathbf{x}_{b}^{j}\right]\right\} _{STF}.\label{eq:contrTi0-1PN}
\end{align}


\subsection{1PN correction to $T^{ii}$\label{subsec:1PN-Tii}}

The leading order $T^{ii}$ component obtained in \cite{Goldberger:2009qd}
has the form
\begin{equation}
T_{0PN}^{ii}\left(t,\mathbf{k}\right)=\left(\sum_{a}m_{a}\mathbf{v}_{a}^{2}-\sum_{a\neq b}\frac{Gm_{a}m_{b}}{2r}+O\left(\mathbf{k}\right)+...\right)e^{-i\mathbf{k}\cdot\mathbf{x}_{a}}.\label{eq:TiiLO}
\end{equation}
Notice that, while $T_{0PN}^{0i}$ in \eqref{eq:T0iLO} is down by  $v^{1}$
 relative to $T_{0PN}^{00}$ in \eqref{eq:T00LO},
the leading order spatial component \eqref{eq:TiiLO} is down by $v^{2}$ compared to
 $T_{0PN}^{00}$, this fixes the PN hierarchy among the components
$T^{00}$, $T^{i0}$ and $T^{ij}$ of the pseudotensor.

To obtain $T_{1PN}^{ii}$ as well as its contributions
to $I_{2PN}^{ij}$ we have to compute all
diagrams that enter at $v^{4}$ with one $\bar{h}^{ii}$ external
momentum.
To compute the spatial component of the pseudotensor and to
extract its contribution to the mass quadrupole moment  we have to
carry out the expansions up to the second order in the radiation momentum.
 $\mathbf{k}^{2}$  may be dropped as 
in section~\ref{subsec:2PN-T00}.

\begin{figure}[H]
\label{Tll_fig7a-7c}
\begin{centering}
\includegraphics[scale=0.35]{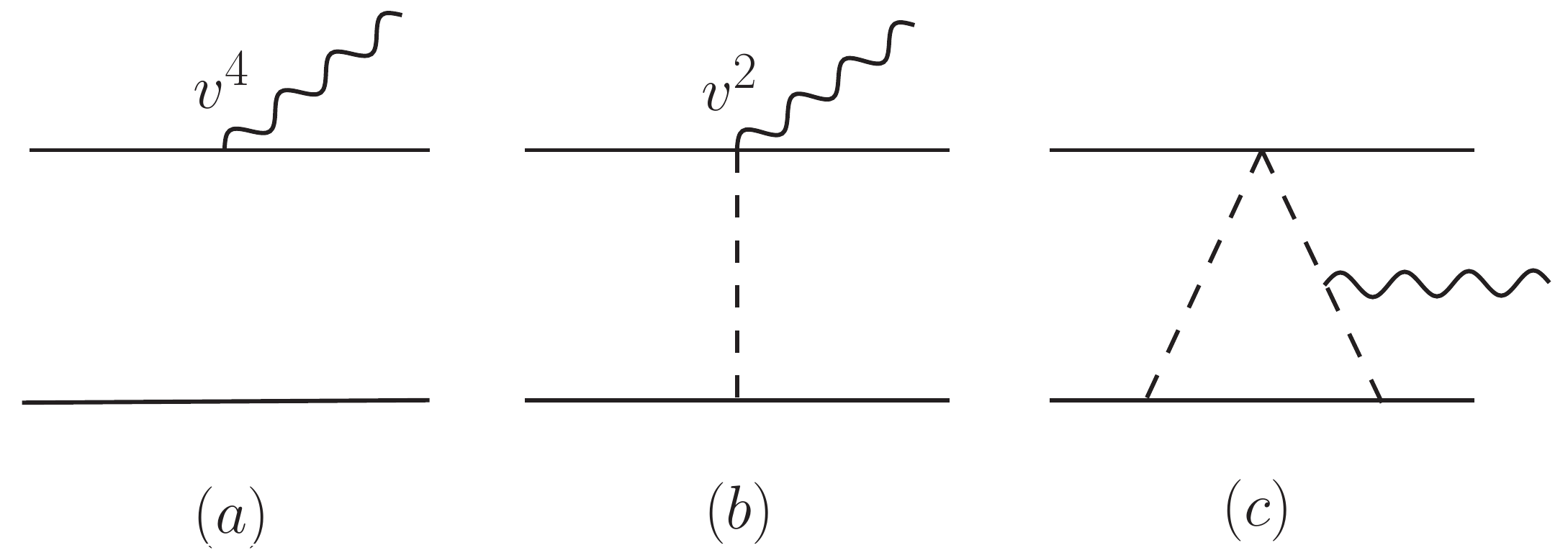}
\par\end{centering}
\caption{Diagrams with $\bar{h}^{ii}$ external momentum.}

\end{figure}

The
diagrams illustrated in Fig.~\ref{Tll_fig7a-7c}, involve  (\ref{eq:ppH00v0},
\ref{eq:ppHH}, \ref{eq:ppHv2hij}, \ref{eq:ppv4hij}),  \eqref{eq:prop}
and \eqref{eq:hHH0-1} give
\begin{align}
T_{Fig7a}^{ll}\left(t,\mathbf{k}\right) & =\sum_{a}\frac{m_{a}}{2}\mathbf{v}_{a}^{4}e^{-i\mathbf{k}\cdot\mathbf{x}_{a}},\\
T_{Fig7b}^{ll}\left(t,\mathbf{k}\right) & =\sum_{a\neq b}\frac{Gm_{a}m_{b}}{r}\mathbf{v}_{a}^{2}e^{-i\mathbf{k}\cdot\mathbf{x}_{a}},\\
T_{Fig7c}^{ll}\left(t,\mathbf{k}\right) & =-\sum_{a\neq b}\frac{G^{2}m_{a}m_{b}m}{2r^{2}}e^{-i\mathbf{k}\cdot\mathbf{x}_{a}}.
\end{align}
It is straightforward to extract the contribution for the mass quadrupole
moment by expanding the exponentials up to the second order in the
radiation momentum,
\begin{equation}
\int d^{3}\mathbf{x}T_{Fig7a-7c}^{ll}\left[\mathbf{x}^{i}\mathbf{x}^{j}\right]_{TF}=\sum_{a}\frac{m_{a}}{2}\mathbf{v}_{a}^{4}\left[\mathbf{x}_{a}^{i}\mathbf{x}_{a}^{j}\right]_{TF}+\sum_{a\neq b}\frac{Gm_{a}m_{b}}{r}\left(\mathbf{v}_{a}^{2}-\frac{Gm}{2r}\right)\left[\mathbf{x}_{a}^{i}\mathbf{x}_{a}^{j}\right]_{TF}.
\end{equation}

\begin{figure}[H]
\label{Tll_fig8a-8e}
\begin{centering}
\includegraphics[scale=0.35]{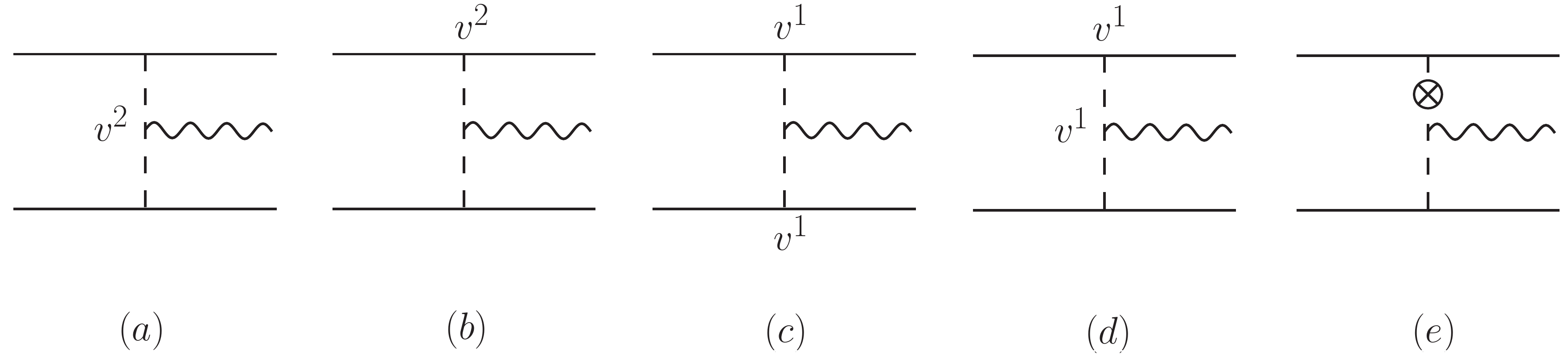}
\par\end{centering}
\caption{One potential graviton exchange with $\bar{h}^{ii}$ external momentum.}
\end{figure}

The computation of $T_{1PN}^{ii}$  follows from the diagrams shown in Fig.~\ref{Tll_fig8a-8e} which involve
(\ref{eq:ppH00v0}-\ref{eq:ppH00Hijv2}) and (\ref{eq:prop}, \ref{eq:corrprop},
\ref{eq:hHH0-1}-\ref{eq:hHH3-1}), %
\begin{align}
T_{Fig8a}^{ll}\left(t,\mathbf{k}\right) & =\sum_{a\neq b}\frac{3Gm_{a}m_{b}}{4r}e^{-i\mathbf{k}\cdot\mathbf{x}_{a}}\left\{ 2\mathbf{v}^{2}+\mathbf{v}_{a}\cdot\mathbf{v}_{b}-2\dot{r}^{2}-\frac{1}{r^{2}}\mathbf{v}_{a}\cdot\mathbf{r}\mathbf{v}_{b}\cdot\mathbf{r}+2\mathbf{a}\cdot\mathbf{r}\right.\nonumber \\
 & +\frac{i}{2}\mathbf{k}^{i}\left[\left(2\mathbf{v}^{2}+\mathbf{v}_{a}\cdot\mathbf{v}_{b}-2\dot{r}^{2}-\frac{1}{r^{2}}\mathbf{v}_{a}\cdot\mathbf{r}\mathbf{v}_{b}\cdot\mathbf{r}+2\mathbf{a}\cdot\mathbf{r}\right)\mathbf{r}^{i}\right.\nonumber \\
 & \left.+\mathbf{v}_{b}^{i}\left(4\mathbf{v}_{b}\cdot\mathbf{r}-3\mathbf{v}_{a}\cdot\mathbf{r}\right)+\mathbf{v}_{a}^{i}\left(3\mathbf{v}_{b}\cdot\mathbf{r}-4\mathbf{v}_{a}\cdot\mathbf{r}\right)-2r^{2}\left(\mathbf{a}_{a}^{i}+\mathbf{a}_{b}^{i}\right)\right]\nonumber \\
 & +\frac{1}{6}\mathbf{k}^{i}\mathbf{k}^{j}\left[\left(-2\mathbf{v}^{2}-\mathbf{v}_{a}\cdot\mathbf{v}_{b}-2\mathbf{a}\cdot\mathbf{r}+2\dot{r}^{2}+\frac{1}{r^{2}}\mathbf{v}_{a}\cdot\mathbf{r}\mathbf{v}_{b}\cdot\mathbf{r}\right)\mathbf{r}^{i}\mathbf{r}^{j}\right.\nonumber \\
 & +\left(6\mathbf{v}_{a}\cdot\mathbf{r}-8\mathbf{v}_{b}\cdot\mathbf{r}\right)\mathbf{v}_{b}^{i}\mathbf{r}^{j}+\left(4\mathbf{v}_{a}\cdot\mathbf{r}-3\mathbf{v}_{b}\cdot\mathbf{r}\right)\mathbf{v}_{a}^{i}\mathbf{r}^{j}\nonumber \\
 & \left.\left.+r^{2}\left(-4\mathbf{v}_{a}^{i}\mathbf{v}_{a}^{j}-3\mathbf{v}_{a}^{i}\mathbf{v}_{b}^{j}-4\mathbf{v}_{b}^{i}\mathbf{v}_{b}^{j}-4\mathbf{a}^{i}\mathbf{r}^{j}+6\mathbf{a}_{a}^{i}\mathbf{r}^{j}\right)\right]\right\} +O\left(\mathbf{k}^{3}\right)+...,
\end{align}
\begin{equation}
T_{Fig8b}^{ll}\left(t,\mathbf{k}\right)=\sum_{a\neq b}\frac{Gm_{a}m_{b}}{r}e^{-i\mathbf{k}\cdot\mathbf{x}_{a}}\left\{ \frac{1}{4}\left(\mathbf{v}_{a}^{2}+\mathbf{v}_{b}^{2}\right)-i\mathbf{v}_{a}^{2}\mathbf{k}^{i}\mathbf{r}^{i}-\mathbf{k}^{i}\mathbf{k}^{j}\left(\mathbf{r}^{2}\mathbf{v}_{a}^{i}\mathbf{v}_{a}^{j}-\frac{1}{2}\mathbf{v}_{a}^{2}\mathbf{r}^{i}\mathbf{r}^{j}\right)\right\} +O\left(\mathbf{k}^{3}\right)+...,
\end{equation}
\begin{align}
T_{Fig8c}^{ll}\left(t,\mathbf{k}\right) & =\sum_{a\neq b}\frac{Gm_{a}m_{b}}{2r}e^{-i\mathbf{k}\cdot\mathbf{x}_{a}}\left[-4\mathbf{v}_{a}\cdot\mathbf{v}_{b}-i\mathbf{k}^{i}\left(2\mathbf{v}_{a}\cdot\mathbf{v}_{b}\mathbf{r}^{i}+4\mathbf{v}_{a}\cdot\mathbf{r}\mathbf{v}_{b}^{i}-4\mathbf{v}_{b}\cdot\mathbf{r}\mathbf{v}_{a}^{i}\right)\right.\nonumber \\
 & \left.+\mathbf{k}^{i}\mathbf{k}^{j}\left(2\mathbf{r}^{2}\mathbf{v}_{a}^{i}\mathbf{v}_{b}^{j}+\mathbf{v}_{a}\cdot\mathbf{v}_{b}\mathbf{r}^{i}\mathbf{r}^{j}-2\mathbf{v}_{b}\cdot\mathbf{r}\mathbf{v}_{a}^{i}\mathbf{r}^{j}+2\mathbf{v}_{a}\cdot\mathbf{r}\mathbf{v}_{b}^{i}\mathbf{r}^{j}\right)\right]+O\left(\mathbf{k}^{3}\right)+...,
\end{align}
\begin{align}
T_{Fig8d}^{ll}\left(t,\mathbf{k}\right) & =\sum_{a\neq b}\frac{Gm_{a}m_{b}}{r}e^{-i\mathbf{k}\cdot\mathbf{x_{a}}}\left\{ -4\left(\mathbf{v}\cdot\mathbf{v}_{a}+\mathbf{a}_{a}\cdot\mathbf{r}-\frac{1}{r^{2}}\mathbf{v}\cdot\mathbf{r}\mathbf{v}_{a}\cdot\mathbf{r}\right)\right.\nonumber \\
 & -2i\mathbf{k}^{i}\left[\mathbf{r}^{i}\left(\mathbf{v}\cdot\mathbf{v}_{a}+\mathbf{a}_{a}\cdot\mathbf{r}-\frac{\dot{r}}{r}\mathbf{v}_{a}\cdot\mathbf{r}\right)+\mathbf{v}_{a}\cdot\mathbf{r}\left(\mathbf{v}^{i}-2\mathbf{v}_{a}^{i}\right)\right]\nonumber \\
 & +\frac{1}{3}\mathbf{k}^{i}\mathbf{k}^{j}\left[\mathbf{r}^{2}\left(\mathbf{v}^{i}\mathbf{v}_{a}^{j}+\mathbf{a}_{a}\mathbf{r}^{i}\right)+2\mathbf{r}^{i}\mathbf{r}^{j}\left(\mathbf{v}\cdot\mathbf{v}_{a}+\mathbf{a}_{a}\cdot\mathbf{r}\right)\right.\nonumber \\
 & \left.\left.+r\dot{r}\mathbf{v}_{a}^{i}\mathbf{r}^{j}+2\mathbf{v}_{a}\cdot\mathbf{r}\left(2\mathbf{v}^{i}\mathbf{r}^{j}-3\mathbf{v}_{a}^{i}\mathbf{r}^{j}\right)-2\frac{\dot{r}}{r}\mathbf{v}_{a}\cdot\mathbf{r}\mathbf{r}^{i}\mathbf{r}^{j}\right]\right\} +O\left(\mathbf{k}^{3}\right)+...,
\end{align}
\begin{align}
T_{Fig8e}^{ll}\left(t,\mathbf{k}\right) & =-\sum_{a\neq b}\frac{Gm_{a}m_{b}}{4r}e^{-i\mathbf{k}\cdot\mathbf{x}_{a}}\left\{ 2\left(-\mathbf{a}_{b}\cdot\mathbf{r}+\mathbf{v}_{b}^{2}-\left(\mathbf{v}_{b}\cdot\mathbf{n}\right)^{2}\right)\right.\nonumber \\
 & -\frac{i}{2}\mathbf{k}^{i}\left[\left(\mathbf{a}_{b}\cdot\mathbf{r}-\mathbf{v}_{b}^{2}+\left(\mathbf{v}_{b}\cdot\mathbf{n}\right)^{2}\right)\mathbf{r}^{i}-2\mathbf{v}_{b}\cdot\mathbf{r}\mathbf{v}_{b}^{i}+\mathbf{r}^{2}\mathbf{a}_{b}^{i}\right]\nonumber \\
 & \left.+\frac{i^{2}}{6}\mathbf{k}^{i}\mathbf{k}^{j}\left[\left(-\mathbf{a}_{b}\cdot\mathbf{r}+\mathbf{v}_{b}^{2}-\left(\mathbf{v}_{b}\cdot\mathbf{n}\right)^{2}\right)\mathbf{r}^{i}\mathbf{r}^{j}+4\mathbf{v}_{b}\cdot\mathbf{r}\mathbf{v}_{b}^{i}\mathbf{r}^{j}-2\mathbf{r}^{2}\mathbf{a}_{b}^{i}\mathbf{r}^{j}+2\mathbf{r}^{2}\mathbf{v}_{b}^{i}\mathbf{v}_{b}^{j}\right]\right\} +O\left(\mathbf{k}^{3}\right)+...,
\end{align}
which provides us with
\begin{align}
\int d^{3}\mathbf{x}T_{Fig8a-8e}^{ll}\left[\mathbf{x}^{i}\mathbf{x}^{j}\right]_{TF} & =\sum_{a\neq b}\frac{Gm_{a}m_{b}}{12r}\left\{ \left(10\mathbf{v}_{a}^{2}-17\mathbf{v}_{b}^{2}-10\mathbf{v}_{a}\cdot\mathbf{v}_{b}\right.\right.\nonumber \\
 & \left.+5\left(\mathbf{v}_{a}\cdot\mathbf{n}\right)^{2}+2\mathbf{v}_{a}\cdot\mathbf{n}\mathbf{v}_{b}\cdot\mathbf{n}-8\left(\mathbf{v}_{b}\cdot\mathbf{n}\right)^{2}-5\mathbf{a}_{a}\cdot\mathbf{r}+8\mathbf{a}_{b}\cdot\mathbf{r}\right)\mathbf{x}_{a}^{i}\mathbf{x}_{a}^{j}\nonumber \\
 & +\left(-5\mathbf{v}_{a}^{2}+7\mathbf{v}_{a}\cdot\mathbf{v}_{b}+5\left(\mathbf{v}_{a}\cdot\mathbf{n}\right)^{2}-7\mathbf{v}_{a}\cdot\mathbf{n}\mathbf{v}_{b}\cdot\mathbf{n}-5\mathbf{a}_{a}\cdot\mathbf{r}\right)\mathbf{x}_{a}^{i}\mathbf{x}_{b}^{j}\nonumber \\
 & +\left(4\mathbf{v}_{a}\cdot\mathbf{r}-44\mathbf{v}_{b}\cdot\mathbf{r}\right)\mathbf{v}_{a}^{i}\mathbf{x}_{a}^{j}+\left(14\mathbf{v}_{a}\cdot\mathbf{r}-58\mathbf{v}_{b}\cdot\mathbf{r}\right)\mathbf{v}_{a}^{i}\mathbf{x}_{b}^{j}\nonumber \\
 & \left.+r^{2}\left(38\mathbf{v}_{a}^{i}\mathbf{v}_{a}^{j}-7\mathbf{v}_{a}^{i}\mathbf{v}_{b}^{j}+14\mathbf{a}_{a}^{i}\mathbf{x}_{a}^{j}+19\mathbf{a}_{a}^{i}\mathbf{x}_{b}^{j}\right)\right\} _{STF}.
\end{align}

\begin{figure}[H]
\label{Tll_fig9a-9c}
\begin{centering}
\includegraphics[scale=0.35]{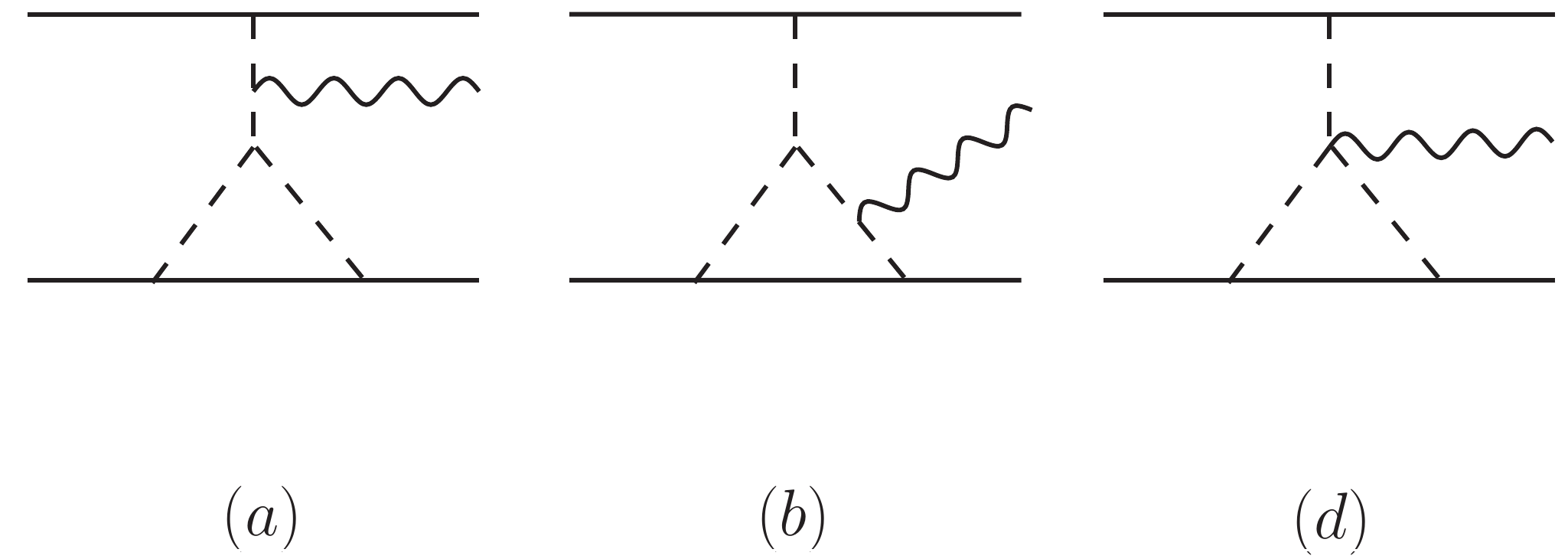}
\par\end{centering}
\caption{Three-potential-graviton exchange with $\bar{h}^{ii}$ external momentum.}
\end{figure}

Finally, the diagrams containing a three-potential-graviton exchange shown in Fig.~\ref{Tll_fig9a-9c}
which involve \eqref{eq:ppH00v0}, \eqref{eq:compij} and \eqref{eq:4gravij} give %
\begin{align}
T_{Fig9a}^{ll}\left(t,\mathbf{k}\right) & =\sum_{a\neq b}\frac{G^{2}m_{a}^{2}m_{b}}{r^{2}}e^{-i\mathbf{k}\cdot\mathbf{x}_{a}}\left(-\frac{5}{2}+\frac{7}{2}i\mathbf{k}^{i}\mathbf{r}^{i}-\frac{4}{3}\mathbf{k}^{i}\mathbf{k}^{j}\mathbf{r}^{i}\mathbf{r}^{j}\right)+O\left(\mathbf{k}^{3}\right)+...\\
T_{Fig9b}^{ll}\left(t,\mathbf{k}\right) & =\sum_{a\neq b}\frac{G^{2}m_{a}^{2}m_{b}}{r^{2}}e^{-i\mathbf{k}\cdot\mathbf{x}_{a}}\left(1-6i\mathbf{k}^{i}\mathbf{r}^{i}+\frac{7}{3}\mathbf{k}^{i}\mathbf{k}^{j}\mathbf{r}^{i}\mathbf{r}^{j}\right)+O\left(\mathbf{k}^{3}\right)+...\\
T_{Fig9c}^{ll}\left(t,\mathbf{k}\right) & =\sum_{a\neq b}\frac{7G^{2}m_{a}m_{b}^{2}}{2r^{2}}e^{-i\mathbf{k}\cdot\mathbf{x}_{a}},
\end{align}
which leads to
\begin{equation}
\int d^{3}\mathbf{x}T_{Fig9a-9c}^{ll}\left[\mathbf{x}^{i}\mathbf{x}^{j}\right]_{TF}=\sum_{a\neq b}\frac{G^{2}m_{a}m_{b}}{r^{2}}\left[\frac{3}{2}m\mathbf{x}_{a}^{i}\mathbf{x}_{a}^{j}-m_{a}\mathbf{x}_{a}^{i}\mathbf{x}_{b}^{j}\right]_{STF}.
\end{equation}
With this, we now write the total contribution of $T_{1PN}^{ll}$
to the mass quadrupole,
\begin{align}
\int d^{3}\mathbf{x}T_{1PN}^{ll}\left[\mathbf{x}^{i}\mathbf{x}^{j}\right]_{TF} & =\sum_{a}\frac{m_{a}}{2}\mathbf{v}_{a}^{4}\left[\mathbf{x}_{a}^{i}\mathbf{x}_{a}^{j}\right]_{TF}+\sum_{a\neq b}\frac{Gm_{a}m_{b}}{12r}\left\{ \left(22\mathbf{v}_{a}^{2}-17\mathbf{v}_{b}^{2}-10\mathbf{v}_{a}\cdot\mathbf{v}_{b}\right.\right.\nonumber \\
 & \left.+5\left(\mathbf{v}_{a}\cdot\mathbf{n}\right)^{2}+2\mathbf{v}_{a}\cdot\mathbf{n}\mathbf{v}_{b}\cdot\mathbf{n}-8\left(\mathbf{v}_{b}\cdot\mathbf{n}\right)^{2}-5\mathbf{a}_{a}\cdot\mathbf{r}+8\mathbf{a}_{b}\cdot\mathbf{r}+12\frac{Gm}{r}\right)\mathbf{x}_{a}^{i}\mathbf{x}_{a}^{j}\nonumber \\
 & +\left(-5\mathbf{v}_{a}^{2}+7\mathbf{v}_{a}\cdot\mathbf{v}_{b}+5\left(\mathbf{v}_{a}\cdot\mathbf{n}\right)^{2}-7\mathbf{v}_{a}\cdot\mathbf{n}\mathbf{v}_{b}\cdot\mathbf{n}-5\mathbf{a}_{a}\cdot\mathbf{r}-12\frac{Gm_{a}}{r}\right)\mathbf{x}_{a}^{i}\mathbf{x}_{b}^{j}\nonumber \\
 & +\left(4\mathbf{v}_{a}\cdot\mathbf{r}-44\mathbf{v}_{b}\cdot\mathbf{r}\right)\mathbf{v}_{a}^{i}\mathbf{x}_{a}^{j}+\left(14\mathbf{v}_{a}\cdot\mathbf{r}-58\mathbf{v}_{b}\cdot\mathbf{r}\right)\mathbf{v}_{a}^{i}\mathbf{x}_{b}^{j}\nonumber \\
 & \left.+r^{2}\left(38\mathbf{v}_{a}^{i}\mathbf{v}_{a}^{j}-7\mathbf{v}_{a}^{i}\mathbf{v}_{b}^{j}+14\mathbf{a}_{a}^{i}\mathbf{x}_{a}^{j}+19\mathbf{a}_{a}^{i}\mathbf{x}_{b}^{j}\right)\right\} _{STF}.\label{eq:contrTll_1PN}
\end{align}


\section{Lower order stress-energy tensors\label{sec:Lower-order}}

 Although $T^{ij}_{0PN}$, $T^{ii}_{0PN}$ and $T^{00}_{1PN}$ have been computed before in
\cite{Goldberger:2009qd}, to write an expression for the mass quadrupole moment at 2PN
order, we need to expand them  in the radiation momentum to
 higher order and terms depending on $\mathbf{k}^{2}$
must be kept.
\begin{figure}[H]
\label{T00_fig11}
\begin{centering}
\includegraphics[scale=0.35]{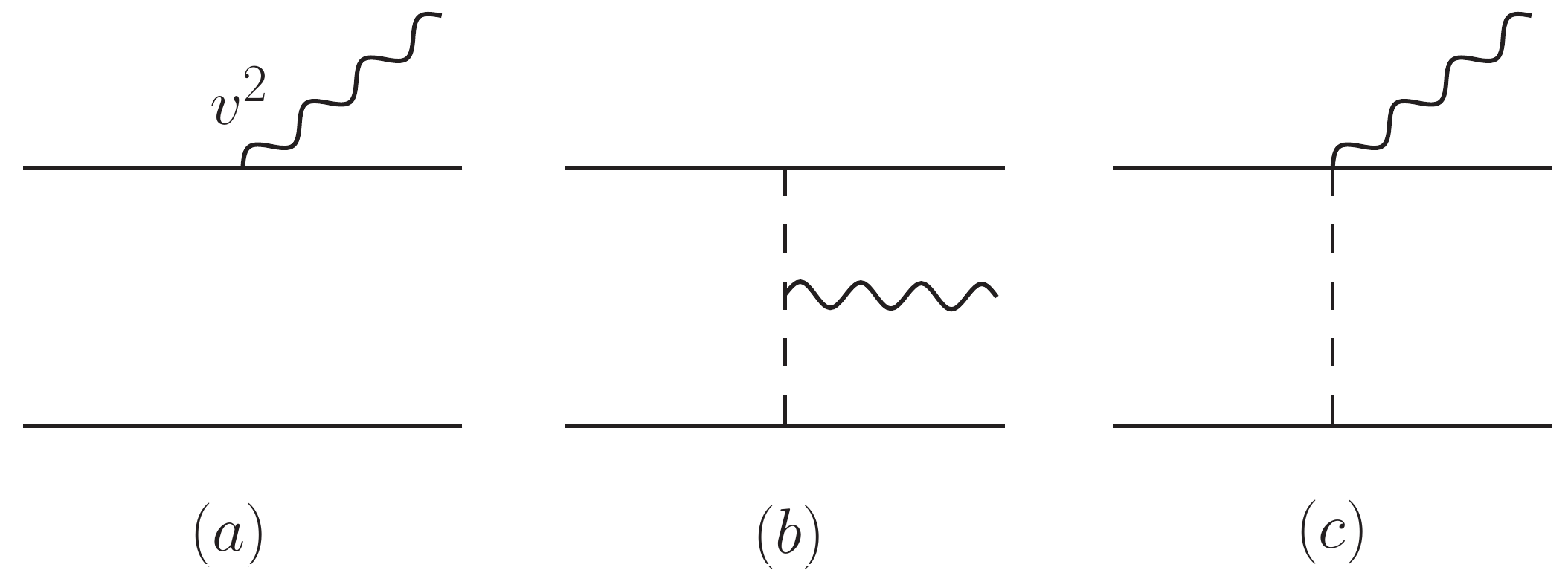}
\par\end{centering}
\caption{Diagrams (a) and (b) contribute to $T_{0PN}^{ij}$ when the external
leg is $\bar{h}^{ij}\left(x\right)$, while diagrams (a), (b) and
(c) contribute to $T_{1PN}^{00}$ when we consider $\bar{h}^{00}\left(x\right)$
as the external leg.}
\end{figure}

To obtain the sixth term of \eqref{eq:I2} we need diagrams in Fig.~\ref{T00_fig11}a-b. This gives us an expression for the leading
order $T^{ij}$, as shown below:
\begin{align}
T_{0PN}^{kl}\left(t,\mathbf{k}\right) & =\sum_{a}m_{a}\mathbf{v}_{a}^{k}\mathbf{v}_{a}^{l}e^{-i\mathbf{k}\cdot\mathbf{x}_{a}}+\sum_{a\neq b}\frac{Gm_{a}m_{b}}{2r}e^{-i\mathbf{k}\cdot\mathbf{x}_{a}}\left\{ -\frac{1}{r^{2}}\mathbf{r}^{k}\mathbf{r}^{l}-\frac{i}{2}\mathbf{k}^{i}\frac{1}{r^{2}}\mathbf{r}^{i}\mathbf{r}^{k}\mathbf{r}^{l}\right.\nonumber \\
 & +\frac{1}{12}\mathbf{k}^{i}\mathbf{k}^{j}\left[10\mathbf{r}^{2}\left(\delta^{kl}\delta^{ij}-\delta^{ik}\delta^{jl}\right)+\delta^{kl}\mathbf{r}^{i}\mathbf{r}^{j}+\delta^{ij}\mathbf{r}^{k}\mathbf{r}^{l}-2\delta^{ik}\mathbf{r}^{j}\mathbf{r}^{l}+\frac{1}{r^{2}}2\mathbf{r}^{i}\mathbf{r}^{j}\mathbf{r}^{k}\mathbf{r}^{l}\right]\nonumber \\
 & -\frac{i}{24}\mathbf{k}^{i}\mathbf{k}^{j}\mathbf{k}^{m}\left[\mathbf{r}^{2}\mathbf{r}^{m}\left(10\delta^{ik}\delta^{jl}-10\delta^{kl}\delta^{ij}\right)-\delta^{ij}\mathbf{r}^{m}\mathbf{r}^{k}\mathbf{r}^{l}+\mathbf{r}^{i}\mathbf{r}^{j}\left(2\delta^{mk}\mathbf{r}^{l}-\delta^{kl}\mathbf{r}^{m}-\frac{1}{r^{2}}\mathbf{r}^{m}\mathbf{r}^{k}\mathbf{r}^{l}\right)\right]\nonumber \\
 & +\frac{1}{240}\mathbf{k}^{i}\mathbf{k}^{j}\mathbf{k}^{m}\mathbf{k}^{n}\left[\frac{16}{3}\mathbf{r}^{4}\delta^{mn}\left(\delta^{kl}\delta^{ij}-\delta^{ik}\delta^{jl}\right)+\mathbf{r}^{2}\delta^{ij}\delta^{mn}\mathbf{r}^{k}\mathbf{r}^{l}-2\mathbf{r}^{2}\delta^{mn}\delta^{ik}\mathbf{r}^{j}\mathbf{r}^{l}\right.\nonumber \\
 & \left.\left.+\mathbf{r}^{m}\mathbf{r}^{n}\left(34\mathbf{r}^{2}\delta^{ik}\delta^{jl}-33\mathbf{r}^{2}\delta^{kl}\delta^{ij}-3\delta^{kl}\mathbf{r}^{i}\mathbf{r}^{j}+6\delta^{ik}\mathbf{r}^{j}\mathbf{r}^{l}-3\delta^{ij}\mathbf{r}^{k}\mathbf{r}^{l}-\frac{2}{r^{2}}\mathbf{r}^{i}\mathbf{r}^{j}\mathbf{r}^{k}\mathbf{r}^{l}\right)\right]\right\} +O\left(\mathbf{k}^{5}\right)+...\label{eq:TklLO}
\end{align}
The first term in the expression above is related to Fig.\ref{T00_fig11}a, which
comes from the simple source action term $-\sum\frac{m_{a}}{2m_{pl}}\int dt_{a}\mathbf{v}_{a}^{i}\mathbf{v}_{a}^{j}\overline{h}^{ij}\left(x_{a}\right).$
The other terms come from Fig.~\ref{T00_fig11}b, which is composed of \eqref{eq:ppH00v0}
and \eqref{eq:hHH} by considering, 
\begin{equation}
F^{\left\langle H^{00}H^{00}\right\rangle }\left[q,k,\bar{h}^{ij}\right]=\bar{h}^{ij}\left[-\frac{1}{2}\mathbf{q}^{i}\mathbf{q}^{j}-\frac{1}{2}\mathbf{q}^{i}\mathbf{k}^{j}-\frac{1}{2}\mathbf{k}^{i}\mathbf{k}^{j}+\delta^{ij}\left(\frac{1}{4}\mathbf{q}^{2}+\frac{1}{4}\mathbf{k}\cdot\mathbf{q}+\frac{1}{2}\mathbf{k}^{2}\right)\right].
\end{equation} 
where $F^{\left\langle H^{00}H^{00}\right\rangle }$ is defined in (\ref{eq:hi0HH}).

Now, expanding the exponentials up to the fourth order in the radiation
momentum, we extract the contribution
\begin{align}
\int d^{3}\mathbf{x}\partial_{0}^{2}T_{0PN}^{kl}\mathbf{x}^{k}\mathbf{x}^{l}\left[\mathbf{x}^{i}\mathbf{x}^{j}\right]_{TF} & =\frac{d^{2}}{dt^{2}}\left[\sum_{a}m_{a}\left(\mathbf{v}_{a}\cdot\mathbf{x}_{a}\right)^{2}\mathbf{x}_{a}^{i}\mathbf{x}_{a}^{j}\right]_{TF}\nonumber \\
 & +\frac{d^{2}}{dt^{2}}\left\{ \sum_{a\neq b}\frac{Gm_{a}m_{b}}{6r}\left[\left(27\mathbf{r}^{2}+\mathbf{x}_{a}^{2}-2\mathbf{x}_{a}\cdot\mathbf{x}_{b}-\frac{2}{r^{2}}\mathbf{r}\cdot\mathbf{x}_{a}\mathbf{r}\cdot\mathbf{x}_{b}\right)\mathbf{x}_{a}^{i}\mathbf{x}_{a}^{j}\right.\right.\nonumber \\
 & \left.\left.+\left(\frac{27}{2}\mathbf{r}^{2}+\mathbf{x}_{a}^{2}-\frac{1}{r^{2}}\left(\mathbf{r}\cdot\mathbf{x}_{a}\right)^{2}\right)\mathbf{x}_{a}^{i}\mathbf{x}_{b}^{j}\right]\right\} _{STF}.\label{eq:contrTkl-0PN}
\end{align}
Taking the trace of \eqref{eq:TklLO}, we get
\begin{align}
T_{0PN}^{ll}\left(t,\mathbf{k}\right) & =\sum_{a}m_{a}\mathbf{v}_{a}^{2}e^{-i\mathbf{k}\cdot\mathbf{x}_{a}}+\sum_{a\neq b}\frac{Gm_{a}m_{b}}{2r}e^{-i\mathbf{k}\cdot\mathbf{x}_{a}}\left[-1-\frac{i}{2}\mathbf{k}^{i}\mathbf{r}^{i}+\frac{1}{4}\mathbf{k}^{i}\mathbf{k}^{j}\left(7\mathbf{r}^{2}\delta^{ij}+\mathbf{r}^{i}\mathbf{r}^{j}\right)\right.\nonumber \\
 & +\frac{i}{24}\mathbf{k}^{i}\mathbf{k}^{j}\mathbf{k}^{m}\left(21\mathbf{r}^{2}\delta^{ij}\mathbf{r}^{m}+2\mathbf{r}^{i}\mathbf{r}^{j}\mathbf{r}^{m}\right)\nonumber \\
 & \left.+\frac{1}{144}\mathbf{k}^{i}\mathbf{k}^{j}\mathbf{k}^{m}\mathbf{k}^{n}\left(7\mathbf{r}^{4}\delta^{ij}\delta^{mn}-42\mathbf{r}^{2}\delta^{ij}\mathbf{r}^{m}\mathbf{r}^{n}-3\mathbf{r}^{i}\mathbf{r}^{j}\mathbf{r}^{m}\mathbf{r}^{n}\right)\right]+O\left(\mathbf{k}^{5}\right)+...,
\end{align}
which contributes to the quadrupole in the form below:
\begin{align}
\int d^{3}\mathbf{x}\partial_{0}^{2}T_{0PN}^{ll}\mathbf{x}^{2}\left[\mathbf{x}^{i}\mathbf{x}^{j}\right]_{TF} & =\frac{d^{2}}{dt^{2}}\left[\sum_{a}m_{a}\mathbf{v}_{a}^{2}\mathbf{x}_{a}^{2}\mathbf{x}_{a}^{i}\mathbf{x}_{a}^{j}\right]_{TF}\nonumber \\
 & +\frac{d^{2}}{dt^{2}}\left\{ \sum_{a\neq b}\frac{Gm_{a}m_{b}}{12r}\left[\left(-104\mathbf{x}_{a}^{2}+196\mathbf{x}_{a}\cdot\mathbf{x}_{b}-98\mathbf{x}_{b}^{2}\right)\mathbf{x}_{a}^{i}\mathbf{x}_{a}^{j}-49\mathbf{r}^{2}\mathbf{x}_{a}^{i}\mathbf{x}_{b}^{j}\right]\right\} _{STF}.\label{eq:contrTll-0PN}
\end{align}

To be able to compute the seventh contribution in \eqref{eq:I2},
we need an expression for $T_{1PN}^{00}$ up to the fourth order in
the radiation momentum. We regard the source action term $-\sum\frac{m_{a}}{4m_{pl}}\int dt_{a}\mathbf{v}_{a}^{2}\overline{h}^{00}\left(x_{a}\right)$
and also \eqref{eq:prop}, \eqref{eq:ppH00v0}, \eqref{eq:ppH00h00}
and \eqref{eq:hHH0} to solve the diagrams at Fig.~\ref{T00_fig11}a-c. With this,
we get an expression for $T_{1PN}^{00}$ and its contribution to
the mass quadrupole at 2PN, respectively:
\begin{align}
T_{1PN}^{00}\left(t,\mathbf{k}\right) & =\sum_{a}\frac{1}{2}m_{a}\mathbf{v}_{a}^{2}e^{-i\mathbf{k}\cdot\mathbf{x}_{a}}+\sum_{a\neq b}\frac{Gm_{a}m_{b}}{r}e^{-i\mathbf{k}\cdot\mathbf{x}_{a}}\left[-\frac{1}{2}\right.\nonumber \\
 & \left.-\frac{3}{8}\mathbf{k}^{2}r^{2}\left(1+\frac{i}{2}\mathbf{k}^{i}\mathbf{r}^{i}-\frac{1}{6}\mathbf{k}^{i}\mathbf{k}^{j}\mathbf{r}^{i}\mathbf{r}^{j}+\frac{r^{2}}{36}\mathbf{k}^{i}\mathbf{k}^{j}\delta^{ij}\right)\right]+O\left(\mathbf{k}^{5}\right)+...,\label{eq:T001PN}\\
\int d^{3}\mathbf{x}\partial_{0}^{2}T_{1PN}^{00}\mathbf{x}^{2}\left[\mathbf{x}^{i}\mathbf{x}^{j}\right]_{TF} & =\frac{d^{2}}{dt^{2}}\left\{ \sum_{a}\frac{1}{2}m_{a}\mathbf{v}_{a}^{2}\mathbf{x}_{a}^{2}\left[\mathbf{x}_{a}^{i}\mathbf{x}_{a}^{j}\right]_{TF}\right\} \nonumber \\
 & +\frac{d^{2}}{dt^{2}}\left\{ \sum_{a\neq b}\frac{Gm_{a}m_{b}}{r}\left[\frac{7}{4}\mathbf{r}^{2}\left(2\mathbf{x}_{a}^{i}\mathbf{x}_{a}^{j}+\mathbf{x}_{a}^{i}\mathbf{x}_{b}^{j}\right)-\frac{1}{2}\mathbf{x}_{a}^{2}\mathbf{x}_{a}^{i}\mathbf{x}_{a}^{j}\right]_{STF}\right\} .\label{eq:contrT00-1PN}
\end{align}

Moreover, considering the expansion up to the fifth and sixth orders
in the radiation momentum at \eqref{eq:T0iLO} and \eqref{eq:T00LO},
respectively, in addition to taking time derivatives, we get
\begin{align}
\int d^{3}\mathbf{x}\partial_{0}^{3}T_{0PN}^{0l}\mathbf{x}^{2}\mathbf{x}^{l}\left[\mathbf{x}^{i}\mathbf{x}^{j}\right]_{TF} & =\frac{d^{3}}{dt^{3}}\left[\sum_{a}m_{a}\mathbf{v}_{a}\cdot\mathbf{x}_{a}\mathbf{x}_{a}^{2}\mathbf{x}_{a}^{i}\mathbf{x}_{a}^{j}\right]_{TF},\label{eq:contrT0l-0PN}\\
\int d^{3}\mathbf{x}\partial_{0}^{4}T_{0PN}^{00}\mathbf{x}^{4}\left[\mathbf{x}^{i}\mathbf{x}^{j}\right]_{TF} & =\frac{d^{4}}{dt^{4}}\left[\sum_{a}m_{a}\mathbf{x}_{a}^{4}\mathbf{x}_{a}^{i}\mathbf{x}_{a}^{j}\right]_{TF}.\label{eq:contrT00-0PN}
\end{align}

Before writing the final expression for the 2PN correction to the
mass quadrupole moment, we still need to write the contribution of
$I_{1PN}^{ij}\left(\mathbf{a}_{1PN}\right)$, which is given by the
two terms
\begin{align}
\left[\int d^{3}\mathbf{x}\partial_{0}T_{0PN}^{0l}\mathbf{x}^{l}\mathbf{x}^{i}\mathbf{x}^{j}\right]_{TF} & \xrightarrow{2PN}\sum_{a}m_{a}\mathbf{a}_{1PNa}\cdot\mathbf{x}_{a}\left[\mathbf{x}_{a}^{i}\mathbf{x}_{a}^{j}\right]_{TF},\\
\left[\int d^{3}\mathbf{x}\partial_{0}^{2}T_{0PN}^{00}\mathbf{x}^{2}\mathbf{x}^{i}\mathbf{x}^{j}\right]_{TF} & \xrightarrow{2PN}\sum_{a}2m_{a}\left[\mathbf{x}_{a}^{2}\mathbf{a}_{1PNa}^{i}\mathbf{x}_{a}^{j}+\mathbf{a}_{1PNa}\cdot\mathbf{x}_{a}\mathbf{x}_{a}^{i}\mathbf{x}_{a}^{j}\right]_{STF},
\end{align}
where the 1PN correction to the acceleration, for instance obtained
in \cite{PhysRevD.86.044029} using the EFT framework, is given by 
\begin{eqnarray}
\mathbf{a}_{1PN(1)}^{i} & = & \frac{Gm_{2}}{2r^{2}}\left\{ \mathbf{n}^{i}\left[\frac{2Gm}{r}-3\left(\mathbf{v}_{1}^{2}+\mathbf{v}_{2}^{2}\right)+7\left(\mathbf{v}_{1}\cdot\mathbf{v}_{2}\right)+3\left(\mathbf{v}_{1}\cdot\mathbf{n}\right)\left(\mathbf{v}_{2}\cdot\mathbf{n}\right)\right]\right.\nonumber \\
 &  & -\mathbf{v}_{1}^{i}\left(\mathbf{v}_{2}\cdot\mathbf{n}\right)-\left(\mathbf{v}_{1}\cdot\mathbf{n}\right)\mathbf{v}_{2}^{i}+\dot{r}\left(6\mathbf{v}_{1}^{i}-7\mathbf{v}_{2}^{i}-\mathbf{n}^{i}\left(\mathbf{v}_{2}\cdot\mathbf{n}\right)\right)\nonumber \\
 &  & \left.-6r\mathbf{a}_{1}^{i}+7r\mathbf{a}_{2}^{i}+\left(\mathbf{v}^{i}-\mathbf{n}^{i}\dot{r}\right)\left(\mathbf{v}_{2}\cdot\mathbf{n}\right)+r\mathbf{n}^{i}\left(\mathbf{a}_{2}\cdot\mathbf{n}\right)+\mathbf{n}^{i}\left(\mathbf{v}_{2}\cdot\left(\mathbf{v}-\mathbf{n}\dot{r}\right)\right)\right\} \nonumber \\
 &  & -\frac{1}{2}\mathbf{a}_{1}^{i}\mathbf{v}_{1}^{2}-\mathbf{v}_{1}^{i}\left(\mathbf{v}_{1}\cdot\mathbf{a}_{1}\right).\label{eq:a1pn1}
\end{eqnarray}


\section{Consistency tests\label{sec:Consistency-tests}}
Here we  check the expressions for the components
$T_{2PN}^{00}$, $T_{1PN}^{0i}$ and $T_{1PN}^{ll}$, which were obtained
here for the first time in EFT approach with previous results derived using different methods.

The results presented in section \ref{subsec:2PN-T00} allow us to
write down an expression for the temporal component of the pseudotensor
up to 2PN order\footnote{To have an expression for $T^{00}$ containing terms of second order
in the radiation momentum, we would have to include $\mathbf{k}^{2}$
terms, but we discarded those terms since they are not needed to extract
the contribution of $T_{2PN}^{00}$ to the mass quadrupole moment.
Nevertheless, it is enough to consider terms up the first order in
the radiation momentum to perform the consistency tests on
$T_{2PN}^{00}$ in this section.},
\begin{align}
T^{00}\left(t,\mathbf{k}\right) & =e^{-i\mathbf{k}\cdot\mathbf{x}_{a}}\left\{ \sum_{a}m_{a}\left(1+\frac{1}{2}\mathbf{v}_{a}^{2}+\frac{3}{8}\mathbf{v}_{a}^{4}\right)+\sum_{a\neq b}\frac{Gm_{a}m_{b}}{2r}\left\{ -1+\mathbf{v}^{2}+\frac{7}{2}\mathbf{v}_{a}^{2}-\frac{5}{2}\mathbf{v}_{b}^{2}+\frac{5}{2}\mathbf{v}_{a}\cdot\mathbf{v}_{b}\right.\right.\nonumber \\
 & -\frac{5}{2}\mathbf{v}_{a}\cdot\mathbf{n}\mathbf{v}_{b}\cdot\mathbf{n}+2\left(\mathbf{v}_{b}\cdot\mathbf{n}\right)^{2}-\dot{r}^{2}+\mathbf{a}\cdot\mathbf{r}+2\mathbf{a}_{b}\cdot\mathbf{r}+\frac{G}{r}\left(4m_{a}-3m_{b}\right)\nonumber \\
 & +\frac{1}{2}i\mathbf{k}^{i}\left[\left(\mathbf{v}^{2}+\frac{1}{2}\mathbf{v}_{b}^{2}+\frac{5}{2}\mathbf{v}_{a}\cdot\mathbf{v}_{b}-\dot{r}^{2}-\frac{5}{2}\mathbf{v}_{a}\cdot\mathbf{n}\mathbf{v}_{b}\cdot\mathbf{n}+\frac{3}{2}\left(\mathbf{v}_{b}\cdot\mathbf{n}\right)^{2}+\mathbf{a}\cdot\mathbf{r}+\frac{3}{2}\mathbf{a}_{b}\cdot\mathbf{r}\right)\mathbf{r}^{i}\right.\nonumber \\
 & \left.\left.\left.+\left(8\mathbf{v}_{a}\cdot\mathbf{r}+\frac{11}{2}\mathbf{v}_{b}\cdot\mathbf{r}-2r\dot{r}\right)\mathbf{v}_{a}^{i}+\left(\frac{9}{2}\mathbf{v}_{b}\cdot\mathbf{r}+\frac{1}{2}r\dot{r}\right)\mathbf{v}_{b}^{i}+r^{2}\left(7\mathbf{a}_{a}^{i}-2\mathbf{a}_{b}^{i}\right)+\frac{6Gm}{r}\right]\right\} +O\left(\mathbf{k}^{2}\right)+...\right\} .\label{eq:finalT00}
\end{align}

We can use \eqref{eq:Texp} to read off different contributions
of $T^{00}$ to the dynamics of the binary system. For instance, at
zeroth order in the radiation momentum, we can read off the mechanical
energy of the system. It is straightforward to see in \eqref{eq:finalT00}
that the leading order terms in the PN approximation reproduce the
total mass of the two-body system, while the next-to-leading order
terms provide us with the Newtonian energy. The terms that account for the
next-to-next-to-leading order (2PN) correction to this pseudotensor,
which were calculated in the section \ref{subsec:2PN-T00} of this
paper, give us the following contribution to the conserved energy,
\begin{align}
E_{1PN} & =\int d^{3}\mathbf{x}T_{2PN}^{00}\left(x\right)\nonumber \\
 & =\frac{3}{8}\sum\limits _{a}m_{a}\mathbf{v}_{a}^{4}+\sum_{a\neq b}\frac{Gm_{a}m_{b}}{4r}\left[6\mathbf{v}_{a}^{2}-7\left(\mathbf{v}_{a}\cdot\mathbf{v}_{b}\right)-\left(\mathbf{v}_{a}\cdot\mathbf{n}\right)\left(\mathbf{v}_{b}\cdot\mathbf{n}\right)+\frac{Gm}{r}\right].
\end{align}
This result is equal to the first correction to the Newtonian energy
presented in Eq.~(205) of \cite{blanchet} and can also be calculated computing the Hamiltonian function using
the Lagrangian obtained by Einstein, Infeld and Hoffman in \cite{eih0}.

Regarding the 2PN terms in Eq.~\eqref{eq:finalT00}, we can read off the correction to the center of mass position%
\begin{align}
\mathbf{G}_{2PN} & =\int d^{3}\mathbf{x}T_{2PN}^{00}\left(x\right)\mathbf{x}\nonumber \\
 & =\frac{3}{8}\sum\limits _{a}m_{a}\mathbf{v}_{a}^{4}\mathbf{x}_{a}+\sum_{a\neq b}\frac{Gm_{a}m_{b}}{4r}\left\{ \left[\frac{19}{2}\mathbf{v}_{a}^{2}-7\mathbf{v}_{a}\cdot\mathbf{v}_{b}-\frac{7}{2}\mathbf{v}_{b}^{2}-\mathbf{v}_{a}\cdot\mathbf{n}\mathbf{v}_{b}\cdot\mathbf{n}\right.\right.\nonumber \\
 & \left.\left.-\frac{1}{2}\left(\mathbf{v}_{a}\cdot\mathbf{n}\right)^{2}+\frac{1}{2}\left(\mathbf{v}_{b}\cdot\mathbf{n}\right)^{2}-5\frac{Gm_{a}}{r}+7\frac{Gm_{b}}{r}\right]\mathbf{x}_{a}-7\left(\mathbf{v}_{a}\cdot\mathbf{r}+\mathbf{v}_{b}\cdot\mathbf{r}\right)\mathbf{v}_{a}\right\} ,\label{eq:G2PN}
\end{align}
which agrees with the result presented in Eq.~(B2c) of \cite{PhysRevD.97.044037},
where $\frac{d\mathbf{G}}{dt}=\mathbf{P}$, the total conserved linear
momentum, such that the center of mass frame is defined by $\mathbf{G}=0$.
By solving this equation iteratively, using the equations of motion
to reduce the second time derivatives of the position, we get the
2PN correction to the center of mass frame,
\begin{align}
\delta\mathbf{r}_{2PN} & =\frac{\nu\delta m}{m}\left\{ \mathbf{r}\left[\left(\frac{3}{8}-\frac{3\nu}{2}\right)\mathbf{v}^{4}+\frac{Gm}{r}\left(\left(\frac{19}{8}+\frac{3\nu}{2}\right)\mathbf{v}^{2}\right)\right.\right.\nonumber \\
 & \left.\left.+\left(-\frac{1}{8}+\frac{3\nu}{4}\right)\dot{r}^{2}+\left(\frac{7}{4}-\frac{\nu}{2}\right)\frac{Gm}{r}\right]-\mathbf{v}\left[\frac{7}{4}Gm\dot{r}\right]\right\} ,\label{eq:dr2PN}
\end{align}
which agrees with (B4a), (B4b) and (B5b) of \cite{PhysRevD.97.044037}.

Let us now consider the results obtained in section \ref{subsec:1PN-T0i}
to write down an expression for $T^{0l}$ up to 1PN order,
\begin{align}
T^{0l}\left(t,\mathbf{k}\right) & =e^{-i\mathbf{k}\cdot\mathbf{x}_{a}}\left\{ \sum_{a}m_{a}\mathbf{v}_{a}^{l}\left(1+\frac{1}{2}\mathbf{v}_{a}^{2}\right)+\sum_{a\neq b}\frac{Gm_{a}m_{b}}{4r}\left[-3\mathbf{v}_{a}^{l}+\mathbf{v}_{b}^{l}-\frac{1}{r^{2}}\left(\mathbf{v}_{a}+\mathbf{v}_{b}\right)\cdot\mathbf{r}\mathbf{r}^{l}\right.\right.\nonumber \\
 & \left.\left.-\frac{i}{2}\mathbf{k}^{i}\left(\mathbf{v}^{i}\mathbf{r}^{l}-16\mathbf{v}_{a}^{i}\mathbf{r}^{l}+15\mathbf{r}^{i}\mathbf{v}_{a}^{l}-\mathbf{r}^{i}\mathbf{v}_{b}^{l}+3r\dot{r}\delta^{il}+\frac{1}{r^{2}}\left(\mathbf{v}_{a}+\mathbf{v}_{b}\right)\cdot\mathbf{r}\mathbf{r}^{i}\mathbf{r}^{l}\right)\right]+O\left(\mathbf{k}^{2}\right)+...\right\} \label{eq:finalT0i}
\end{align}
Taking into account only terms of order zero in the radiation momentum,
we obtain the 1PN correction to the linear momentum of the binary
system,
\begin{align}
\mathbf{P}_{1PN}=\int d^{3}\mathbf{x}T_{1PN}^{0l}\left(x\right) & =-\left[\frac{Gm_{1}m_{2}}{2r^{3}}\left(\mathbf{v}_{1}+\mathbf{v}_{2}\right)\cdot\mathbf{r}\right]\mathbf{x}_{1}^{l}+\left[\frac{m_{1}}{2}\mathbf{v}_{1}^{2}-\frac{Gm_{1}m_{2}}{2r}\right]\mathbf{v}_{1}^{l}+1\leftrightarrow2.
\end{align}
The result above agrees with Eq.~(B1) and Eq.~(B2b)
of reference \cite{PhysRevD.97.044037}. Considering all linear terms in
the radiation momentum in \eqref{eq:finalT0i},  we are able to obtain the 1PN correction to the angular
momentum of the binary system,
\begin{align}
\mathbf{L}_{1PN}^{i}=-\frac{1}{2}\epsilon^{ilk}\int d^{3}\mathbf{x}\left(T_{1PN}^{0l}\mathbf{x}^{k}-T_{1PN}^{0k}\mathbf{x}^{l}\right) & =\frac{1}{2}\nu m\left(\mathbf{r}\times\mathbf{v}\right)^{i}\left[\left(1-3\nu\right)\mathbf{v}^{2}+\frac{Gm}{r}\left(6+2\nu\right)\right],
\end{align}
which agrees with Eq.~(2.9b) of reference \cite{kidder}.

Furthermore, considering the result obtained in section \ref{subsec:1PN-Tii},
we provide an expression for $T^{ll}\left(t,\mathbf{k}\right)$ up
to 1PN order:
\begin{align}
T^{ll}\left(t,\mathbf{k}\right) & =e^{-i\mathbf{k}\cdot\mathbf{x}_{a}}\left\{ \sum_{a}m_{a}\mathbf{v}_{a}^{2}\left(1+\frac{1}{2}\mathbf{v}_{a}^{2}\right)+\sum_{a\neq b}\frac{Gm_{a}m_{b}}{4r}\left\{ -2-5\mathbf{v}_{a}^{2}+5\mathbf{v}_{b}^{2}-\mathbf{v}_{a}\cdot\mathbf{v}_{b}-6\dot{r}^{2}\right.\right.\nonumber \\
 & -3\mathbf{v}_{a}\cdot\mathbf{n}\mathbf{v}_{b}\cdot\mathbf{n}+2\left(\mathbf{v}_{b}\cdot\mathbf{n}\right)^{2}+16\dot{r}\mathbf{v}_{a}\cdot\mathbf{n}-10\mathbf{a}_{a}\cdot\mathbf{r}-4\mathbf{a}_{b}\cdot\mathbf{r}+\frac{G}{r}\left(-8m_{a}+12m_{b}\right)\nonumber \\
 & +i\mathbf{k}^{i}\left[\mathbf{r}^{i}\left(-9\mathbf{v}_{a}^{2}+\frac{5}{2}\mathbf{v}_{b}^{2}-\frac{1}{2}\mathbf{v}_{a}\cdot\mathbf{v}_{b}-\frac{3}{2}\mathbf{v}_{a}\cdot\mathbf{n}\mathbf{v}_{b}\cdot\mathbf{n}+\frac{1}{2}\left(\mathbf{v}_{b}\cdot\mathbf{n}\right)^{2}\right.\right.\nonumber \\
 & \left.-3\dot{r}^{2}+8\dot{r}\mathbf{v}_{a}\cdot\mathbf{n}-5\mathbf{a}_{a}\cdot\mathbf{r}-\frac{5}{2}\mathbf{a}_{b}\cdot\mathbf{r}-\frac{10Gm_{a}}{r}\right)+\left(2\mathbf{v}_{a}\cdot\mathbf{r}+\frac{25}{2}\mathbf{v}_{b}\cdot\mathbf{r}\right)\mathbf{v}_{a}^{i}\nonumber \\
 & \left.\left.\left.+\left(-\frac{9}{2}\mathbf{v}_{a}\cdot\mathbf{r}+5\mathbf{v}_{b}\cdot\mathbf{r}\right)\mathbf{v}_{b}^{i}-\mathbf{r}^{2}\left(3\mathbf{a}_{a}^{i}+\frac{5}{2}\mathbf{a}_{b}^{i}\right)\right]\right\} +O\left(\mathbf{k}^{2}\right)+...\right\} .\label{eq:finalTll}
\end{align}
We can use the moment relation
\begin{equation}
\int d^{3}\mathbf{x}T^{ll}=\frac{1}{2}\frac{d^{2}}{dt^{2}}\int d^{3}\mathbf{x}T^{00}\mathbf{x}^{2}.\label{eq:CE}
\end{equation}
 to prove the self-consistency of our results.
At leading order in the PN expansion, it is trivial to prove that
this relation holds using \eqref{eq:finalT00} and \eqref{eq:finalTll}, while at next-to-leading order  more computation
is required. From \eqref{eq:finalTll} we can read off up to 1PN,
\begin{equation}
\int d^{3}\mathbf{x}T^{ll}=\sum_{a}m_{a}\mathbf{v}_{a}^{2}\left(1+\frac{1}{2}\mathbf{v}_{a}^{2}\right)+\sum_{a\neq b}\frac{Gm_{a}m_{b}}{r}\left[-\frac{1}{2}-\frac{1}{4}\mathbf{v}_{a}\cdot\mathbf{v}_{b}+\frac{3}{2}\left(\mathbf{v}_{a}\cdot\mathbf{n}\right)^{2}-\frac{7}{4}\mathbf{v}_{a}\cdot\mathbf{n}\mathbf{v}_{b}\cdot\mathbf{n}+\frac{5}{2}\frac{Gm_{a}}{r}\right].\label{eq:Tll-zeroth}
\end{equation}
To check if the result above satisfies \eqref{eq:CE}, we need a complete
expression for $T^{00}\left(t,\mathbf{k}\right)$ up to 1PN order
and which contains all terms up to the quadratic order in the radiation
momentum. In other words, we cannot discard terms proportional to $\mathbf{k}^{2}$
as we did in section \ref{subsec:2PN-T00}, where we drppped these terms that would not contribute to the trace-free quadrupole moment. Therefore, the expression that we need for $T^{00}\left(t,\mathbf{k}\right)$
is the sum of \eqref{eq:T00LO} with \eqref{eq:T001PN}, which provides
us with the following result up to 1PN order:
\begin{equation}
\frac{1}{2}\frac{d^{2}}{dt^{2}}\int d^{3}\mathbf{x}T^{00}\mathbf{x}^{2}=\frac{1}{2}\frac{d^{2}}{dt^{2}}\left[\sum_{a}m_{a}\left(1+\frac{1}{2}\mathbf{v}_{a}^{2}\right)\mathbf{x}_{a}^{2}+\sum_{a\neq b}\frac{Gm_{a}m_{b}}{r}\left(-\frac{1}{2}\mathbf{x}_{a}^{2}+\frac{9}{4}r^{2}\right)\right].\label{eq:T00-second}
\end{equation}
At this point, it is straightforward to show that, after taking the
second order time derivative and imposing the leading and next-to-leading
order equations of motion that \eqref{eq:CE}
holds, as we expected.

\section{Mass quadrupole moment at 2PN order\label{sec:Mass-quadrupole}}

We are now ready to sum the contributions \eqref{eq:contrT00-2PN},
\eqref{eq:contrTi0-1PN}, \eqref{eq:contrTll_1PN}, \eqref{eq:contrTkl-0PN},
\eqref{eq:contrTll-0PN}, \eqref{eq:contrT00-1PN}, \eqref{eq:contrT0l-0PN},
\eqref{eq:contrT00-0PN} and to write down the expression for the
2PN correction to the mass quadrupole moment  in a general orbit,
\begin{align}
I_{2PN}^{ij} & =\sum_{a}m_{a}f_{1(a)}^{ij}+\sum_{a\neq b}\frac{Gm_{a}m_{b}}{r}f_{2(a,b)}^{ij}+\frac{d}{dt}\left[\sum_{a}m_{a}f_{3(a)}^{ij}+\sum_{a\neq b}\frac{Gm_{a}m_{b}}{r}f_{4(a,b)}^{ij}\right]\nonumber \\
 & +\frac{d^{2}}{dt^{2}}\left[\sum_{a}m_{a}f_{5(a)}^{ij}+\sum_{a\neq b}\frac{Gm_{a}m_{b}}{r}f_{6(a,b)}^{ij}\right]+\frac{d^{3}}{dt^{3}}\left[\sum_{a}m_{a}f_{7(a)}^{ij}\right]+\frac{d^{4}}{dt^{4}}\left[\sum_{a}m_{a}f_{8(a)}^{ij}\right],\label{eq:I2pn-notCM}
\end{align}
where we have defined the following quantities for convenience:
\begin{equation}
f_{1(a)}^{ij}\equiv\left[\frac{7}{8}\mathbf{v}_{a}^{4}\mathbf{x}_{a}^{i}\mathbf{x}_{a}^{j}+\frac{11}{21}\mathbf{x}_{a}^{2}\mathbf{a}_{1PNa}^{i}\mathbf{x}_{a}^{j}-\frac{17}{21}\mathbf{a}_{1PNa}\cdot\mathbf{x}_{a}\mathbf{x}_{a}^{i}\mathbf{x}_{a}^{j}\right]_{STF},\label{eq:f1}
\end{equation}
\begin{align}
f_{2(a,b)}^{ij} & \equiv\frac{1}{12}\left[\left(50\mathbf{v}_{a}^{2}-28\mathbf{v}_{b}^{2}-32\mathbf{v}_{a}\cdot\mathbf{v}_{b}-4\dot{r}^{2}-24\mathbf{v}_{a}\cdot\mathbf{n}\mathbf{v}_{b}\cdot\mathbf{n}\right.\right.\nonumber \\
 & \left.+8\left(\mathbf{v}_{a}\cdot\mathbf{n}\right)^{2}+14\left(\mathbf{v}_{b}\cdot\mathbf{n}\right)^{2}-4\mathbf{a}_{a}\cdot\mathbf{r}+10\mathbf{a}_{b}\cdot\mathbf{r}+24\frac{Gm_{a}}{r}+18\frac{Gm_{b}}{r}\right)\mathbf{x}_{a}^{i}\mathbf{x}_{a}^{j}\nonumber \\
 & +\left(-4\mathbf{v}_{a}^{2}+8\mathbf{v}_{a}\cdot\mathbf{v}_{b}-12\mathbf{v}_{a}\cdot\mathbf{n}\mathbf{v}_{b}\cdot\mathbf{n}+8\left(\mathbf{v}_{a}\cdot\mathbf{n}\right)^{2}-2\dot{r}^{2}-4\mathbf{a}_{a}\cdot\mathbf{r}-24\frac{Gm_{1}}{r}\right)\mathbf{x}_{a}^{i}\mathbf{x}_{b}^{j}\nonumber \\
 & +\mathbf{v}_{a}^{i}\mathbf{x}_{a}^{j}\left(-16\mathbf{v}_{a}\cdot\mathbf{r}-64\mathbf{v}_{b}\cdot\mathbf{r}\right)+\mathbf{v}_{a}^{i}\mathbf{x}_{b}^{j}\left(40\mathbf{v}_{a}\cdot\mathbf{r}-32\mathbf{v}_{b}\cdot\mathbf{r}\right)\nonumber \\
 & \left.+\mathbf{r}^{2}\left(40\mathbf{v}_{a}^{i}\mathbf{v}_{a}^{j}-8\mathbf{v}_{a}^{i}\mathbf{v}_{b}^{j}-8\mathbf{a}_{a}^{i}\mathbf{x}_{a}^{j}-4\mathbf{a}_{a}^{i}\mathbf{x}_{b}^{j}\right)\right]_{STF},
\end{align}
\begin{equation}
f_{3(a)}^{ij}\equiv-\frac{2}{3}\mathbf{v}_{a}^{2}\mathbf{v}_{a}\cdot\mathbf{x}_{a}\left[\mathbf{x}_{a}^{i}\mathbf{x}_{a}^{j}\right]_{TF},
\end{equation}
\begin{align}
f_{4(a,b)}^{ij} & \equiv-\frac{1}{9}\left[\left(8\mathbf{r}^{2}-20\mathbf{r}\cdot\mathbf{x}_{b}\right)\mathbf{v}_{a}^{i}\mathbf{x}_{a}^{j}+\left(20\mathbf{r}^{2}-22\mathbf{r}\cdot\mathbf{x}_{b}\right)\mathbf{v}_{a}^{i}\mathbf{x}_{b}^{j}\right.\nonumber \\
 & +\left(22\mathbf{v}_{a}\cdot\mathbf{x}_{a}-30\mathbf{v}_{b}\cdot\mathbf{x}_{a}-8\mathbf{v}_{a}\cdot\mathbf{x}_{b}+8\mathbf{v}_{b}\cdot\mathbf{x}_{b}-\frac{2}{r^{2}}\left(\mathbf{v}_{a}+\mathbf{v}_{b}\right)\cdot\mathbf{r}\mathbf{r}\cdot\mathbf{x}_{b}\right)\mathbf{x}_{a}^{i}\mathbf{x}_{a}^{j}\nonumber \\
 & \left.+\left(9\mathbf{v}_{a}\cdot\mathbf{x}_{a}-7\mathbf{v}_{a}\cdot\mathbf{x}_{b}-\frac{1}{r^{2}}\left(\mathbf{v}_{a}+\mathbf{v}_{b}\right)\cdot\mathbf{r}\mathbf{r}\cdot\mathbf{x}_{b}\right)\mathbf{x}_{a}^{i}\mathbf{x}_{b}^{j}\right]_{STF},
\end{align}
\begin{equation}
f_{5(a)}^{ij}\equiv\left(\frac{1}{6}\left(\mathbf{v}_{a}\cdot\mathbf{x}_{a}\right)^{2}+\frac{19}{84}\mathbf{v}_{a}^{2}\mathbf{x}_{a}^{2}\right)\left[\mathbf{x}_{a}^{i}\mathbf{x}_{a}^{j}\right]_{TF},
\end{equation}
\begin{align}
f_{6(a,b)}^{ij} & \equiv\left[\left(\frac{31}{42}\mathbf{x}_{a}^{2}-\frac{11}{6}\mathbf{x}_{a}\cdot\mathbf{x}_{b}+\frac{8}{9}\mathbf{x}_{b}^{2}-\frac{1}{18}\frac{1}{r^{2}}\mathbf{r}\cdot\mathbf{x}_{a}\mathbf{r}\cdot\mathbf{x}_{b}\right)\mathbf{x}_{a}^{i}\mathbf{x}_{a}^{j}\right.\nonumber \\
 & \left.+\left(\frac{4}{9}\mathbf{r}^{2}+\frac{1}{36}\mathbf{x}_{a}^{2}-\frac{1}{36}\frac{1}{r^{2}}\left(\mathbf{r}\cdot\mathbf{x}_{a}\right)^{2}\right)\mathbf{x}_{a}^{i}\mathbf{x}_{b}^{j}\right]_{STF},
\end{align}
\begin{equation}
f_{7(a)}^{ij}\equiv-\frac{1}{7}\mathbf{v}_{a}\cdot\mathbf{x}_{a}\mathbf{x}_{a}^{2}\left[\mathbf{x}_{a}^{i}\mathbf{x}_{a}^{j}\right]_{TF},
\end{equation}
\begin{equation}
f_{8(a)}^{ij}\equiv\frac{23}{1512}\mathbf{x}_{a}^{4}\left[\mathbf{x}_{a}^{i}\mathbf{x}_{a}^{j}\right]_{TF}.
\end{equation}
With the exception of the accelerations in \eqref{eq:f1} which are
of 1PN order, all other accelerations in $I_{2PN}^{ij}$ should be
taken as the Newtonian acceleration.

In order to write the 2PN correction of the mass quadrupole moment
in the center of mass frame, we must have in mind that the positions
of the compact bodies in this frame are given by
\begin{align}
\mathbf{x}_{1} & =\frac{m_{2}}{m}\mathbf{r}+\delta\mathbf{r}_{1PN}+...,\label{eq:x1}\\
\mathbf{x}_{2} & =-\frac{m_{1}}{m}\mathbf{r}+\delta\mathbf{r}_{1PN}+...,\label{eq:x2}
\end{align}
where $\delta\mathbf{r}_{1PN}$ accounts for the 1PN correction to
the center of mass frame, which can be obtained following the procedure
presented through \eqref{eq:G2PN} and \eqref{eq:dr2PN} but this
time using \eqref{eq:T001PN}. Thus, the corrections to the center of frame
necessary to write the 2PN mass quadrupole are
\begin{align}
\delta\mathbf{r}_{1PN} & =\frac{\nu\delta m}{2m}\mathbf{r}\left(\mathbf{v}^{2}-\frac{Gm}{r}\right),\label{eq:dr1PN}\\
\delta\mathbf{r}_{2PN} & =\frac{\nu\delta m}{2m}\left\{ \mathbf{r}\left[\left(\frac{3}{4}-3\nu\right)\mathbf{v}^{4}+\frac{Gm}{r}\left(\left(\frac{19}{4}+3\nu\right)\mathbf{v}^{2}\right)\right.\right.\nonumber \\
 & \left.\left.+\left(-\frac{1}{4}+\frac{3\nu}{2}\right)\dot{r}^{2}+\left(\frac{7}{2}-\nu\right)\frac{Gm}{r}\right]-\mathbf{v}\left[\frac{7}{2}Gm\dot{r}\right]\right\} .\label{eq:dr2PN-2}
\end{align}
Applying \eqref{eq:x1} and \eqref{eq:x2} to \eqref{eq:mq0PN} and
\eqref{eq:mq1PN}, we obtain the following contributions at 2PN order:
\begin{equation}
I_{0PN+2PN}^{ij}=\frac{\nu^{2}\delta m^{2}}{4m}\left(\mathbf{v}^{4}-2\mathbf{v}^{2}\frac{Gm}{r}+\frac{G^{2}m^{2}}{r^{2}}\right)\left[\mathbf{r}^{i}\mathbf{r}^{j}\right]_{TF},
\end{equation}
\begin{align}
I_{1PN+1PN}^{ij} & =\frac{\nu^{2}\delta m^{2}}{21m}\left\{ \left[-29\mathbf{v}^{4}+\frac{Gm}{r}\left(41\mathbf{v}^{2}+\frac{17}{2}\dot{r}^{2}-12\frac{Gm}{r}\right)\right]\mathbf{r}^{i}\mathbf{r}^{j}\right.\nonumber \\
 & \left.+\left(24\mathbf{v}^{2}-19\frac{Gm}{r}\right)r\dot{r}\mathbf{v}^{i}\mathbf{r}^{j}+\left(-22\mathbf{v}^{2}+22\frac{Gm}{r}\right)r^{2}\mathbf{v}^{i}\mathbf{v}^{j}\right\} _{STF}.
\end{align}
Adding these two contributions to \eqref{eq:I2pn-notCM} after
applying \eqref{eq:x1} and \eqref{eq:x2}, we finally obtain the
expression for the 2PN correction to the mass quadrupole moment in
the center of mass frame,
\begin{align}
I_{2PN}^{ij} & =m\nu\left\{ \mathbf{r}^{i}\mathbf{r}^{j}\left[\frac{1}{252}\left(653-1906\nu+337\nu^{2}\right)\frac{G^{2}m^{2}}{r^{2}}+\frac{1}{756}\left(2021-5947\nu-4883\nu^{2}\right)\frac{Gm}{r}v^{2}\right.\right.\nonumber \\
 & -\frac{1}{756}\left(131-907\nu+1273\nu^{2}\right)\frac{Gm}{r}\dot{r}^{2}+\frac{1}{504}\left(253-1835\nu+3545\nu^{2}\right)v^{4}\nonumber \\
 & -r\dot{r}\mathbf{r}^{i}\mathbf{v}^{j}\left[\frac{1}{378}\left(1085-4057\nu-1463\nu^{2}\right)\frac{Gm}{r}+\frac{1}{63}\left(26-202\nu+418\nu^{2}\right)v^{2}\right]\nonumber \\
 & \left.+\mathbf{v}^{i}\mathbf{v}^{j}\left[\frac{1}{189}\left(742-335\nu-985\nu^{2}\right)\frac{Gm}{r}+\frac{1}{126}\left(41-337\nu+733\nu^{2}\right)v^{2}+\frac{5}{63}\left(1-5\nu+5\nu^{2}\right)\dot{r}^{2}\right]\right\} _{STF}.\label{eq:I2PN}
\end{align}

We can use the result above to compute, for instance, the 2PN correction
to the power loss, whose expression in terms of the multipole
moments is given by \cite{andirad2}
\begin{equation}
P=-\frac{G}{5}\left\{ I_{ij}^{(3)}I_{ij}^{(3)}-\frac{5}{189}I_{ijk}^{(4)}I_{ijk}^{(4)}+\frac{5}{9072}I_{ijkl}^{(5)}I_{ijkl}^{(5)}+\frac{16}{9}J_{ij}^{(3)}J_{ij}^{(3)}-\frac{5}{84}J_{ijk}^{(4)}J_{ijk}^{(4)}+...\right\} .
\end{equation}
The expressions for these multipole moments below 2PN order are known
and can be found for instance in \cite{Porto:2016pyg}. Considering all
terms which contribute to the power loss at 2PN order in the expression
above, making use of \eqref{eq:I2PN} and the 2PN acceleration \eqref{eq:A2PN} obtained
in the appendix
\ref{sec:AppB}, we get
\begin{align}
P_{EFT}^{2PN} & =-\frac{8}{15}\frac{G^{3}m^{4}\nu^{2}}{r^{4}}\left\{ \frac{2}{3}\left(-253+1026\nu-56\nu^{2}\right)\frac{G^{3}m^{3}}{r^{3}}+\left[\frac{1}{756}\left(245185+81828\nu+4368\nu^{2}\right)v^{2}\right.\right.\nonumber \\
 & \left.-\frac{1}{252}\left(97247+9798\nu+5376\nu^{2}\right)\dot{r}^{2}\right]\frac{G^{2}m^{2}}{r^{2}}+\left[\frac{1}{21}\left(-4446+5237\nu-1393\nu^{2}\right)v^{4}\right.\nonumber \\
 & \left.+\frac{1}{7}\left(4987-8513\nu+2165\nu^{2}\right)v^{2}\dot{r}^{2}-\frac{1}{63}\left(33510-60971\nu+14290\nu^{2}\right)\right]\frac{Gm}{r}\nonumber \\
 & +\frac{1}{42}\left(1692-5497\nu+4430\nu^{2}\right)v^{6}-\frac{1}{14}\left(1719-10278\nu+6292\nu^{2}\right)v^{4}\dot{r}^{2}\nonumber \\
 & \left.+\frac{1}{14}\left(2018-15207\nu+7572\nu^{2}\right)v^{2}\dot{r}^{4}-\frac{1}{42}\left(2501-20234\nu+8404\nu^{2}\right)\dot{r}^{6}\right\} .\label{eq:P2PN}
\end{align}
At this point we can see that \eqref{eq:I2PN} and \eqref{eq:P2PN}
seem to be in disagreement with the results presented at \cite{PhysRevD.54.4813} and \cite{PhysRevD.56.7708} where the Epstein-Wagoner formalism and multipolar post-Minkowskian approach of Blanchet, Damour, and Iyer (BDI) were used, respectively. For instance, the mass
quadrupole moment presented in this paper and ones in the mentioned
references differ by a factor of $-\frac{4G^{2}m^{2}}{r^{2}}\left[m\nu\mathbf{r}^{i}\mathbf{r}^{j}\right]_{TF}$.
The power loss shown above and the energy fluxes at (6.13d) in \cite{PhysRevD.54.4813} and at (3.5d) in \cite{PhysRevD.56.7708} differ
by a global minus sign, as well as by the numerical factors on terms
depending on $\frac{G^{5}m^{6}\nu^{2}v^{2}}{r^{6}}$ and $\frac{G^{5}m^{6}\nu^{2}\dot{r}^{2}}{r^{6}}$.
The difference in the global sign comes from the relation $P=-\frac{dE}{dt}$,
which is actually a matter of convention on how the energy flux is
defined. For this reason, we consider instead $\left|P\right|=\left|\frac{dE}{dt}\right|$
and compare the result for the power loss obtained here against the
ones in the literature, and we find the following difference
\begin{equation}
P_{EFT}-\bar{P}=\frac{32}{5}\frac{G^{5}m^{6}\nu^{2}}{r^{6}}\left(4v^{2}-3\dot{r}^{2}\right),
\end{equation}
where $\bar{P}$ is the modulus of the energy flux computed via the Epstein-Wagoner and BDI approaches \footnote{If the power is expressed in terms of the gauge invariant frequency of a circular orbit $P=\bar P$.}.

Furthermore, the 2PN
acceleration obtained in the appendix
\ref{sec:AppB} is also different from the one presented in \cite{PhysRevD.56.7708}, which was
computed via the BDI formalism. It turns out
that these differences should not be a surprise since the gauge choice
adopted here and in other formalisms are not the same: in the BDI and in the
Epstein-Wagoner approaches the harmonic gauge is used, while in the
EFT approach we use the linearized harmonic gauge \eqref{eq:GF},
which depends on the background field metric. The different gauge
choices for fixing the gravity action imply different coordinate systems.
In fact, the difference between the mass quadrupole moments suggests
a coordinate transformation of the form
\begin{equation}
\mathbf{r}_{EFT}\to\bar{\mathbf{r}}-\frac{2G^{2}m^{2}}{r^{2}}\bar{\mathbf{r}},
\end{equation}
where $\bar{\mathbf{r}}$ is the coordinate used in the BDI and Epstein-Wagoner
approaches. When this transformation is applied to the power loss
\eqref{eq:P2PN}, we can verify that
\begin{equation}
P_{EFT}\left(\bar{\mathbf{r}}\right)=\bar{P}.
\end{equation}
An analogous comparison holds for the mass quadrupole moment and the 2PN acceleration,
showing the agreement between our results and the literature. It should
also be noticed that this coordinate transformation was already brought
to attention in \cite{nrgr} when the authors used NRGR  to calculate the spacetime metric generated by a point
mass at rest.


\section{Final remarks\label{sec:Final-remarks}}

In this paper, we provided an independent computation of the 2PN correction to the mass quadrupole
moment of a binary system of compact bodies moving in general orbits,
using the EFT approach in the linearized harmonic gauge. We calculated
high order corrections to the components of the pseudo-stress-energy
tensor, which were used to obtain the mass quadrupole moment correction
as well as the 1PN correction to the conserved energy and to the linear
and angular momenta of the system and the 2PN correction to the center
of mass frame. We used these quantities to perform tests that confirmed
the consistency of our results within the EFT formalism itself and
with results presented in the literature computed using different
formalisms. Therefore, we not only extracted the contributions of the stress-energy pseudotensor to the 2PN correction to the mass quadrupole, but we provided the expressions for the components of the pseudotensor with higher order corrections that will be useful for future calculations on the dynamics of compact binary system.

We also calculated the 2PN correction to the equation of motion in
the linearized harmonic gauge that was used, together with the mass
quadrupole moment obtained in this paper, to write down the power
loss due to the emission of gravitational waves. We thus compared our results
against the literature and we showed that the 2PN correction to the
mass quadrupole moment, to the relative acceleration of the two-body
system and to the power loss obtained in this paper are in agreement
with the results computed via the BDI and in the Epstein-Wagoner formalisms
once a coordinate transformation is performed.

Although the 2PN correction to the mass quadrupole and to the equation of motion of compact binary systems obtained here were known in the literature, this derivation  establishes the ground work for  higher order calculations in the EFT formalism. Finally, these are the final missing ingredients necessary for the analysis of the radiation reaction of the binary system at the next-to-next-to-leading order in the EFT approach, which will be presented in a future
paper.


\section{Acknowledgements}

A.K.L., N.T.M, and Z.Y. are supported in part by the National Science Foundation under Grant No. PHY-1820760. 

\appendix

\section{}\label{sec:AppA}

In this appendix we show the ingredients used to compute the components
of the pseudotensor. We used the package xAct \cite{xact} from Mathematica
for the extraction of the vertices from the action.

\subsection*{Source terms}

The source action terms needed to compute the contributions to $T_{2PN}^{00}$
are given below: 
\begin{align}
S^{\mathbf{v}^{0}} & =-\sum_{a}\frac{m_{a}}{2m_{Pl}}\int dt_{a}H^{00}\left(x_{a}\right),\label{eq:ppH00v0}\\
S^{\mathbf{v}^{1}} & =\sum_{a}\frac{m_{a}}{m_{Pl}}\int dt_{a}\mathbf{v}_{a}^{i}H^{0i}\left(x_{a}\right),\label{eq:ppH0iv1}\\
S^{\mathbf{v}^{2}} & =-\sum_{a}\frac{m_{a}}{2m_{Pl}}\int dt_{a}\left(\frac{\mathbf{v}_{a}^{2}}{2}H^{00}\left(x_{a}\right)+\mathbf{v}_{a}^{i}\mathbf{v}_{a}^{j}H^{ij}\left(x_{a}\right)\right),\label{eq:ppH00Hijv2}\\
S_{\bar{h}^{00}}^{\mathbf{v}^{0}} & =\sum_{a}\frac{m_{a}}{4m_{Pl}^{2}}\int dt_{a}H^{00}\left(x_{a}\right)\bar{h}^{00}\left(x_{a}\right),\label{eq:ppH00h00}\\
S_{\bar{h}^{00}}^{\mathbf{v}^{1}} & =-\sum_{a}\frac{m_{a}}{2m_{Pl}^{2}}\int dt_{a}\mathbf{v}_{a}^{i}H^{0i}\left(x_{a}\right)\bar{h}^{00}\left(x_{a}\right),\label{eq:ppH0iv1h00}\\
S_{\bar{h}^{00}}^{\mathbf{v}^{2}} & =\sum_{a}\frac{m_{a}}{8m_{Pl}^{2}}\int dt_{a}\left(3\mathbf{v}_{a}^{2}H^{00}\left(x_{a}\right)+2\mathbf{v}_{a}^{i}\mathbf{v}_{a}^{j}H^{ij}\left(x_{a}\right)\right)\bar{h}^{00}\left(x_{a}\right),\label{eq:pph00v2}\\
S_{\bar{h}^{00}}^{\mathbf{v}^{4}} & =-\sum_{a}\frac{3m_{a}}{16m_{Pl}}\int dt_{a}\mathbf{v}_{a}^{4}\bar{h}^{00}\left(x_{a}\right),\label{eq:pph00v4}\\
S_{H^{2}}^{\mathbf{v}^{0}} & =\sum_{a}\frac{m_{a}}{8m_{Pl}^{2}}\int dt_{a}H^{00}\left(x_{a}\right)H^{00}\left(x_{a}\right),\label{eq:ppHH}\\
S_{H^{2}\bar{h}^{00}}^{\mathbf{v}^{0}} & =-\sum_{a}\frac{3m_{a}}{16m_{Pl}^{3}}\int dt_{a}H^{00}\left(x_{a}\right)H^{00}\left(x_{a}\right)\bar{h}^{00}\left(x_{a}\right).\label{eq:ppHHh}
\end{align}
In addition, to write down the contributions for $T_{1PN}^{0i}$ we
must to consider
\begin{align}
S_{\bar{h}^{0i}}^{\mathbf{v}^{1}} & =-\sum_{a}\frac{m_{a}}{2m_{Pl}^{2}}\int dt_{a}\mathbf{v}_{a}^{i}H^{00}\left(x_{a}\right)\bar{h}^{0i}\left(x_{a}\right),\label{eq:ppHv1h0i}\\
S_{\bar{h}^{0i}}^{\mathbf{v}^{3}} & =\sum_{a}\frac{m_{a}}{2m_{Pl}}\int dt\mathbf{v}_{a}^{2}\mathbf{v}_{a}^{i}\bar{h}^{0i}\left(x_{a}\right),\label{eq:ppv3h0i}
\end{align}
whereas for $T_{1PN}^{ll}$ the following terms are also necessary,
\begin{align}
S_{\bar{h}^{ij}}^{\mathbf{v}^{2}} & =\sum_{a}\frac{m_{a}}{4m_{pl}^{2}}\int dt_{a}\mathbf{v}_{a}^{i}\mathbf{v}_{a}^{j}H^{00}\left(x_{a}\right)\overline{h}^{ij}\left(x_{a}\right),\label{eq:ppHv2hij}\\
S_{\bar{h}^{ij}}^{\mathbf{v}^{4}} & =-\sum_{a}\frac{m_{a}}{4m_{pl}}\int dt_{a}\mathbf{v}_{a}^{2}\mathbf{v}_{a}^{i}\mathbf{v}_{a}^{j}\overline{h}^{ij}\left(x_{a}\right).\label{eq:ppv4hij}
\end{align}

Although all the sources terms above are conveniently expressed in
position space, effectively we perform the partial Fourier transform\footnote{We consider the partial Fourier transform for the radiation field
as well.}
\begin{equation}
H^{\mu\nu}\left(t,\mathbf{q}\right)=\int d^{3}xH^{\mu\nu}\left(t,\mathbf{x}\right)e^{-i\mathbf{q}\cdot\mathbf{x}},
\end{equation}
to carry out the Feynman diagrams in momentum space.

\subsection*{Vertices}

From the EH action expanded in the radiation and potential fields
and fixed with the background gauge, we obtain the propagator
\begin{equation}
\left\langle H_{\mu\nu}\left(t,\mathbf{q}\right)H_{\alpha\beta}\left(t',\mathbf{q}'\right)\right\rangle =-i\left(2\pi\right)^{3}P_{\mu\nu\alpha\beta}\delta\left(t-t'\right)\delta^{3}\left(\mathbf{q}+\mathbf{q}'\right)\frac{1}{\mathbf{q}^{2}},\label{eq:prop}
\end{equation}
as well as its correction
\begin{equation}
\left\langle H_{\mu\nu}\left(t,\mathbf{q}\right)H_{\alpha\beta}\left(t',\mathbf{q}'\right)\right\rangle _{v^{2}}=-i\left(2\pi\right)^{3}P_{\mu\nu\alpha\beta}\frac{d^{2}}{dtdt'}\delta\left(t-t'\right)\delta^{3}\left(\mathbf{q}+\mathbf{q}'\right)\frac{1}{\mathbf{q}^{4}}.\label{eq:corrprop}
\end{equation}

The two-potential-one-radiation vertex regarded inside the momentum
integrals of the internal potential momenta coupled to the particles
has the form 
\begin{equation}
\int\frac{d^{3}\mathbf{q}}{\left(2\pi\right)^{3}}\int\frac{d^{3}\mathbf{q}'}{\left(2\pi\right)^{3}}e^{-i\mathbf{q}\cdot\mathbf{x}_{1}}e^{-i\mathbf{q}'\cdot\mathbf{x}_{2}}\left\langle iS_{\bar{h}H^{2}}\right\rangle =-\frac{i}{m_{Pl}}\delta\left(t-t'\right)\int\frac{d^{3}\mathbf{q}}{\left(2\pi\right)^{3}}e^{-i\mathbf{q}\cdot\mathbf{x}}\frac{F\left[q,k,\bar{h}\right]}{\mathbf{q}^{2}\left(\mathbf{q}+\mathbf{k}\right)^{2}},\label{eq:hHH}
\end{equation}
for which the different contractions necessary to write down the contributions
to $T_{2PN}^{00}$ are
\begin{align}
F^{\left\langle H^{00}H^{00}\right\rangle }\left[q,k,\bar{h}^{00}\right] & =\bar{h}^{00}\left[\frac{3}{4}\left(\mathbf{q}^{2}+\mathbf{k}\cdot\mathbf{q}\right)-\frac{5}{4}q_{0}^{2}-\frac{5}{4}k_{0}q_{0}-\frac{1}{2}k_{0}^{2}\right],\label{eq:hHH0}\\
F^{\left\langle \mathbf{v}_{1}^{k}H^{0k}H^{00}\right\rangle }\left[q,k,\bar{h}^{00}\right] & =\bar{h}^{00}\mathbf{v}_{1}^{k}\left[-\mathbf{q}^{k}\left(q_{0}+\frac{1}{2}k_{0}\right)\right],\label{eq:hHH1}\\
F^{\left\langle \mathbf{v}_{1}^{k}\mathbf{v}_{1}^{l}H^{kl}H^{00}\right\rangle }\left[q,k,\bar{h}^{00}\right] & =\bar{h}^{00}\mathbf{v}_{1}^{k}\mathbf{v}_{1}^{l}\left[\frac{1}{4}\delta^{kl}\left(\mathbf{q}^{2}+3\mathbf{k}\cdot\mathbf{q}\right)-\frac{1}{2}\mathbf{k}^{k}\mathbf{k}^{l}\right],\label{eq:hHH2}\\
F^{\left\langle \mathbf{v}_{1}^{k}H^{0k}H^{0l}\mathbf{v}_{2}^{l}\right\rangle }\left[q,k,\bar{h}^{00}\right] & =\bar{h}^{00}\mathbf{v}_{1}^{k}\mathbf{v}_{2}^{l}\left[-\frac{1}{4}\delta^{kl}\left(\mathbf{q}^{2}+\mathbf{k}\cdot\mathbf{q}\right)+\frac{1}{4}\mathbf{k}^{k}\mathbf{k}^{l}\right].\label{eq:hHH3}
\end{align}
On the other hand, to compute the contributions to $T_{1PN}^{0i}$,
the contractions required are 
\begin{align}
F^{\left\langle H^{00}H^{00}\right\rangle }\left[q,k,\bar{h}^{0i}\right] & =\bar{h}^{0i}\left[q_{0}\left(\mathbf{q}^{i}+\frac{1}{2}\mathbf{k}^{i}\right)+k_{0}\left(\frac{1}{2}\mathbf{q}^{i}+\mathbf{k}^{i}\right)\right],\label{eq:hi0HH}\\
F^{\left\langle \mathbf{v}_{1}^{k}H^{0k}H^{00}\right\rangle }\left[q,k,\bar{h}^{0i}\right] & =\bar{h}^{0i}\mathbf{v}_{1}^{k}\left[-\delta^{ik}\left(\frac{1}{2}\mathbf{q}^{2}+\mathbf{k}\cdot\mathbf{q}\right)+\mathbf{q}^{i}\mathbf{k}^{k}+\frac{1}{2}\mathbf{k}^{i}\mathbf{k}^{k}\right],\label{eq:hi0HHv1}
\end{align}
whereas for $T_{1PN}^{ll}$ we need
\begin{align}
F^{\left\langle H^{00}H^{00}\right\rangle }\left[q,k,\bar{h}^{ll}\right] & =\bar{h}^{ll}\left[\frac{1}{4}\mathbf{q}^{2}+\frac{1}{4}\mathbf{k}\cdot\mathbf{q}-\frac{3}{4}\left(q_{0}^{2}+k_{0}q_{0}+2k_{0}^{2}\right)\right],\label{eq:hHH0-1}\\
F^{\left\langle \mathbf{v}_{1}^{k}H^{0k}H^{00}\right\rangle }\left[q,k,\bar{h}^{ll}\right] & =\bar{h}^{ll}\mathbf{v}_{1}^{k}\left[-k_{0}\left(\mathbf{q}^{k}+\frac{1}{2}\mathbf{k}^{k}\right)\right],\label{eq:hHH1-1}\\
F^{\left\langle \mathbf{v}_{1}^{k}\mathbf{v}_{1}^{l}H^{kl}H^{00}\right\rangle }\left[q,k,\bar{h}^{ll}\right] & =\bar{h}^{ll}\mathbf{v}_{1}^{k}\mathbf{v}_{1}^{l}\left[\delta^{kl}\left(-\frac{1}{4}\mathbf{q}^{2}+\frac{1}{4}\mathbf{k}\cdot\mathbf{q}\right)-\frac{1}{2}\mathbf{k}^{k}\mathbf{k}^{l}\right],\label{eq:hHH2-1}\\
F^{\left\langle \mathbf{v}_{1}^{k}H^{0k}H^{0l}\mathbf{v}_{2}^{l}\right\rangle }\left[q,k,\bar{h}^{ll}\right] & =\bar{h}^{ll}\mathbf{v}_{1}^{k}\mathbf{v}_{2}^{l}\left[\frac{1}{4}\delta^{kl}\left(\mathbf{q}^{2}+\mathbf{k}\cdot\mathbf{q}\right)-\frac{1}{2}\mathbf{q}^{l}\mathbf{k}^{k}+\frac{1}{2}\mathbf{q}^{k}\mathbf{k}^{l}+\frac{1}{4}\mathbf{k}^{k}\mathbf{k}^{l}\right].\label{eq:hHH3-1}
\end{align}

The three-graviton vertex, in turn, comes naturally in a simple form
even not integrated on the internal momenta:
\begin{align}
\left\langle H_{\mathbf{q}_{1}}^{00}H_{\mathbf{q}_{2}}^{00}H_{\mathbf{q}_{3}}^{00}\right\rangle  & =-\frac{\left(2\pi\right)^{3}}{4m_{Pl}}\delta\left(t_{2}-t_{1}\right)\delta\left(t_{3}-t_{1}\right)\delta^{3}\left(\boldsymbol{q}_{1}+\boldsymbol{q}_{2}+\boldsymbol{q}_{3}\right)\frac{\left(\boldsymbol{q}_{1}^{2}+\boldsymbol{q}_{2}^{2}+\boldsymbol{q}_{3}^{2}\right)}{\boldsymbol{q}_{1}^{2}\boldsymbol{q}_{2}^{2}\boldsymbol{q}_{3}^{2}}.\label{eq:3pot}
\end{align}

The composition of the three-potential-graviton vertex with two-potential-one-radiation-graviton
vertex, after integrating in the third momentum, the integrand takes
the form
\begin{equation}
\frac{F\left[\mathbf{q}_{1},\mathbf{q}_{2},\mathbf{k},\bar{h}\right]}{\mathbf{q}_{1}^{2}\mathbf{q}_{2}^{2}\left(\mathbf{q}_{1}+\mathbf{k}\right)^{2}\left(\mathbf{q}_{1}+\mathbf{q}_{2}+\mathbf{k}\right)^{2}},
\end{equation}
in which the numerators for the contractions needed to compute the
contributions for $T_{2PN}^{00}$ and $T_{1PN}^{ll}$ are, respectively,
\begin{align}
F^{\left\langle H^{00}H^{00}H^{00}\right\rangle }\left[\mathbf{q}_{1},\mathbf{q}_{2},\mathbf{k},\bar{h}^{00}\right] & =\frac{1}{4}\bar{h}_{00}\left[\mathbf{q}_{1}^{4}+\frac{5}{2}\mathbf{q}_{1}^{2}\left(\mathbf{q}_{1}\cdot\mathbf{q}_{2}\right)+\frac{5}{2}\mathbf{q}_{1}^{2}\mathbf{q}_{2}^{2}+\left(\mathbf{q}_{1}+\mathbf{q}_{2}\right)^{2}\left(\mathbf{q}_{1}\cdot\mathbf{k}\right)\right.\nonumber \\
 & \left.+\frac{5}{2}\mathbf{q}_{1}^{2}\left(\mathbf{q}_{2}\cdot\mathbf{k}\right)+3\left(\mathbf{q}_{1}\cdot\mathbf{k}\right)\left(\mathbf{q}_{2}\cdot\mathbf{k}\right)-\left(\mathbf{q}_{1}\cdot\mathbf{k}\right)^{2}+\left(\mathbf{q}_{2}\cdot\mathbf{k}\right)^{2}\right],\label{eq:comp00}\\
F^{\left\langle H^{00}H^{00}H^{00}\right\rangle }\left[\mathbf{q}_{1},\mathbf{q}_{2},\mathbf{k},\bar{h}^{ll}\right] & =-\frac{\bar{h}^{ll}}{8}\left[2\mathbf{q}_{1}^{4}-\mathbf{q}_{1}^{2}\mathbf{q}_{1}\cdot\mathbf{q}_{2}+10\mathbf{q}_{1}^{2}\mathbf{q}_{1}\cdot\mathbf{k}+10\left(\mathbf{q}_{1}\cdot\mathbf{k}\right)^{2}-\mathbf{q}_{1}^{2}\mathbf{q}_{2}^{2}\right.\nonumber \\
 & \left.-\mathbf{q}_{1}^{2}\mathbf{q}_{2}\cdot\mathbf{k}-2\left(\mathbf{q}_{1}\cdot\mathbf{k}\right)\left(\mathbf{q}_{2}\cdot\mathbf{k}\right)-2\left(\mathbf{q}_{2}\cdot\mathbf{k}\right)^{2}\right].\label{eq:compij}
\end{align}

The three-potential-one-radiation-graviton vertex integrated in the
internal momenta can be expressed in this way:
\begin{equation}
\prod_{i=1}^{3}\int\frac{d^{3}\mathbf{q}_{i}}{\left(2\pi\right)^{9}}e^{i\mathbf{q}_{i}\cdot\mathbf{x}_{i}}\left\langle iS_{\bar{h}H^{3}}\right\rangle =-\frac{1}{m_{Pl}^{2}}\delta\left(t_{2}-t_{1}\right)\delta\left(t_{3}-t_{1}\right)\int\frac{d^{3}\mathbf{q}_{2}}{\left(2\pi\right)^{3}}\int\frac{d^{3}\mathbf{q}_{3}}{\left(2\pi\right)^{3}}\frac{e^{i\left(\mathbf{q}_{2}+\mathbf{q}_{3}\right)\cdot\mathbf{x}}F\left[\mathbf{q}_{2},\mathbf{q}_{3},\mathbf{k},\bar{h}\right]}{\mathbf{q}_{2}^{2}\mathbf{q}_{3}^{2}\left(\mathbf{q}_{2}+\mathbf{q}_{3}+\mathbf{k}\right)^{2}},\label{eq:4grav}
\end{equation}
where we have chosen to integrate on $\mathbf{q}_{1}$, for instance
coupled to particle 1, and leaving the momenta $\mathbf{q}_{2}$ and
\textbf{$\mathbf{q}_{3}$}, both coupled to particle 2, to be integrated
in the process of solving the diagrams. For this case, the contractions
required to write down the contribution for $T_{2PN}^{00}$ and $T_{1PN}^{ll}$,
respectively, are given by
\begin{align}
F^{\left\langle H^{00}H^{00}H^{00}\right\rangle }\left[\mathbf{q}_{2},\mathbf{q}_{3},\mathbf{k},\bar{h}^{00}\right] & =-\frac{1}{8}\bar{h}^{00}\left(\mathbf{q}_{2}^{2}+\mathbf{q}_{3}^{2}+\mathbf{q}_{2}\cdot\mathbf{q}_{3}+\mathbf{q}_{2}\cdot\mathbf{k}+\mathbf{q}_{3}\cdot\mathbf{k}\right),\label{eq:4grav00}\\
F^{\left\langle H^{00}H^{00}H^{00}\right\rangle }\left[\mathbf{q}_{2},\mathbf{q}_{3},\mathbf{k},\bar{h}^{ll}\right] & =-\frac{7}{8}\bar{h}^{ll}\left(\mathbf{q}_{2}^{2}+\mathbf{q}_{3}^{2}+\mathbf{q}_{2}\cdot\mathbf{q}_{3}+\mathbf{q}_{2}\cdot\mathbf{k}+\mathbf{q}_{3}\cdot\mathbf{k}\right).\label{eq:4gravij}
\end{align}

\subsection*{Integrals}

To solve integrals in the momentum space, it is helpful to use some
general relations that can be obtained by using Feynman parameters \cite{Peskin:1995ev}.
If we consider a spacetime of $d$ dimensions, then
for $D=d-1$ we have
\begin{align}
\int\frac{d^{D}\mathbf{k}}{\left(2\pi\right)^{D}}\frac{e^{-i\mathbf{k}\cdot\mathbf{r}}}{\left(\mathbf{k}^{2}\right)^{a}} & =\frac{1}{\left(4\pi\right)^{\frac{D}{2}}}\frac{\Gamma\left(\frac{D}{2}-a\right)}{\Gamma\left(a\right)}\left(\frac{r^{2}}{4}\right)^{a-\frac{D}{2}},\\
\int\frac{d^{D}\mathbf{k}}{\left(2\pi\right)^{D}}\frac{1}{\left[\mathbf{k}^{2}\right]^{a}\left[\left(\mathbf{k}-\mathbf{p}\right)^{2}\right]^{b}} & =\frac{\left(\mathbf{p}^{2}\right)^{\frac{D}{2}-a-b}}{\left(4\pi\right)^{\frac{D}{2}}}\frac{\Gamma\left(a+b-\frac{D}{2}\right)}{\Gamma\left(a\right)\Gamma\left(b\right)}\frac{\Gamma\left(\frac{D}{2}-a\right)\Gamma\left(\frac{D}{2}-b\right)}{\Gamma\left(D-a-b\right)},\\
\int\frac{d^{D}\mathbf{k}}{\left(2\pi\right)^{D}}\frac{\mathbf{k}^{i}}{\left[\mathbf{k}^{2}\right]^{a}\left[\left(\mathbf{k}-\mathbf{p}\right)^{2}\right]^{b}} & =\frac{\mathbf{p}^{i}\left(\mathbf{p}^{2}\right)^{\frac{D}{2}-a-b}}{\left(4\pi\right)^{\frac{D}{2}}}\frac{\Gamma\left(a+b-\frac{D}{2}\right)}{\Gamma\left(a\right)\Gamma\left(b\right)}\frac{\Gamma\left(\frac{D}{2}-a+1\right)\Gamma\left(\frac{D}{2}-b\right)}{\Gamma\left(D-a-b+1\right)},\\
\int\frac{d^{D}\mathbf{k}}{\left(2\pi\right)^{D}}\frac{\mathbf{k}^{i}\mathbf{k}^{j}}{\left[\mathbf{k}^{2}\right]^{a}\left[\left(\mathbf{k}-\mathbf{p}\right)^{2}\right]^{b}} & =\frac{1}{\left(4\pi\right)^{\frac{D}{2}}}\frac{\left(\mathbf{p}^{2}\right)^{\frac{D}{2}-a-b}}{\Gamma\left(a\right)\Gamma\left(b\right)\Gamma\left(D-a-b+2\right)}\nonumber \\
 & \times\left\{ \frac{g^{ij}\mathbf{p}^{2}}{2}\Gamma\left(a+b-1-\frac{D}{2}\right)\Gamma\left(\frac{D}{2}-a+1\right)\Gamma\left(\frac{D}{2}-b+1\right)\right.\nonumber \\
 & \left.+\mathbf{p}^{i}\mathbf{p}^{j}\Gamma\left(a+b-\frac{D}{2}\right)\Gamma\left(\frac{D}{2}-b\right)\Gamma\left(\frac{D}{2}-a+2\right)\right\} .
\end{align}
These integrals are especially important to solve diagrams that has
a composition of the three-potential-graviton vertex with the two-potential-one-radiation
vertex, where an analysis of the integrals in an arbitrary dimension
$D$ is required to handle divergences.

\section{}\label{sec:AppB}

In this appendix we present the result for the 2PN acceleration computed
via the EFT approach in the linearized harmonic gauge.

To write down the equation of motion of the binary system at 2PN order,
we need to obtain the Lagrangian by integrating out the potential
modes of the gravitational fields in the action \eqref{eq:Sinitial}.
Below the diagrams which contribute to the dynamics at 2PN order are
presented.

\begin{figure}[H]
\label{2PNLag_1}
\begin{centering}
\includegraphics[scale=0.35]{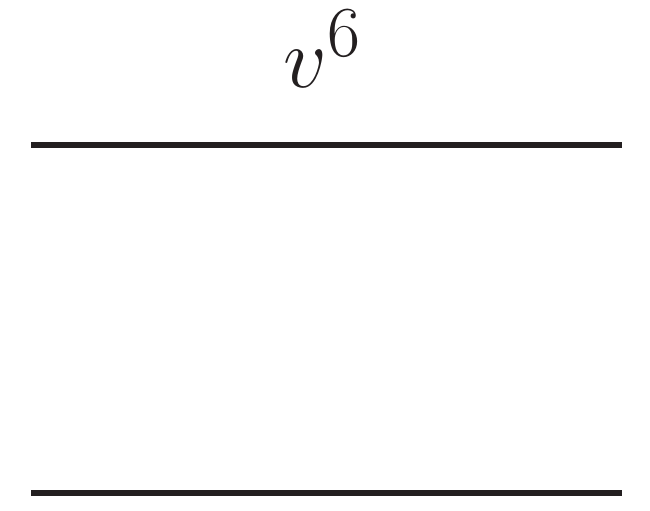}
\par\end{centering}
\caption{Diagram with no graviton exchange.}
\end{figure}

The simplest contribution to the 2PN Lagrangian comes from the diagram
show in Fig.~\ref{2PNLag_1}, which gives the following contribution:
\begin{equation}
L_{Fig11}=\sum_{a}\frac{1}{16}m_{a}\mathbf{v}_{a}^{6}.
\end{equation}

\begin{figure}[H]
\label{2PNLag_2a_2f}
\begin{centering}
\includegraphics[scale=0.35]{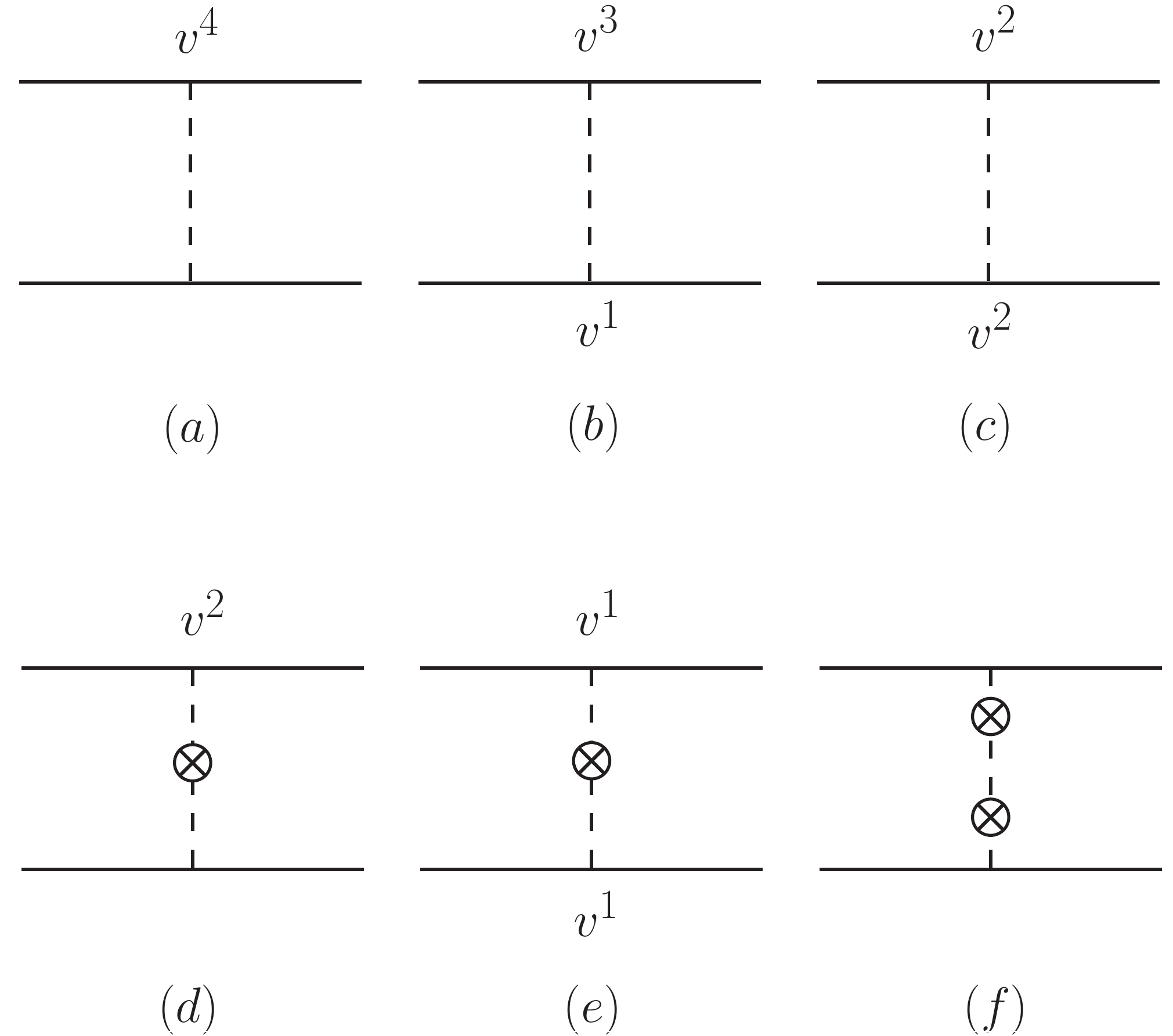}
\par\end{centering}
\caption{Diagrams with one-graviton exchange.}
\end{figure}

Next, we have the diagrams with one-graviton exchange illustrated
in Fig.~\ref{2PNLag_2a_2f}. Summing those diagrams together yields
\begin{align}
L_{Fig12} & =\sum_{a\neq b}\frac{Gm_{a}m_{b}}{16r^{3}}\left\{ 15r^{4}\mathbf{a}_{a}\cdot\mathbf{a}_{b}+r^{2}\left[14\mathbf{v}_{a}^{2}-20\mathbf{v}_{a}^{2}\mathbf{v}_{a}\cdot\mathbf{v}_{b}+2\left(\mathbf{v}_{a}\cdot\mathbf{v}_{b}\right)^{2}\right.\right.\nonumber \\
 & \left.+3\mathbf{v}_{a}^{2}\mathbf{v}_{b}^{2}+2\mathbf{v}_{b}^{2}\mathbf{a}_{a}\cdot\mathbf{r}-\mathbf{a}_{a}\cdot\mathbf{r}\mathbf{a}_{b}\cdot\mathbf{r}+28\mathbf{a}_{b}\cdot\mathbf{v}_{a}\mathbf{v}_{a}\cdot\mathbf{r}+24\mathbf{a}_{a}\cdot\mathbf{v}_{a}\mathbf{v}_{b}\cdot\mathbf{r}\right]\nonumber \\
 & \left.+2\left(\mathbf{a}_{b}\cdot\mathbf{r}-\mathbf{v}_{b}^{2}\right)\left(\mathbf{v}_{a}\cdot\mathbf{r}\right)^{2}+12\left(\mathbf{v}_{a}\cdot\mathbf{v}_{b}-\mathbf{v}_{a}^{2}\right)\mathbf{v}_{a}\cdot\mathbf{r}\mathbf{v}_{b}\cdot\mathbf{r}+\frac{3}{r^{2}}\left(\mathbf{v}_{a}\cdot\mathbf{r}\right)^{2}\left(\mathbf{v}_{b}\cdot\mathbf{r}\right)^{2}\right\} .
\end{align}

\begin{figure}[H]
\label{2PNLag_3a_3e}
\begin{centering}
\includegraphics[scale=0.35]{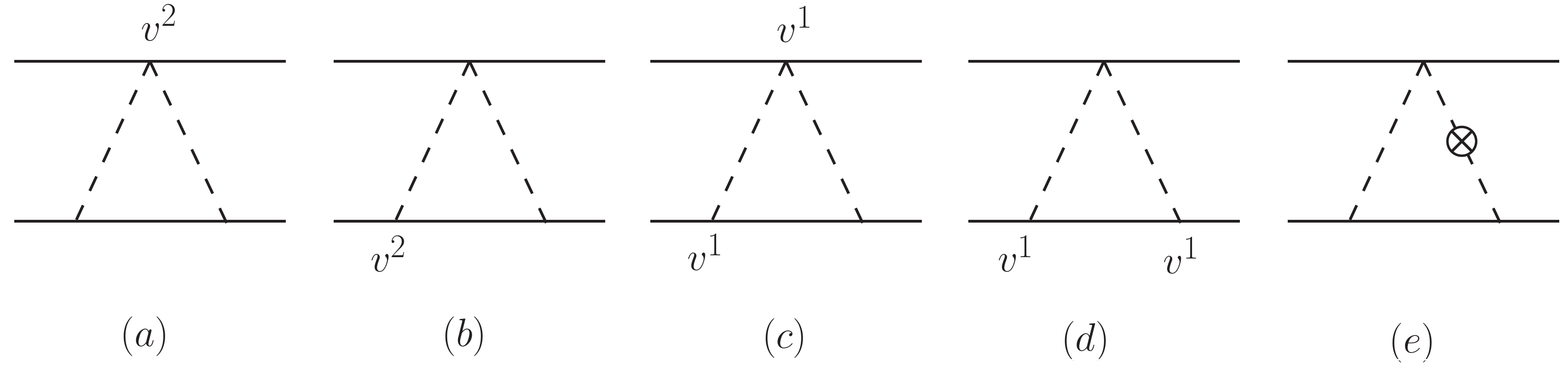}
\par\end{centering}
\caption{Diagrams with two-graviton exchange.}

\end{figure}

In Fig.~\ref{2PNLag_3a_3e} we show all diagrams with two-graviton exchange that enter
at the second PN order. The sum of those diagrams is
\begin{equation}
L_{Fig13}=\sum_{a\neq b}\frac{G^{2}m_{a}^{2}m_{b}}{4r^{4}}\left(6r^{2}\mathbf{v}_{a}^{2}+7r^{2}\mathbf{v}_{b}^{2}-14r^{2}\mathbf{v}_{a}\cdot\mathbf{v}_{b}+2\dot{r}\mathbf{v}_{a}\cdot\mathbf{r}-2\mathbf{v}_{a}\cdot\mathbf{r}\mathbf{v}_{b}\cdot\mathbf{r}\right).
\end{equation}

\begin{figure}[H]
\label{2PNLag_4a_4d}
\begin{centering}
\includegraphics[scale=0.35]{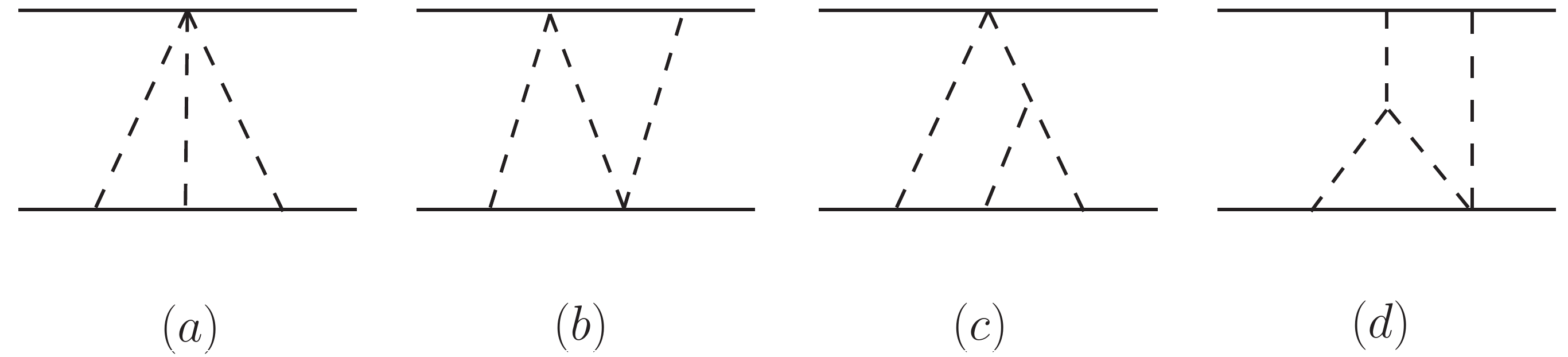}
\par\end{centering}
\caption{(a) three-graviton emission from one of the bodies; (b) symmetric
three-graviton exchange; (c) composition of a three-graviton vertex
with a two-graviton vertex in the source term. }

\end{figure}

There is also the diagram with a three-graviton source term as well
as other two diagrams with combinations of the two-graviton source,
as shown in Fig.~\ref{2PNLag_4a_4d}. Their contribution to the Lagrangian is
\begin{equation}
L_{Fig14}=-\sum_{a\neq b}\frac{G^{3}m_{a}^{2}m_{b}}{2r^{3}}\left(m_{a}+3m_{b}\right).
\end{equation}

\begin{figure}[H]
\label{2PNLag_5a_5i}
\begin{centering}
\includegraphics[scale=0.35]{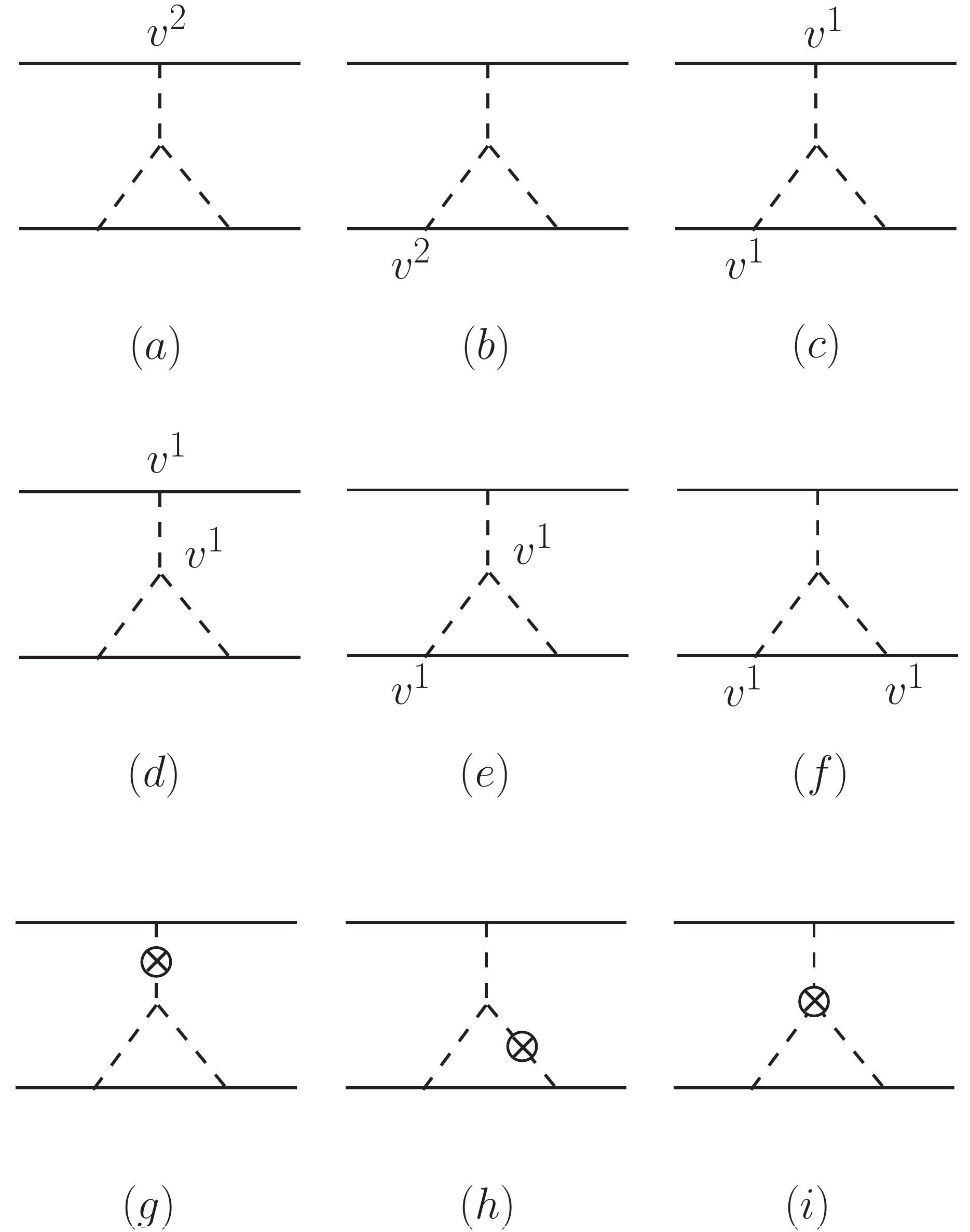}
\par\end{centering}
\caption{Diagrams with three-graviton exchange.}

\end{figure}

The diagrams which contain three-graviton vertices are illustrated
in Fig.~\ref{2PNLag_5a_5i} and give
\begin{equation}
L_{Fig15}=\sum_{a\neq b}\frac{G^{2}m_{a}^{2}m_{b}}{2r^{4}}\left[r^{2}\left(5\mathbf{v}_{a}^{2}-6\mathbf{v}_{a}\cdot\mathbf{v}_{b}+2\mathbf{v}_{b}^{2}+2\mathbf{a}_{b}\cdot\mathbf{r}\right)-9\left(\mathbf{v}_{a}\cdot\mathbf{r}\right)^{2}+14\mathbf{v}_{a}\cdot\mathbf{r}\mathbf{v}_{b}\cdot\mathbf{r}-3\left(\mathbf{v}_{b}\cdot\mathbf{r}\right)^{2}\right].
\end{equation}

\begin{figure}[H]
\label{2PNLag_6a_6b}
\begin{centering}
\includegraphics[scale=0.35]{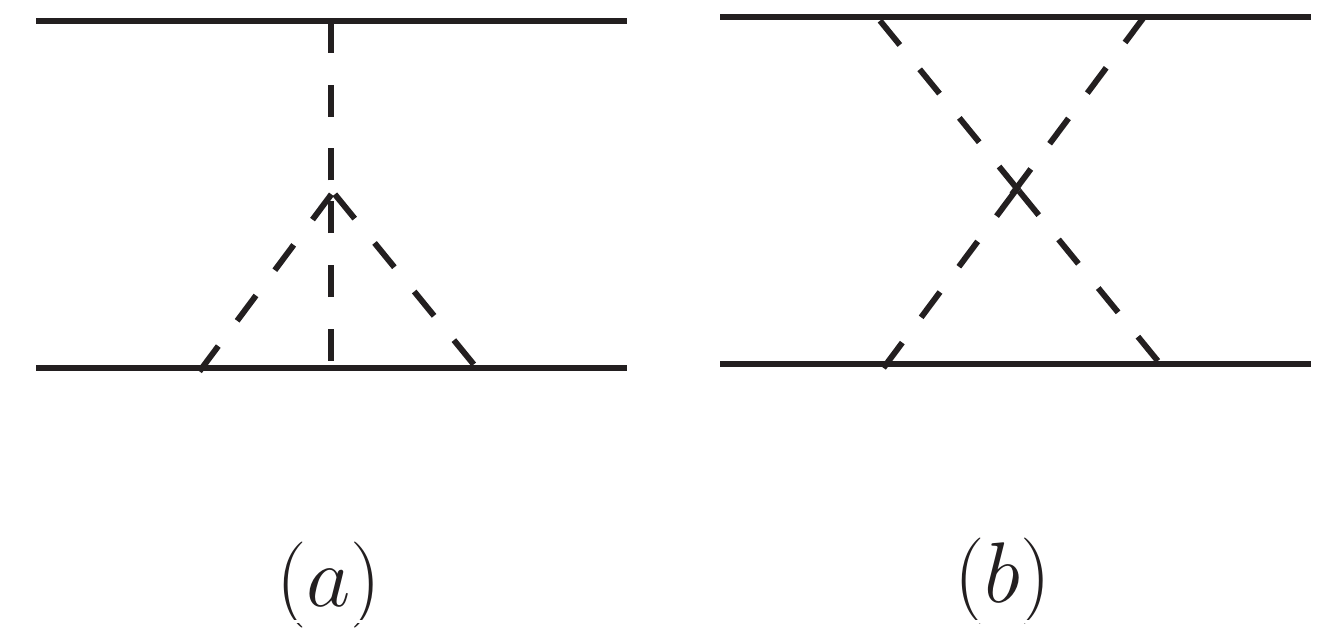}
\par\end{centering}
\centering{}\caption{Diagrams with four-graviton vertex.}
\end{figure}

In Fig.~\ref{2PNLag_6a_6b}, we show diagrams with a four-graviton vertex that enter at the
2PN order and, together, yield the result
\begin{equation}
L_{Fig16}=\sum_{a\neq b}\frac{G^{3}m_{a}^{3}m_{b}}{r^{3}}.
\end{equation}

\begin{figure}[H]
\label{2PNLag_7a_7c}
\begin{centering}
\includegraphics[scale=0.35]{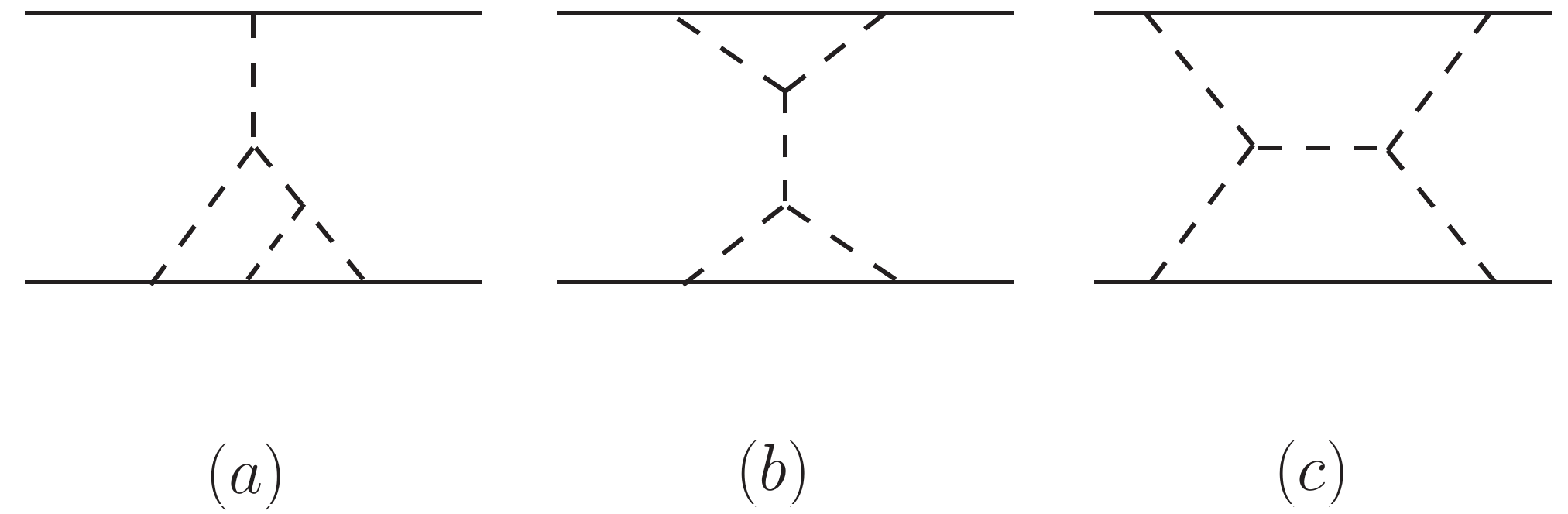}
\par\end{centering}
\caption{Diagrams with five propagators.}
\end{figure}

Lastly, the diagrams with five propagators are shown in Fig.~\ref{2PNLag_7a_7c} and
provide us with the following result:
\begin{equation}
L_{Fig17}=\sum_{a\neq b}\frac{G^{3}m_{a}^{2}m_{b}}{r^{3}}\left(m_{b}-2m_{a}\right).
\end{equation}

Summing up all contributions from Fig.~\ref{2PNLag_1} to Fig.~\ref{2PNLag_7a_7c}, we write down
the Lagrangian at 2PN order in the linearized harmonic gauge:
\begin{align}
L_{2PN} & =\frac{1}{16}m_{1}\mathbf{v}_{1}^{6}-\frac{G^{3}m_{1}m_{2}}{2r^{3}}\left(3m_{1}^{2}+m_{1}m_{2}\right)+\frac{G^{2}m_{1}m_{2}}{4r^{2}}\left[\left(16m_{1}+11m_{2}\right)\mathbf{v}_{1}^{2}\right.\nonumber \\
 & \left.-13m\mathbf{v}_{1}\cdot\mathbf{v}_{2}-4m_{2}\mathbf{a}_{1}\cdot\mathbf{r}-\frac{2}{r^{2}}\left(8m_{1}+3m_{2}\right)\left(\mathbf{v}_{1}\cdot\mathbf{r}\right)^{2}+\frac{12}{r^{2}}m\mathbf{v}_{1}\cdot\mathbf{r}\mathbf{v}_{2}\cdot\mathbf{r}\right]\nonumber \\
 & +\frac{Gm_{1}m_{2}}{8r}\left[\frac{15}{2}r^{2}\mathbf{a}_{1}\cdot\mathbf{a}_{2}+7\mathbf{v}_{1}^{4}-10\mathbf{v}_{1}^{2}\mathbf{v}_{1}\cdot\mathbf{v}_{2}+\left(\mathbf{v}_{1}\cdot\mathbf{v}_{2}\right)^{2}+\frac{3}{2}\mathbf{v}_{1}^{2}\mathbf{v}_{2}^{2}\right.\nonumber \\
 & +\mathbf{a}_{1}\cdot\mathbf{r}\mathbf{v}_{2}^{2}-14\mathbf{a}_{1}\cdot\mathbf{v}_{2}\mathbf{v}_{2}\cdot\mathbf{r}+12\mathbf{a}_{1}\cdot\mathbf{v}_{1}\mathbf{v}_{2}\cdot\mathbf{r}-\frac{1}{2}\mathbf{a}_{1}\cdot\mathbf{r}\mathbf{a}_{2}\cdot\mathbf{r}-\frac{1}{r^{2}}\mathbf{a}_{1}\cdot\mathbf{r}\left(\mathbf{v}_{2}\cdot\mathbf{r}\right)^{2}\nonumber \\
 & \left.+\frac{1}{r^{2}}\left(6\mathbf{v}_{1}\cdot\mathbf{r}\mathbf{v}_{2}\cdot\mathbf{r}\mathbf{v}_{1}\cdot\mathbf{v}_{2}-\left(\mathbf{v}_{1}\cdot\mathbf{r}\right)^{2}\mathbf{v}_{2}^{2}-6\mathbf{v}_{1}\cdot\mathbf{r}\mathbf{v}_{2}\cdot\mathbf{r}\mathbf{v}_{2}^{2}+\frac{3}{2r^{2}}\left(\mathbf{v}_{1}\cdot\mathbf{r}\right)^{2}\left(\mathbf{v}_{2}\cdot\mathbf{r}\right)^{2}\right)\right]\nonumber \\
 & +1\leftrightarrow2.
\end{align}

We use the Lagrangian above to determine the equations of motion of
the two-body system at the second PN order. Below we show the acceleration
for one of the objects in the binary:
\begin{align}
\mathbf{a}_{1}^{2PN} & =\frac{1}{8}\frac{Gm_{2}}{r^{3}}\mathbf{r}\left\{ \frac{G^{2}}{r^{2}}\left(-2m_{1}^{2}-20m_{1}m_{2}+16m_{2}^{2}\right)+\frac{G}{r}\left[\left(18m_{1}+56m_{2}\right)\mathbf{v}_{1}^{2}\right.\right.\nonumber \\
 & -\left(84m_{1}+128m_{2}\right)\mathbf{v}_{1}\cdot\mathbf{v}_{2}+\left(58m_{1}+64m_{2}\right)\mathbf{v}_{2}^{2}+30m_{1}\mathbf{a}_{1}\cdot\mathbf{r}-12m\mathbf{a}_{2}\cdot\mathbf{r}\nonumber \\
 & \left.+\frac{28}{r^{2}}\left(m_{1}-4m_{2}\right)\mathbf{v}_{1}\cdot\mathbf{r}\left(\mathbf{v}_{1}\cdot\mathbf{r}-2\mathbf{v}_{2}\cdot\mathbf{r}\right)-\frac{1}{r^{2}}\left(56m_{1}+176m_{2}\right)\left(\mathbf{v}_{2}\cdot\mathbf{r}\right)^{2}\right]\nonumber \\
 & +2\mathbf{v}_{1}^{4}-16\left(\mathbf{v}_{1}\cdot\mathbf{v}_{2}\right)^{2}-16\mathbf{v}_{2}^{4}+32\mathbf{v}_{1}\cdot\mathbf{v}_{2}\mathbf{v}_{2}^{2}-2\mathbf{v}_{1}^{2}\mathbf{a}_{2}\cdot\mathbf{r}-2\mathbf{v}_{2}^{2}\mathbf{a}_{2}\cdot\mathbf{r}\nonumber \\
 & \left.-4\mathbf{a}_{2}\cdot\mathbf{v}_{2}\mathbf{v}_{2}\cdot\mathbf{r}+\frac{\left(\mathbf{v}_{2}\cdot\mathbf{r}\right)^{2}}{r^{2}}\left(12\mathbf{v}_{1}^{2}-48\mathbf{v}_{1}\cdot\mathbf{v}_{2}+36\mathbf{v}_{2}^{2}\right)-15\frac{\left(\mathbf{v}_{2}\cdot\mathbf{r}\right)^{4}}{r^{4}}\right\} \nonumber \\
 & +\frac{1}{4}\frac{Gm_{2}}{r^{3}}\mathbf{v}_{1}\left\{ \frac{G}{r}\left[\left(48m_{2}-15m_{1}\right)\mathbf{v}_{1}\cdot\mathbf{r}+\left(23m_{1}-40m_{2}\right)\mathbf{v}_{2}\cdot\mathbf{r}\right]\right.\nonumber \\
 & +\mathbf{v}_{2}\cdot\mathbf{r}\left(4\mathbf{v}_{1}^{2}+16\mathbf{v}_{1}\cdot\mathbf{v}_{2}-20\mathbf{v}_{2}^{2}\right)-24\frac{\mathbf{v}_{1}\cdot\mathbf{r}\left(\mathbf{v}_{2}\cdot\mathbf{r}\right)^{2}}{r^{2}}+18\frac{\left(\mathbf{v}_{2}\cdot\mathbf{r}\right)^{3}}{r^{2}}\nonumber \\
 & \left.+\mathbf{v}_{1}\cdot\mathbf{r}\left(8\mathbf{v}_{1}^{2}-16\mathbf{v}_{1}\cdot\mathbf{v}_{2}+16\mathbf{v}_{2}^{2}-2\mathbf{a}_{2}\cdot\mathbf{r}\right)+2r^{2}\left(12\mathbf{a}_{1}-7\mathbf{a}_{2}\right)\cdot\mathbf{v}_{1}\right\} \nonumber \\
 & +2\mathbf{a}_{1}\cdot\mathbf{v}_{1}\mathbf{v}_{1}^{2}\mathbf{v}_{1}+\frac{1}{4}\mathbf{a}_{1}\left(49\frac{G^{2}m_{1}m_{2}}{r^{2}}+36\frac{G^{2}m_{2}^{2}}{r^{2}}+12\frac{Gm_{2}}{r}\mathbf{v}_{1}^{2}+\mathbf{v}_{1}^{4}\right)\nonumber \\
 & +\frac{1}{4}\frac{Gm_{2}}{r^{3}}\mathbf{v}_{2}\left\{ \frac{G}{r}\left[\left(31m_{1}-24m_{2}\right)\mathbf{v}_{1}\cdot\mathbf{r}+\left(40m_{2}-9m_{1}\right)\mathbf{v}_{2}\cdot\mathbf{r}\right]\right.\nonumber \\
 & +\mathbf{v}_{2}\cdot\mathbf{r}\left(-4\mathbf{v}_{1}^{2}-16\mathbf{v}_{1}\cdot\mathbf{v}_{2}+20\mathbf{v}_{2}^{2}\right)+24\frac{\mathbf{v}_{1}\cdot\mathbf{r}\left(\mathbf{v}_{2}\cdot\mathbf{r}\right)^{2}}{r^{2}}-18\frac{\left(\mathbf{v}_{2}\cdot\mathbf{r}\right)^{3}}{r^{2}}\nonumber \\
 & \left.+\mathbf{v}_{1}\cdot\mathbf{r}\left(16\mathbf{v}_{1}\cdot\mathbf{v}_{2}-16\mathbf{v}_{2}^{2}\right)-14r^{2}\mathbf{a}_{2}\cdot\mathbf{v}_{2}\right\} \nonumber \\
 & -\frac{7}{4}\frac{Gm_{2}}{r}\mathbf{a}_{2}\left(6\frac{Gm}{r}+\mathbf{v}_{1}^{2}+\mathbf{v}_{2}^{2}\right).\label{eq:a2pn1}
\end{align}
All accelerations in the right hand side of the equality above should
be regarded as Newtonian accelerations if we want the entire expression
to be of definite 2PN order. To write the acceleration in the center
of mass frame, we have to consider, in addition to \eqref{eq:a2pn1},
the reduced contribution from applying the equation of motion inside
the \eqref{eq:a1pn1} as well as the PN corrections to the center
of mass frame \eqref{eq:x1} and \eqref{eq:x2}. Adding these contributions
together, we finally obtain the expression for the relative acceleration
of the two-body system in the center of mass frame, at the second
PN order, in the linearized harmonic gauge:
\begin{align}
\mathbf{a}_{2PN} & =-\frac{Gm}{8r^{3}}\left\{ \mathbf{r}\left[\left(56+174\nu\right)\frac{G^{2}m^{2}}{r^{2}}-\left(32+52\nu-16\nu^{2}\right)\frac{Gm}{r}v^{2}+\left(112-200\nu-16\nu^{2}\right)\frac{Gm}{r}\dot{r}^{2}\right.\right.\nonumber \\
 & \left.+\left(24\nu-32\nu^{2}\right)v^{4}-\left(36\nu-48\nu^{2}\right)v^{2}\dot{r}^{2}+\left(15\nu-45\nu^{2}\right)\dot{r}^{4}\right]\nonumber \\
 & \left.+4r\dot{r}\mathbf{v}\left[\left(-12+41\nu+8\nu^{2}\right)\frac{Gm}{r}-\left(15\nu+4\nu^{2}\right)v^{2}+\left(9\nu+6\nu^{2}\right)\dot{r}^{2}\right]\right\} . \label{eq:A2PN}
\end{align}


\bibliographystyle{unsrt}
\bibliography{Ref2PN}

\begin{thebibliography}{10}

\bibitem{Abbott:2016blz}
B.~P. Abbott et~al.
\newblock {Observation of Gravitational Waves from a Binary Black Hole Merger}.
\newblock {\em Phys. Rev. Lett.}, 116(6):061102, 2016.

\bibitem{2016htt}
B.~P. Abbott et~al.
\newblock {Astrophysical Implications of the Binary Black-Hole Merger
  GW150914}.
\newblock {\em Astrophys. J.}, 818(2):L22, 2016.

\bibitem{2016pea}
B.~P. Abbott et~al.
\newblock {Binary Black Hole Mergers in the first Advanced LIGO Observing Run}.
\newblock {\em Phys. Rev.}, X6(4):041015, 2016.

\bibitem{PhysRevLett.118.221101}
B.~P. Abbott et~al.
\newblock Gw170104: Observation of a 50-solar-mass binary black hole
  coalescence at redshift 0.2.
\newblock {\em Phys. Rev. Lett.}, 118:221101, Jun 2017.

\bibitem{PhysRevLett.119.141101}
B.~P. Abbott et~al.
\newblock Gw170814: A three-detector observation of gravitational waves from a
  binary black hole coalescence.
\newblock {\em Phys. Rev. Lett.}, 119:141101, Oct 2017.

\bibitem{PhysRevLett.119.161101}
B.~P. Abbott et~al.
\newblock Gw170817: Observation of gravitational waves from a binary neutron
  star inspiral.
\newblock {\em Phys. Rev. Lett.}, 119:161101, Oct 2017.

\bibitem{Abbott_2017}
B.~P. Abbott et~al.
\newblock {GW}170608: Observation of a 19 solar-mass binary black hole
  coalescence.
\newblock {\em The Astrophysical Journal}, 851(2):L35, dec 2017.

\bibitem{PhysRevX.9.031040}
B.~P. Abbott et~al.
\newblock Gwtc-1: A gravitational-wave transient catalog of compact binary
  mergers observed by ligo and virgo during the first and second observing
  runs.
\newblock {\em Phys. Rev. X}, 9:031040, Sep 2019.

\bibitem{Abbott_2017_mult}
B.~P. Abbott et~al.
\newblock Gravitational waves and gamma-rays from a binary neutron star merger:
  {GW}170817 and {GRB} 170817a.
\newblock {\em The Astrophysical Journal}, 848(2):L13, oct 2017.

\bibitem{Abbott_2017_multi}
B.~P. Abbott et~al.
\newblock Multi-messenger observations of a binary neutron star merger.
\newblock {\em The Astrophysical Journal}, 848(2):L12, oct 2017.

\bibitem{Abbott_2019}
B.~P. Abbott et~al.
\newblock Low-latency gravitational-wave alerts for multimessenger astronomy
  during the second advanced {LIGO} and virgo observing run.
\newblock {\em The Astrophysical Journal}, 875(2):161, apr 2019.

\bibitem{nrgr}
Walter~D. Goldberger and Ira~Z. Rothstein.
\newblock {An Effective field theory of gravity for extended objects}.
\newblock {\em Phys.Rev.}, D73:104029, 2006.

\bibitem{nrgrLH}
Walter~D. Goldberger.
\newblock {Les Houches lectures on effective field theories and gravitational
  radiation}.
\newblock In {\em {Les Houches Summer School - Session 86: Particle Physics and
  Cosmology: The Fabric of Spacetime Les Houches, France.}}, 2007.

\bibitem{Rothstein:2014sra}
Ira~Z. Rothstein.
\newblock {Progress in effective field theory approach to the binary inspiral
  problem}.
\newblock {\em Gen. Rel. Grav.}, 46:1726, 2014.

\bibitem{Foffa:2013qca}
Stefano Foffa and Riccardo Sturani.
\newblock {Effective field theory methods to model compact binaries}.
\newblock {\em Class. Quant. Grav.}, 31(4):043001, 2014.

\bibitem{Porto:2016pyg}
Rafael~A. Porto.
\newblock {{The effective field theorist's approach to gravitational
  dynamics}}.
\newblock {\em Phys. Rept.}, 633:1--104, 2016.

\bibitem{Levi:2018nxp}
Michele Levi.
\newblock {Effective Field Theories of Post-Newtonian Gravity: A comprehensive
  review}.
\newblock 2018.

\bibitem{Foffa:2019rdf}
Stefano Foffa and Riccardo Sturani.
\newblock {Conservative dynamics of binary systems to fourth Post-Newtonian
  order in the EFT approach I: Regularized Lagrangian}.
\newblock {\em Phys. Rev.}, D100(2):024047, 2019.

\bibitem{Foffa:2019yfl}
Stefano Foffa, Rafael~A. Porto, Ira Rothstein, and Riccardo Sturani.
\newblock {Conservative dynamics of binary systems to fourth Post-Newtonian
  order in the EFT approach II: Renormalized Lagrangian}.
\newblock {\em Phys. Rev.}, D100(2):024048, 2019.

\bibitem{Bini:2013zaa}
Donato Bini and Thibault Damour.
\newblock {Analytical determination of the two-body gravitational interaction
  potential at the fourth post-Newtonian approximation}.
\newblock {\em Phys. Rev.}, D87(12):121501, 2013.

\bibitem{Damour:2014jta}
Thibault Damour, Piotr Jaranowski, and Gerhard Schafer.
\newblock {Nonlocal-in-time action for the fourth post-Newtonian conservative
  dynamics of two-body systems}.
\newblock {\em Phys. Rev.}, D89(6):064058, 2014.

\bibitem{Bernard:2015njp}
Laura Bernard, Luc Blanchet, Alejandro Bohe, Guillaume Faye, and Sylvain
  Marsat.
\newblock {Fokker action of nonspinning compact binaries at the fourth
  post-Newtonian approximation}.
\newblock {\em Phys. Rev.}, D93(8):084037, 2016.

\bibitem{Bernard:2016wrg}
Laura Bernard, Luc Blanchet, Alejandro Bohe, Guillaume Faye, and Sylvain
  Marsat.
\newblock {Energy and periastron advance of compact binaries on circular orbits
  at the fourth post-Newtonian order}.
\newblock {\em Phys. Rev.}, D95(4):044026, 2017.

\bibitem{Porto:2010zg}
Rafael~A. Porto, Andreas Ross, and Ira~Z. Rothstein.
\newblock {Spin induced multipole moments for the gravitational wave flux from
  binary inspirals to third Post-Newtonian order}.
\newblock {\em JCAP}, 1103:009, 2011.

\bibitem{Porto:2012as}
Rafael~A. Porto, Andreas Ross, and Ira~Z. Rothstein.
\newblock {Spin induced multipole moments for the gravitational wave amplitude
  from binary inspirals to 2.5 Post-Newtonian order}.
\newblock {\em JCAP}, 1209:028, 2012.

\bibitem{Goldberger:2009qd}
Walter~D. Goldberger and Andreas Ross.
\newblock {Gravitational radiative corrections from effective field theory}.
\newblock {\em Phys. Rev.}, D81:124015, 2010.

\bibitem{Blanchet:2001ax}
Luc Blanchet, Guillaume Faye, Bala~R. Iyer, and Benoit Joguet.
\newblock {Gravitational wave inspiral of compact binary systems to 7/2
  postNewtonian order}.
\newblock {\em Phys. Rev.}, D65:061501, 2002.
\newblock [Erratum: Phys. Rev.D71,129902(2005)].

\bibitem{andirad2}
Andreas Ross.
\newblock {Multipole expansion at the level of the action}.
\newblock {\em Phys. Rev.}, D85:125033, 2012.

\bibitem{nrgr2pn}
James~B. Gilmore and Andreas Ross.
\newblock {Effective field theory calculation of second post-Newtonian binary
  dynamics}.
\newblock {\em Phys. Rev.}, D78:124021, 2008.

\bibitem{KS}
Barak Kol and Michael Smolkin.
\newblock Non-relativistic gravitation: from newton to einstein and back.
\newblock {\em Classical and Quantum Gravity}, 25(14):145011, jun 2008.

\bibitem{Porto:2017dgs}
Rafael~A. Porto and Ira~Z. Rothstein.
\newblock {Apparent ambiguities in the post-Newtonian expansion for binary
  systems}.
\newblock {\em Phys. Rev.}, D96(2):024062, 2017.

\bibitem{benira}
Benjamin Grinstein and Ira~Z. Rothstein.
\newblock {Effective field theory and matching in nonrelativistic gauge
  theories}.
\newblock {\em Phys. Rev.}, D57:78--82, 1998.

\bibitem{Cheung:2018wkq}
Clifford Cheung, Ira~Z. Rothstein, and Mikhail~P. Solon.
\newblock {From Scattering Amplitudes to Classical Potentials in the
  Post-Minkowskian Expansion}.
\newblock {\em Phys. Rev. Lett.}, 121(25):251101, 2018.

\bibitem{PhysRevD.86.044029}
Chad~R. Galley and Adam~K. Leibovich.
\newblock Radiation reaction at 3.5 post-newtonian order in effective field
  theory.
\newblock {\em Phys. Rev. D}, 86:044029, Aug 2012.

\bibitem{blanchet}
Luc Blanchet.
\newblock {Gravitational Radiation from Post-Newtonian Sources and Inspiralling
  Compact Binaries}.
\newblock {\em Living Rev.Rel.}, 17(1):2, 2014.

\bibitem{eih0}
A.~Einstein, L.~Infeld, and B.~Hoffman.
\newblock {Progress in effective field theory approach to the binary inspiral
  problem}.
\newblock {\em Annals Math.}, 39:65, 1938.

\bibitem{PhysRevD.97.044037}
Laura Bernard, Luc Blanchet, Guillaume Faye, and Tanguy Marchand.
\newblock Center-of-mass equations of motion and conserved integrals of compact
  binary systems at the fourth post-newtonian order.
\newblock {\em Phys. Rev. D}, 97:044037, Feb 2018.

\bibitem{kidder}
Lawrence~E. Kidder.
\newblock {Coalescing binary systems of compact objects to postNewtonian 5/2
  order. 5. Spin effects}.
\newblock {\em Phys. Rev.}, D52:821--847, 1995.

\bibitem{PhysRevD.54.4813}
Clifford~M. Will and Alan~G. Wiseman.
\newblock Gravitational radiation from compact binary systems: Gravitational
  waveforms and energy loss to second post-newtonian order.
\newblock {\em Phys. Rev. D}, 54:4813--4848, Oct 1996.

\bibitem{PhysRevD.56.7708}
A.~Gopakumar and Bala~R. Iyer.
\newblock Gravitational waves from inspiraling compact binaries: Angular
  momentum flux, evolution of the orbital elements, and the waveform to the
  second post-newtonian order.
\newblock {\em Phys. Rev. D}, 56:7708--7731, Dec 1997.

\bibitem{xact}
{J. M. Martin-Garcia}.
\newblock {xAct: Efficient tensor computer algebra for the Wolfram Language.
  https://www.xAct.es}.

\bibitem{Peskin:1995ev}
Michael~E. Peskin and Daniel~V. Schroeder.
\newblock {\em {An Introduction to quantum field theory}}.
\newblock Addison-Wesley, Reading, USA, 1995.

\end{thebibliography}

\end{document}